\documentclass[aps,prd,twocolumn,showpacs,superscriptaddress,groupedaddress]{revtex4}  
\usepackage{graphicx}
\usepackage{epsfig}
\usepackage{dcolumn}
\usepackage{bm}
\usepackage[usenames]{color} 
\usepackage{subfigure}
\usepackage{amsmath}
\usepackage{amssymb}
\usepackage{float}

\newcommand{\beq}{\begin{equation}}
\newcommand{\eeq}{\end{equation}}
\newcommand{\barr}{\begin{eqnarray}}
\newcommand{\earr}{\end{eqnarray}}

\def\apar{A_{\|}}
\def\aperp{A_{\perp}}

\def\bq{\begin{quote}}
\def\eq{\end{quote}}

\def\spose#1{\hbox to 0pt{#1\hss}}
\def\lsim{\mathrel{\spose{\lower 3pt\hbox{$\mathchar"218$}}
 \raise 2.0pt\hbox{$\mathchar"13C$}}}
\def\gsim{\mathrel{\spose{\lower 3pt\hbox{$\mathchar"218$}}
 \raise 2.0pt\hbox{$\mathchar"13E$}}}
 
\def \phis{$\phi_s^{J/\psi \phi}$}

\def\bs{${B_s^0}$}
\def\bd{${B^0_d}$}
\def\bsdec{${B_s^0 \rightarrow J/\psi \phi}$}

\def\bddec{${B_d^0 \rightarrow J/\psi K^*}$}

\def\D0{D\O }

\def\GeVp{ {\ifmmode \;{{\mbox{\mathrm GeV}} / {\mbox\mathrm c}} \else
${{\mbox{\mathrm GeV}} / {\mbox\mathrm c}}$ \fi }}
\def\MeVp{ {\ifmmode \;{{\mbox{\mathrm MeV}} / {\mbox\mathrm c}} \else
${{\mbox{\mathrm MeV}} / {\mbox\mathrm c}}$ \fi }}
\def\MeV{ {\ifmmode \;{{\mbox{\mathrm MeV}} / {\mbox\mathrm c}^2} \else
${{\mbox{\mathrm MeV}} / {\mbox\mathrm c}^2}$ \fi }}
\def\GeV{ {\ifmmode \;{{\mbox{\mathrm GeV}} / {\mbox\mathrm c}^2} \else
${{\mbox{\mathrm GeV}} / {\mbox\mathrm c}^2}$ \fi }}

\begin{document}

\hspace{5.2in} \mbox{FERMILAB-PUB-11-463-E}

\title{ Measurement of the {\boldmath $CP$}-violating  phase  {\boldmath $\phi_s^{J/\psi \phi}$}
using the flavor-tagged decay {\boldmath ${B_s^0 \rightarrow J/\psi \phi}$}
in 8 fb{\boldmath $^{-1}$} of  {\boldmath $p \overline p$} collisions }   

\date{September 14, 2011}
\affiliation{Universidad de Buenos Aires, Buenos Aires, Argentina}
\affiliation{LAFEX, Centro Brasileiro de Pesquisas F{\'\i}sicas, Rio de Janeiro, Brazil}
\affiliation{Universidade do Estado do Rio de Janeiro, Rio de Janeiro, Brazil}
\affiliation{Universidade Federal do ABC, Santo Andr\'e, Brazil}
\affiliation{Instituto de F\'{\i}sica Te\'orica, Universidade Estadual Paulista, S\~ao Paulo, Brazil}
\affiliation{University of Science and Technology of China, Hefei, People's Republic of China}
\affiliation{Universidad de los Andes, Bogot\'{a}, Colombia}
\affiliation{Charles University, Faculty of Mathematics and Physics, Center for Particle Physics, Prague, Czech Republic}
\affiliation{Czech Technical University in Prague, Prague, Czech Republic}
\affiliation{Center for Particle Physics, Institute of Physics, Academy of Sciences of the Czech Republic, Prague, Czech Republic}
\affiliation{Universidad San Francisco de Quito, Quito, Ecuador}
\affiliation{LPC, Universit\'e Blaise Pascal, CNRS/IN2P3, Clermont, France}
\affiliation{LPSC, Universit\'e Joseph Fourier Grenoble 1, CNRS/IN2P3, Institut National Polytechnique de Grenoble, Grenoble, France}
\affiliation{CPPM, Aix-Marseille Universit\'e, CNRS/IN2P3, Marseille, France}
\affiliation{LAL, Universit\'e Paris-Sud, CNRS/IN2P3, Orsay, France}
\affiliation{LPNHE, Universit\'es Paris VI and VII, CNRS/IN2P3, Paris, France}
\affiliation{CEA, Irfu, SPP, Saclay, France}
\affiliation{IPHC, Universit\'e de Strasbourg, CNRS/IN2P3, Strasbourg, France}
\affiliation{IPNL, Universit\'e Lyon 1, CNRS/IN2P3, Villeurbanne, France and Universit\'e de Lyon, Lyon, France}
\affiliation{III. Physikalisches Institut A, RWTH Aachen University, Aachen, Germany}
\affiliation{Physikalisches Institut, Universit{\"a}t Freiburg, Freiburg, Germany}
\affiliation{II. Physikalisches Institut, Georg-August-Universit{\"a}t G\"ottingen, G\"ottingen, Germany}
\affiliation{Institut f{\"u}r Physik, Universit{\"a}t Mainz, Mainz, Germany}
\affiliation{Ludwig-Maximilians-Universit{\"a}t M{\"u}nchen, M{\"u}nchen, Germany}
\affiliation{Fachbereich Physik, Bergische Universit{\"a}t Wuppertal, Wuppertal, Germany}
\affiliation{Panjab University, Chandigarh, India}
\affiliation{Delhi University, Delhi, India}
\affiliation{Tata Institute of Fundamental Research, Mumbai, India}
\affiliation{University College Dublin, Dublin, Ireland}
\affiliation{Korea Detector Laboratory, Korea University, Seoul, Korea}
\affiliation{CINVESTAV, Mexico City, Mexico}
\affiliation{Nikhef, Science Park, Amsterdam, the Netherlands}
\affiliation{Radboud University Nijmegen, Nijmegen, the Netherlands and Nikhef, Science Park, Amsterdam, the Netherlands}
\affiliation{Joint Institute for Nuclear Research, Dubna, Russia}
\affiliation{Institute for Theoretical and Experimental Physics, Moscow, Russia}
\affiliation{Moscow State University, Moscow, Russia}
\affiliation{Institute for High Energy Physics, Protvino, Russia}
\affiliation{Petersburg Nuclear Physics Institute, St. Petersburg, Russia}
\affiliation{Instituci\'{o} Catalana de Recerca i Estudis Avan\c{c}ats (ICREA) and Institut de F\'{i}sica d'Altes Energies (IFAE), Barcelona, Spain}
\affiliation{Stockholm University, Stockholm and Uppsala University, Uppsala, Sweden}
\affiliation{Lancaster University, Lancaster LA1 4YB, United Kingdom}
\affiliation{Imperial College London, London SW7 2AZ, United Kingdom}
\affiliation{The University of Manchester, Manchester M13 9PL, United Kingdom}
\affiliation{University of Arizona, Tucson, Arizona 85721, USA}
\affiliation{University of California Riverside, Riverside, California 92521, USA}
\affiliation{Florida State University, Tallahassee, Florida 32306, USA}
\affiliation{Fermi National Accelerator Laboratory, Batavia, Illinois 60510, USA}
\affiliation{University of Illinois at Chicago, Chicago, Illinois 60607, USA}
\affiliation{Northern Illinois University, DeKalb, Illinois 60115, USA}
\affiliation{Northwestern University, Evanston, Illinois 60208, USA}
\affiliation{Indiana University, Bloomington, Indiana 47405, USA}
\affiliation{Purdue University Calumet, Hammond, Indiana 46323, USA}
\affiliation{University of Notre Dame, Notre Dame, Indiana 46556, USA}
\affiliation{Iowa State University, Ames, Iowa 50011, USA}
\affiliation{University of Kansas, Lawrence, Kansas 66045, USA}
\affiliation{Kansas State University, Manhattan, Kansas 66506, USA}
\affiliation{Louisiana Tech University, Ruston, Louisiana 71272, USA}
\affiliation{Boston University, Boston, Massachusetts 02215, USA}
\affiliation{Northeastern University, Boston, Massachusetts 02115, USA}
\affiliation{University of Michigan, Ann Arbor, Michigan 48109, USA}
\affiliation{Michigan State University, East Lansing, Michigan 48824, USA}
\affiliation{University of Mississippi, University, Mississippi 38677, USA}
\affiliation{University of Nebraska, Lincoln, Nebraska 68588, USA}
\affiliation{Rutgers University, Piscataway, New Jersey 08855, USA}
\affiliation{Princeton University, Princeton, New Jersey 08544, USA}
\affiliation{State University of New York, Buffalo, New York 14260, USA}
\affiliation{Columbia University, New York, New York 10027, USA}
\affiliation{University of Rochester, Rochester, New York 14627, USA}
\affiliation{State University of New York, Stony Brook, New York 11794, USA}
\affiliation{Brookhaven National Laboratory, Upton, New York 11973, USA}
\affiliation{Langston University, Langston, Oklahoma 73050, USA}
\affiliation{University of Oklahoma, Norman, Oklahoma 73019, USA}
\affiliation{Oklahoma State University, Stillwater, Oklahoma 74078, USA}
\affiliation{Brown University, Providence, Rhode Island 02912, USA}
\affiliation{University of Texas, Arlington, Texas 76019, USA}
\affiliation{Southern Methodist University, Dallas, Texas 75275, USA}
\affiliation{Rice University, Houston, Texas 77005, USA}
\affiliation{University of Virginia, Charlottesville, Virginia 22901, USA}
\affiliation{University of Washington, Seattle, Washington 98195, USA}
\author{V.M.~Abazov} \affiliation{Joint Institute for Nuclear Research, Dubna, Russia}
\author{B.~Abbott} \affiliation{University of Oklahoma, Norman, Oklahoma 73019, USA}
\author{B.S.~Acharya} \affiliation{Tata Institute of Fundamental Research, Mumbai, India}
\author{M.~Adams} \affiliation{University of Illinois at Chicago, Chicago, Illinois 60607, USA}
\author{T.~Adams} \affiliation{Florida State University, Tallahassee, Florida 32306, USA}
\author{G.D.~Alexeev} \affiliation{Joint Institute for Nuclear Research, Dubna, Russia}
\author{G.~Alkhazov} \affiliation{Petersburg Nuclear Physics Institute, St. Petersburg, Russia}
\author{A.~Alton$^{a}$} \affiliation{University of Michigan, Ann Arbor, Michigan 48109, USA}
\author{G.~Alverson} \affiliation{Northeastern University, Boston, Massachusetts 02115, USA}
\author{G.A.~Alves} \affiliation{LAFEX, Centro Brasileiro de Pesquisas F{\'\i}sicas, Rio de Janeiro, Brazil}
% ...
%
%
\collaboration{The D0 Collaboration\footnote{with visitors from
$^{a}$Augustana College, Sioux Falls, SD, USA,
$^{b}$The University of Liverpool, Liverpool, UK,
$^{c}$UPIITA-IPN, Mexico City, Mexico,
$^{d}$SLAC, Menlo Park, CA, USA,
$^{e}$University College London, London, UK,
$^{f}$Centro de Investigacion en Computacion - IPN, Mexico City, Mexico,
$^{g}$ECFM, Universidad Autonoma de Sinaloa, Culiac\'an, Mexico,
and 
$^{h}$Universit{\"a}t Bern, Bern, Switzerland.
$^{\ddag}$Deceased.
}} \noaffiliation
\vskip 0.25cm

\begin{abstract}
We report an updated measurement of the {\sl CP}-violating phase, \phis,
and the  decay-width difference for the two mass eigenstates, $\Delta \Gamma_s$,
from the flavor-tagged decay ${B_s^0 \rightarrow J/\psi \phi}$.
The data sample corresponds 
to an integrated luminosity of 8.0 fb$^{-1}$
accumulated with the D0 detector using $p \overline p$ collisions at $\sqrt{s} = 1.96$ TeV produced 
at the Fermilab Tevatron collider.
The 68\% bayesian credibility intervals, including systematic uncertainties, are
$\Delta \Gamma_s =  0.163  ^{+0.065} _{-0.064}$ ps$^{-1}$ and
 $\phi_s^{J/\psi \phi} =  -0.55 ^{+0.38} _{-0.36}$. 
The $p$-value for the Standard Model point 
is 29.8\%.
\end{abstract}
\pacs{13.25.Hw, 11.30.Er}

\maketitle

\section{\label{sec:intro}Introduction}

The meson-antimeson mixing
and the phenomenon of charge-conjugation-parity {\sl (CP)} violation 
in  neutral 
mesons systems are  key problems of particle physics.
In the standard model (SM), the light ($L$) and heavy ($H$) mass eigenstates 
of the  $B_s^0$ system  are expected
to have  sizeable 
mass and decay width differences: $\Delta M_s \equiv M_H - M_L$ and
$\Delta \Gamma_s \equiv \Gamma_L - \Gamma_H$. 
The two mass eigenstates are expected to be almost pure {\sl CP}
eigenstates.  
The {\sl CP}-violating  phase
that appears in $b \rightarrow c \overline c  s$ decays is
due to the interference of the decay with and without mixing, and  it
is predicted~\cite{UTfit}
to be $\phi_s^{J/\psi \phi} = -2\beta_s^{SM} = 2\arg[-V_{tb}V^*_{ts}/V_{cb}V^*_{cs}]
 = -0.038 \pm 0.002$, where
$V_{ij}$ are elements of the  Cabibbo-Kobayashi-Maskawa 
quark-mixing matrix~\cite{ckm}. 
New phenomena~\cite{NPmkal,NPdrobnak,NPwang,NPaka,NPjs,NPajaltouni,NPnandisoni,NPdatta,NPjg,NPab1,NPzl,NPab2,NPbd,NPjps,NPab3,NPnelson,NPgi,NPaa,NPlle,NPfb,NPcwc} 
 may alter the observed phase
to $\phi_s^{J/\psi \phi} \equiv -2\beta_s \equiv -2\beta_s^{SM} +\phi_s^\Delta$.
A significant deviation of \phis\ from its small SM value would
indicate the presence of processes beyond SM.

The analysis of the decay chain \bsdec, $J/\psi \rightarrow \mu ^+ \mu ^-$, $\phi \rightarrow K^+ K^-$
separates the {\sl CP}-even and {\sl CP}-odd states using the
angular distributions of the decay products.
It is a unique feature of the decay \bsdec\ that because of the
sizeable lifetime difference between the two mass eigenstates,
there is a sensitivity  to $\phi_s^{J/\psi \phi}$ even in the absence of the flavor tagging information.
The first direct constraint on \phis\ ~\cite{prl07,cdf08} was derived by analysing 
${B_s^0 \rightarrow J/\psi \phi}$ decays where the flavor
(i.e., $B^0_s$ or $\overline{B}^0_s$) at the time of production was not 
determined (``tagged'').
It was followed by an improved analysis~\cite{prl08},
based on  2.8 fb$^{-1}$  of integrated luminosity,
that included the  information on the $B_s^0$ flavor  at production.
In addition, the CDF collaboration has performed a measurement~\cite{cdfprl} of \phis\
using 1.35  fb$^{-1}$ of data. After the submission of this Article,
new measurements of the $CP$ violation parameters in the 
$B^0_s\to J/\psi \phi$ decay have been published by the CDF~\cite{cdf2011} and
the LHCb~\cite{lhcb2011} Collaborations.

In this Article,  we present new results from the time-dependent
amplitude analysis of the decay \bsdec\  using a
data sample corresponding to an integrated luminosity 
of 8.0 fb$^{-1}$ collected  with the  D0 detector~\cite{run2det}
at the Fermilab Tevatron Collider.  In addition to the increase in the size 
of the data sample used in the analysis, we also take into account
the $\cal{S}$-wave $K^+K^-$ under the $\phi$ peak that has been 
suggested~\cite{Stone} to contribute between 5-10\%.
We measure $\Delta \Gamma_s$;
the average lifetime of the \bs\ system, 
$\overline \tau_s =1/\overline \Gamma_s$, where
$\overline\Gamma_s \equiv(\Gamma_H+\Gamma_L)/2$; and
the {\sl CP}-violating phase \phis. 
Section~\ref{sec:detector} briefly describes the D0 detector.
Section~\ref{sec:event} presents the event reconstruction and the data set.
Sections~\ref{sec:multivar} and ~\ref{sec:flavor} describe the event selection 
requirements and the procedure of determining the
flavor of the initial state of the $B_s^0$ candidate.
In Sec.~\ref{sec:fitting} we describe the analysis formalism and the fitting 
method, present fit results, and discuss systematic uncertainties
in the results. We obtain the bayesian credibility intervals
for physics parameters using a procedure based on the 
 Markov Chain Monte Carlo (MCMC) technique, presented in Sec.~\ref{sec:mcmc}.
We summarize and discuss the results in Sec.~\ref{sec:conclusions}.

\section{\label{sec:detector}Detector}

  The D0 detector consists of a central tracking system, calorimetry system and
  muon detectors, as detailed in Refs.~\cite{run2det,layer0,muon}. The central
  tracking system comprises  a silicon microstrip tracker (SMT) and a central
   fiber tracker (CFT), both located inside a 1.9~T superconducting solenoidal
   magnet.  The tracking system is designed to optimize tracking and vertexing
   for pseudorapidities $|\eta|<3$,
   where  $\eta = -\ln[\tan(\Theta/2)]$, and  $\Theta$ is the 
    polar angle with respect to the proton beam direction.
   
   The SMT can reconstruct the $p\overline{p}$ interaction vertex (PV) 
   for interactions   with at least three tracks with a precision
   of 40 $\mu$m in the plane transverse to the beam direction and determine
   the impact parameter of any track relative to the PV with a precision between
    20 and 50 $\mu$m, depending on the the number of hits in the SMT.
    
   
   The muon detector is positioned outside the calorimeter.
   It consists of a central muon system covering the pseudorapidity region $|\eta|<1$ 
   and a forward muon system covering the pseudorapidity region $1<|\eta|<2$.
   Both central and forward systems consist of a layer of drift tubes
   and scintillators inside
   1.8~T toroidal magnets and two similar layers outside the toroids.
   
   The trigger and data acquisition systems are designed to accommodate the high 
   instantaneous luminosities of  Tevatron Run II.

\section{\label{sec:event} Data Sample and Event Reconstruction}

The analysis presented here is based on data accumulated
between February 2002 and June 2010.
Events are collected with a mixture of single- and dimuon triggers. Some
triggers require track displacement with respect to the primary vertex 
(large track impact parameter). Since this condition biases the $B^0_s$ 
lifetime measurement, the events selected exclusively by these triggers are 
removed from our sample.


Candidate 
$B^0_s \rightarrow J/\psi \phi$, $J/\psi \rightarrow \mu^+ \mu^-$, $\phi\rightarrow K^+K^-$ events
are required to include two opposite charge muons accompanied by two
opposite charge tracks.
Both muons are required to be detected in the muon chambers inside the toroid magnet,
and at least one of the muons is required to be also detected
outside the toroid.
 Each of the four final-state tracks 
is required to have at least one SMT hit.

To form $B_s^0$ candidates, muon pairs 
 in the invariant mass range $3.096 \pm 0.350$ GeV, consistent with $J/\psi$ decay,
are combined with pairs of opposite charge tracks (assigned the kaon mass) consistent with production at a common vertex, and with an invariant mass in the range $1.019 \pm 0.030$ GeV.
A  kinematic fit under the
$B_s^0$ decay hypothesis  constrains 
the dimuon invariant 
mass to the world-average   $J/\psi$ mass~\cite{PDG}
and  constrains the four-track system to a common vertex.
 
Trajectories of the four $B_s^0$ decay products are
adjusted according to the decay-vertex kinematic fit.
The re-adjusted track parameters are used in the
calculation of  the $B_s^0$ candidate mass and  
decay time, and  of the three angular variables characterising 
the decay as defined later. 
$B_s^0$ candidates are required to
have  an invariant  mass in the range $5.37 \pm 0.20$ GeV.
In events where multiple candidates satisfy these requirements, we select
the candidate with the best decay vertex fit probability.

To reconstruct the PV, we select tracks that 
do not originate from the candidate $B_s^0$ decay, 
and apply a constraint to the average beam-spot position in the transverse plane.
We define the signed  decay length of a \bs\ meson, $L^B_{xy}$, 
as the vector pointing
from the PV to the decay vertex, projected on the
\bs\ transverse momentum $p_T$.
The proper decay time of a \bs\ candidate is given by
 $t = M_{B_s}\vec L_{xy}^B \cdot \vec{p}/(p_T^2)$
where $M_{B_s}$ is the world-average \bs\ mass~\cite{PDG},
and $\vec p$ is the particle momentum.
The distance 
in the beam direction between the PV
and the $B_s^0$ vertex is required to be less than 5 cm. 
Approximately 5 million events are accepted after the selection described in this
section. 

\section{\label{sec:multivar} Background Suppression}

The selection criteria are designed to optimimize the measurement of
$\phi_s^{J/\psi \phi}$ and $\Delta \Gamma_s$.
Most of the background is  due to directly 
produced $J/\psi$ mesons accompanied by  tracks arising from 
hadronization.  This ``prompt'' background is distinguished from 
the ``non-prompt'', or ``inclusive $B \rightarrow J/\psi +X$''  background, 
where the $J/\psi$ meson is a product 
of a $b$-hadron decay  while the tracks forming the $\phi$ candidate 
emanate from a multi-body  decay of a $b$ hadron or from hadronization.
Two different event selection approaches are used, one based on a multi-variate
technique, and one based on simple limits on kinematic and event quality
parameters.

\subsection{ Signal and background simulation}
   Three Monte Carlo simulated samples are used to study background suppression: 
signal, prompt background, and non-prompt background. All three
  are generated with 
{\sc pythia}~\cite{pythia}. Hadronization is also done in {\sc pythia}, but
all hadrons carrying
 heavy flavors are passed on to {\sc EvtGen}~\cite{evtgen}  to
model their decays.
   The prompt background MC sample consists of $J/\psi \to \mu^+ \mu^-$ 
decays produced in  $gg \to J/\psi g$, $gg \to J/\psi \gamma$, 
and $g \gamma \to J/\psi g$ processes.
The signal and non-prompt background samples are generated from primary
  $b \bar{b}$  pair production with all $b$ hadrons being produced inclusively and
the $J/\psi$ mesons forced into $\mu^+ \mu^-$  decays. For the signal sample, events
with a \bs\  are selected,  their  decays to $J/\psi \phi$
are implemented without
mixing and with uniform angular distributions, and the \bs\ mean 
lifetime is set to  $\overline \tau_s$ = 1.464 ps. 
There are approximately 10$^6$ events in each background and the signal MC samples.
All events are passed through a full  {\sc geant}-based~\cite{Geant}
detector simulation. To take into account the effects of multiple
interactions at high luminosity, hits from randomly triggered
$p\bar{p}$ collisions are overlayed on the digitized hits from MC.
These  events are reconstructed with the same program as used for
data.
The  three  samples are corrected so that the $p_T$ distributions of the final state  
particles in $B^0_s \to J/\psi \phi$ decays match those in data (see Appendix~\ref{appmcdata}).   

\subsection{Multivariate event selection}

To discriminate the signal from background events, we use the TMVA 
package~\cite{refroot}.
In preliminary studies using MC simulation, the
Boosted Decision Tree (BDT) algorithm was found to demonstrate the best performance.
Since prompt and non-prompt backgrounds have different kinematic behavior, 
we train two discriminants, one for each type of background.
We use  a set of 33 variables for the prompt background
and 35 variables  for the non-prompt background. The variables and more details of 
the BDT method are given in Appendix \ref{appbdt}.

\begin{figure}[h!tb]
\begin{center}
 \includegraphics*[width=0.4\textwidth]{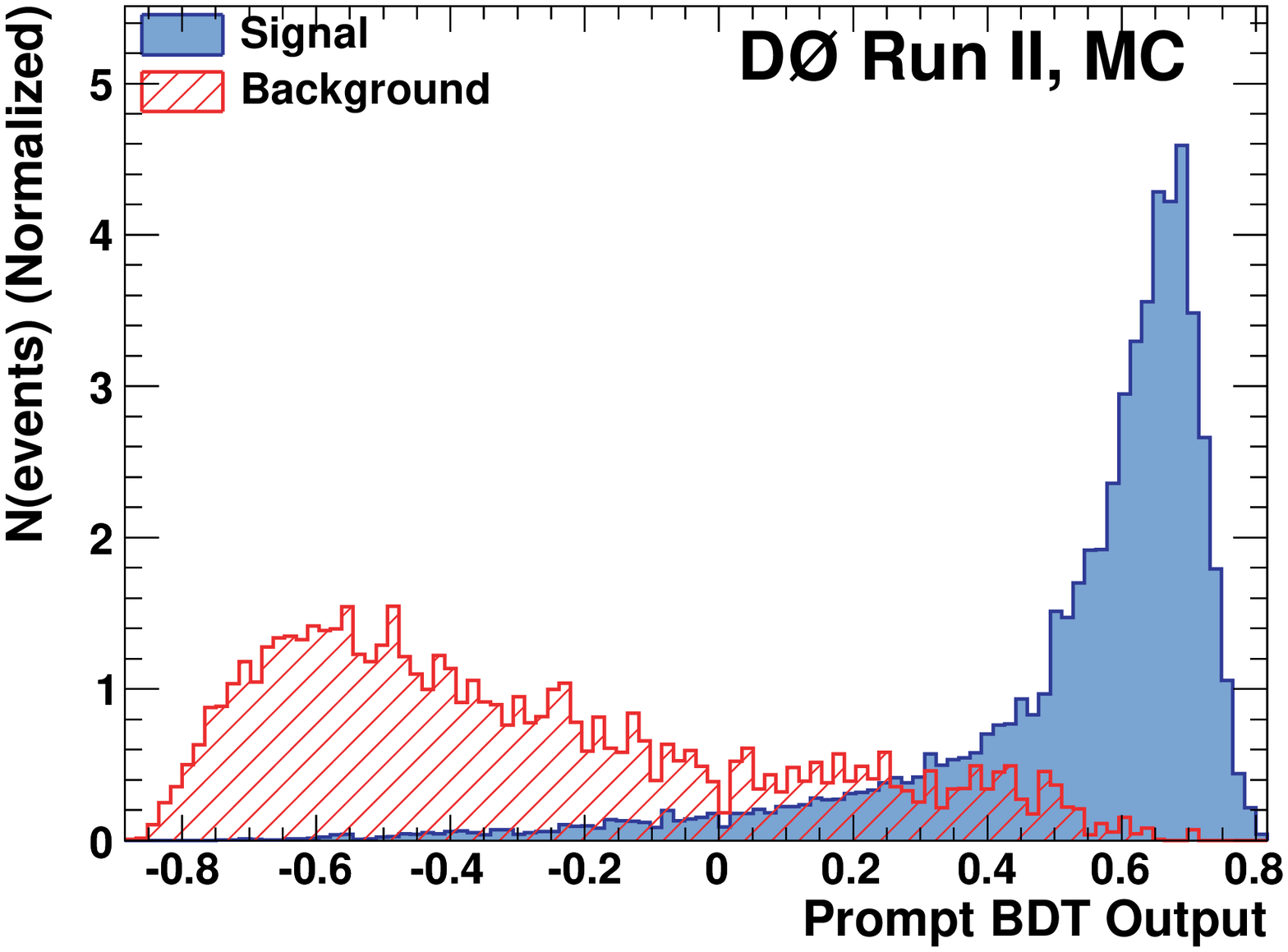}
 \includegraphics*[width=0.4\textwidth]{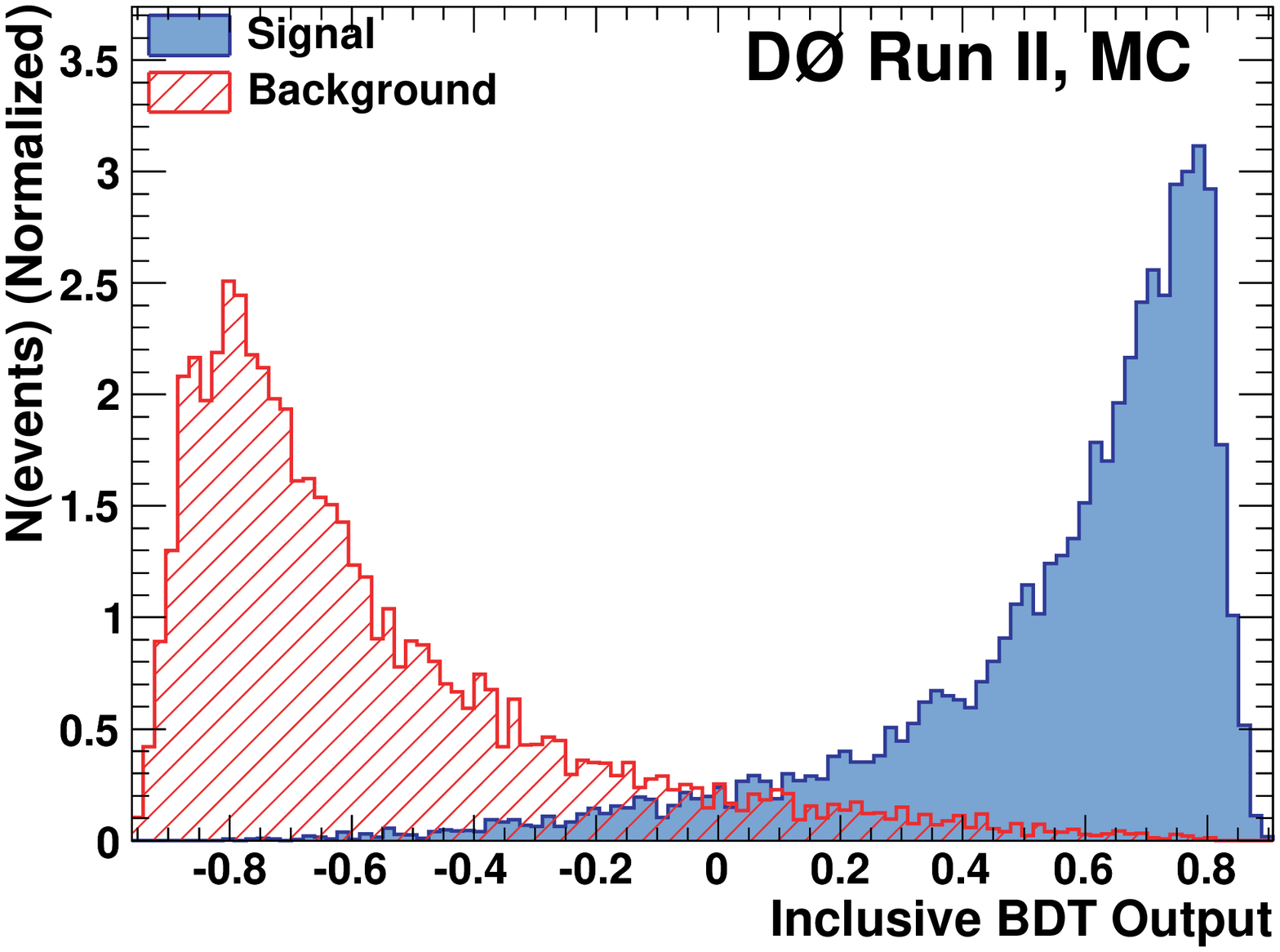}
 \caption{(color online). BDT discriminant output for the  prompt (top)
and  non-prompt (bottom) classifiers.
The signal and background events are taken from simulation.
Events used for BDT training are excluded from these samples.
  }
\label{bdt_prompt}
\end{center}
\end{figure}

The BDT training is performed using a subset of the MC samples, and the 
remaining events are used to test the training. The signal MC sample has 
about 84k events, the prompt background has 
29k events, 
and the non-prompt background has 39k events.  Figure~\ref{bdt_prompt} 
shows the BDT output discriminant for 
the prompt and non-prompt cases. 


\subsection{ Selection Criteria}

\begin{figure}[h!tb]
\begin{center}
 \includegraphics*[width=0.42\textwidth]{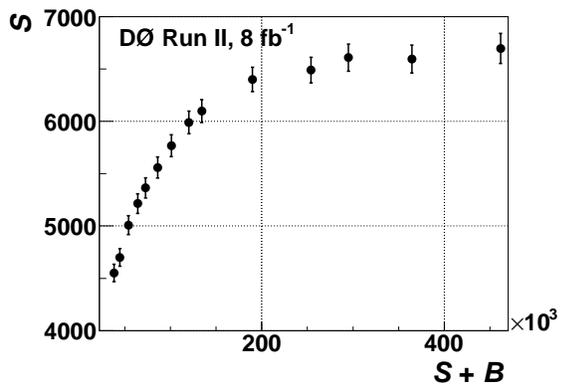}
 \caption{Number of \bsdec\ signal events as a function of the 
total number of events for the 14 criteria sets considered.
}
\label{fig:sigvstot}
\end{center}
\end{figure}

To choose the best set of criteria for the two BDT discriminants, we first step 
through the values of both BDT discriminants from $-0.4$ to 0.8 in increments 
of 0.01 and measure  the $B^0_s$ signal yield for each choice of cuts.
Next, we define 14 signal yield regions between 4000 and 7000 events,
and for each region choose the pair of BDT cuts which gives the highest
significance  $S/\sqrt{S+B}$, where $S$ ($B$) is the number of signal (background)
events in the data sample. 
 The 14 points, in increasing order of the 
signal size $S$, are shown in Table \ref{tPreliCuts}. Figure~\ref{fig:sigvstot} shows the 
number of signal events as a function of the total number of events for the 14 points. As the BDT 
criteria are loosened, the total number of events increases by a factor of ten, while the number of 
signal events increases by about 50\%.

As a test of possible detrimental effects of training on
variables with low separation power,
we have repeated the above procedure using only the variables
whose importance (see Appendix \ref{appbdt}) exceeds 0.01, giving
18 variables for the prompt background and 13 variables for the
non-prompt background.
The resulting number of background events for a given number
of signal events  is larger by about  10\%. Therefore, we proceed with 
the original number of variables.

\begin{table}[hb!]
 \begin{tabular}{ccccc} \hline \hline
Criteria  & $S$ & $S+B$  & Non-prompt  & ~~Prompt   \\ 

Set     &     &     &        BDT  &  BDT  \\ \hline

0  & 4550   &  ~~38130   &   ~~0.45   &  ~~0.42  \\  
1  & 4699   &  ~~44535   &   ~~0.45   &  ~~0.29  \\  
2  & 5008   &  ~~53942   &   ~~0.39   &  ~~0.35  \\  
3  & 5213   &  ~~64044   &   ~~0.36   &  ~~0.30  \\  
4  & 5364   &  ~~72602   &   ~~0.33   &  ~~0.28  \\  
5  & 5558   &  ~~85848   &   ~~0.13   &  ~~0.41  \\  
6  & 5767   &  ~~100986   &   ~~0.21   &  ~~0.29  \\  
7  & 5988   &  ~~120206   &   ~~0.13   &  ~~0.29  \\  
8  & 6097   &  ~~134255   &   ~~0.07   &  ~~0.29  \\  
9  & 6399   &  ~~189865   &   ~~0.04   &  ~~0.10  \\  
10  & 6489   &  ~~254022   &  $ -0.05$   & $ -0.01$  \\  
11  & 6608   &  ~~294949   &  $ -0.13$   &  ~~0.00  \\  
12  & 6594   &  ~~364563   &   $-0.18$   &  $-0.14$  \\  
13  & 6695   &  ~~461744   &   $-0.35$   &  $-0.08$  \\   \hline \hline

 \end{tabular}\caption{Numbers of signal and signal-plus-background 
events for different sets of BDT criteria, shown in the last two columns,
that give the largest value of 
$S/\sqrt{S+B}$ for a given $S$.  }
\label{tPreliCuts}
\end{table}

\begin{figure}[h!tb]
\begin{center}
\subfigure[]
 {\includegraphics*[width=0.37\textwidth]{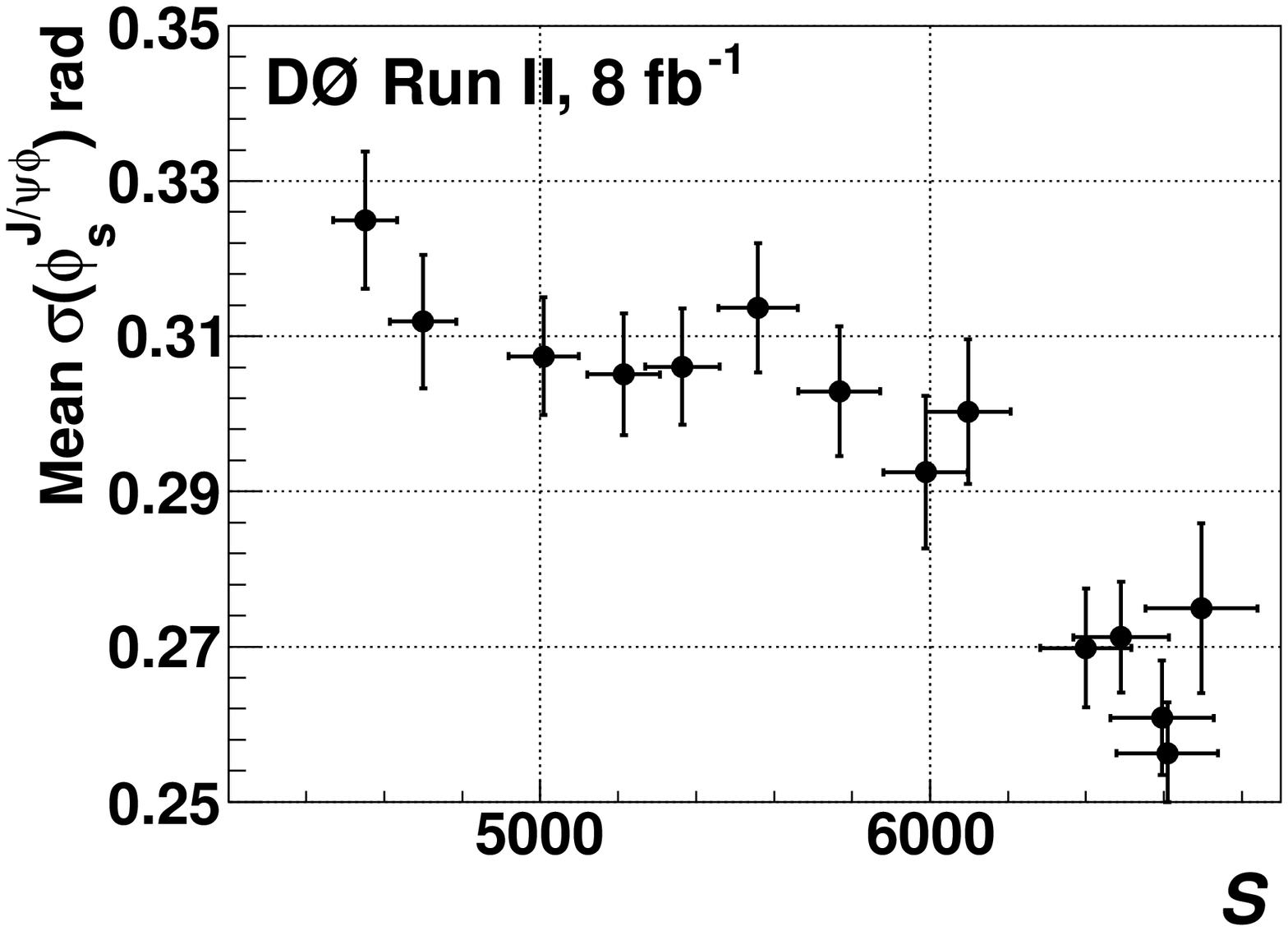}}
\subfigure[]
{\includegraphics*[width=0.37\textwidth]{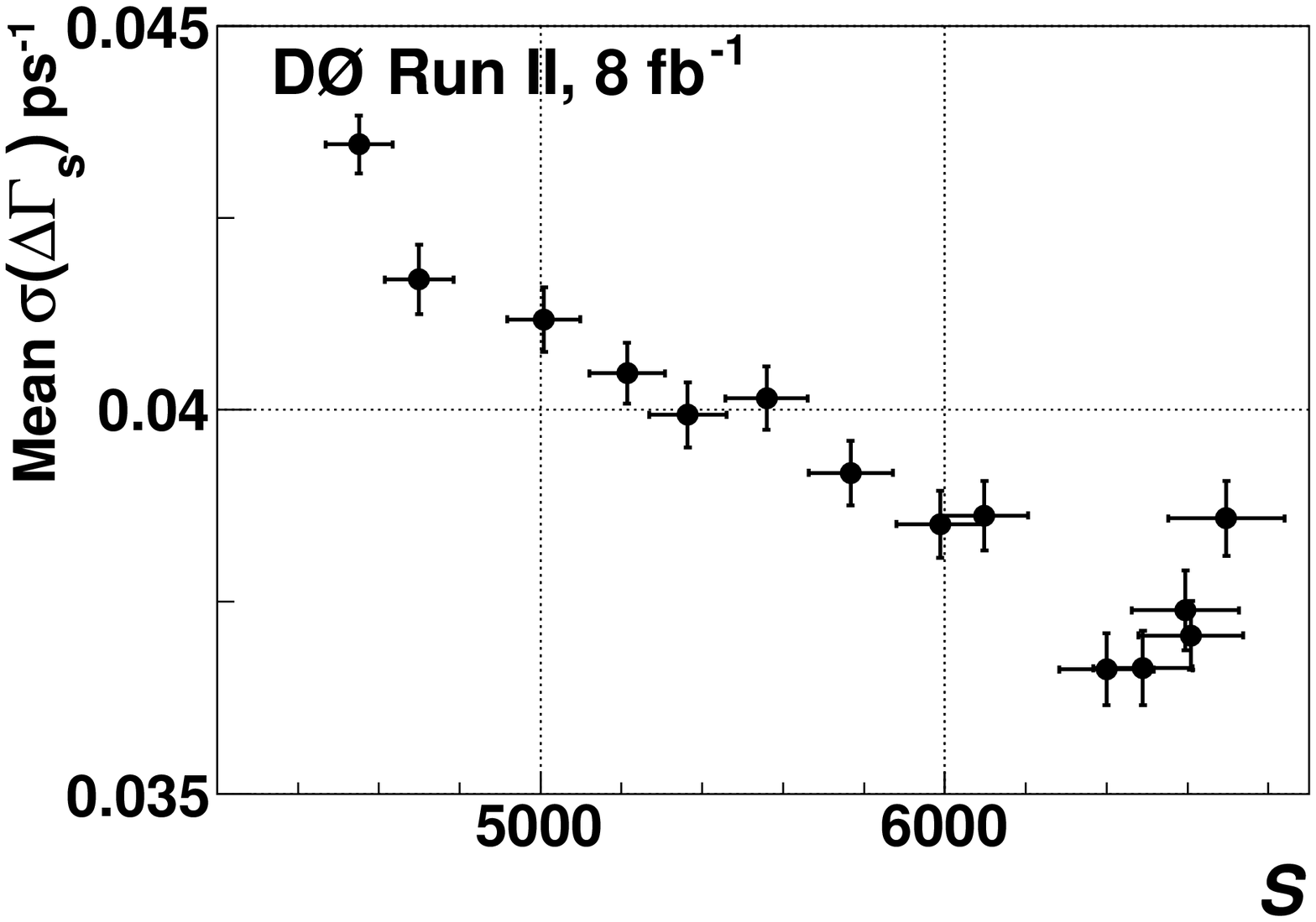}}
 \caption{Ensemble study results for (a) mean value of $\sigma(\phi_{s})$ 
   as a function of the number of signal events and (b) mean value of
   $\sigma(\Delta\Gamma_{s})$ as a function of the 
number of signal events. }
\label{fig:phivssig}
\end{center}
\end{figure}

The choice of the final cut on the BDT output is based on an ensemble study.  For each 
point in Table~\ref{tPreliCuts}, we perform a maximum-likelihood fit to the event 
distribution in the 2-dimensional (2D) space of $B_s^0$ candidate mass and 
proper time. This 2D fit provides a 
parametrization of the background mass and proper time distribution. We then 
generate pseudo-experiments in the 5D space of $B_s^0$ candidate mass, proper time, and 
three independent angles of decay products, 
using as input the parameters as obtained 
in a preliminary study, 
and the background from the 2D fit. 
We perform a 5D maximum likelihood fit on the ensembles and compare the  
distributions of the statistical uncertainties of
 $\phi_s^{J/\psi\phi}$ ($\sigma(\phi_s^{J/\psi \phi})$) 
and $\Delta \Gamma_s$ ($\sigma(\Delta \Gamma_s)$) for the different sets of criteria. 
The dependence of the mean values of $\sigma(\phi_s^{J/\psi \phi})$ and $\sigma(\Delta 
\Gamma_s)$ on the number of signal events is shown in Figs.~\ref{fig:phivssig}(a) and 
\ref{fig:phivssig}(b). The mean statistical uncertainties of both $\phi_s^{J/\psi \phi}$ and 
$\Delta \Gamma_s$ systematically decrease with increasing signal, favoring looser 
cuts. The gain in the parameter resolution is slower for the three loosest criteria, while 
the total number of events doubles from about 0.25$\times 10^6$ to 0.5$\times 10^6$.
The fits used for these ensemble tests were simplified, 
therefore the magnitude of the predicted uncertainty is expected to underestimate the
final measured precision. However, the general trends should be valid.

 Based on 
these results, we choose the sample that contains about 6500 signal events, (labeled 
``Set 10'' in Table~\ref{tPreliCuts}) as a final selection and refer to it as  
the ``BDT selection''. 
Figure \ref{eff} in Appendix \ref{appbdt} shows the ratios of the
normalized distributions of the 
three angles (see Section \ref{sec:fitting}) and the lifetime
before and after the BDT selection.
The ratios are consistent with unity, which means that the BDT requirements 
do not significantly alter these distributions.

%

\subsection{ Simple Selection}

We select a second event sample 
by applying criteria on event quality and kinematic quantities.
We use the consistency of the results obtained for the BDT and for this sample
as a measure of systematic effects related to imperfect modeling of the
detector acceptance and of the selection requirements. 

The criteria are the same
as in Refs.~\cite{prl07} and \cite{prl08}.
Each of the four tracks is required to have at least two SMT
hits and at least eight hits in SMT or CFT.
We require minimum momentum in the transverse plane $p_T$ for $B_{s}^0$, 
$\phi$, and $K$ meson candidates of 6.0 GeV, 1.5 GeV, and 0.7 GeV, respectively.
Muons are required to have $p_T$ above 1.5 GeV.
For events in the central 
rapidity region (an event is considered to be central if the higher $p_T$
muon has $|\eta(\mu_{\rm leading})|<1$), 
we require the
transverse momentum of the $J/\psi$ meson to exceed 4 GeV. 
In addition, $J/\psi$ candidates are accepted if the invariant mass of the muon pair
is in the range 3.1 $\pm$ 0.2 GeV.
Events  are required to satisfy the condition  $\sigma(t)<0.2$ ps
where $\sigma(t)$ is the uncertainty on the decay proper time obtained
from the propagation of the uncertainties in the decay-vertex kinematic
fit, the primary vertex position, and the $B_s^0$ candidate transverse 
momentum.
We refer to this second sample  as the ``Square-cuts'' sample.

\section{\label{sec:flavor} Flavor Tagging }

At the Tevatron, $b$ quarks are mostly produced in $b \overline b$ pairs.
The flavor of the initial state of the $B_s^0$ candidate is determined by 
exploiting  properties of particles produced by the other $b$ hadron  
(``opposite-side tagging'', or OST). The OST-discriminating variables $x_i$ 
are based primarily on the presence of a  muon or an electron from the 
semi-leptonic decay or the decay vertex charge of the other $b$ hadron 
produced in the $p\overline p$ interaction.

For the initial $b$ quark, the probability density function (PDF) for a given 
variable $x_i$ is denoted as $f_i^b(x_i)$, while for the initial $\overline b$
quark it is denoted as $f_i^{\overline b}(x_i)$. The combined tagging variable
$y$ is defined as:

\begin{equation}
y = \prod_{i=1}^{n}y_i;\,\,  y_i = \frac{f_i^{\overline b}(x_i)}{f_i^b(x_i)}.
\end{equation} 

A given variable $x_i$ can be undefined for some events. For example, there 
are events that don't contain an identified muon from the opposite side. In 
this case, the corresponding variable $y_i$ is set to 1.

In this way the OST algorithm  assigns to each event a value of the 
predicted tagging parameter  $d=(1-y)/(1+y)$ in the range [$-1$,1], with
$d>0$ tagged as an initial $b$ quark and  $d<0$ tagged as an initial 
$\overline b$ quark. Larger  $|d|$ values correspond to higher tagging 
confidence. In events where no tagging information is available  $d$ is set to
zero. The efficiency $\epsilon$ of the OST, defined as fraction of the number
of candidates with $d \ne 0$, is  18\%.
The OST-discriminating variables and algorithm are described in detail in 
Ref.~\cite{bflavor}.

The  tagging  dilution ${\cal D}$ is defined as
\begin{equation}
{\cal D} = \frac{N_{\rm cor} - N_{\rm wr}}{N_{\rm cor} + N_{\rm wr}},
\label{dilution}
\end{equation}
where $N_{\rm cor}$ ($N_{\rm wr}$) is the number of events with correctly 
(wrongly) identified initial $B$-meson flavor.
The dependence of the tagging dilution on the tagging parameter 
$d$ is calibrated 
 with data for which the flavor ($B$ or $\overline B$) is known.

\subsection{OST calibration}

The dilution calibration is based on four
independent $B_d^0 \to \mu \nu D^{*\pm}$   data samples corresponding 
to different time periods,
denoted IIa, IIb1, IIb2, and IIb3,
with different detector configurations and different distributions of 
instantaneous luminosity. The Run IIa sample was used in Ref.~\cite{bflavor}.

\begin{figure}[H]
\begin{center}
\includegraphics*[width=0.45\textwidth]{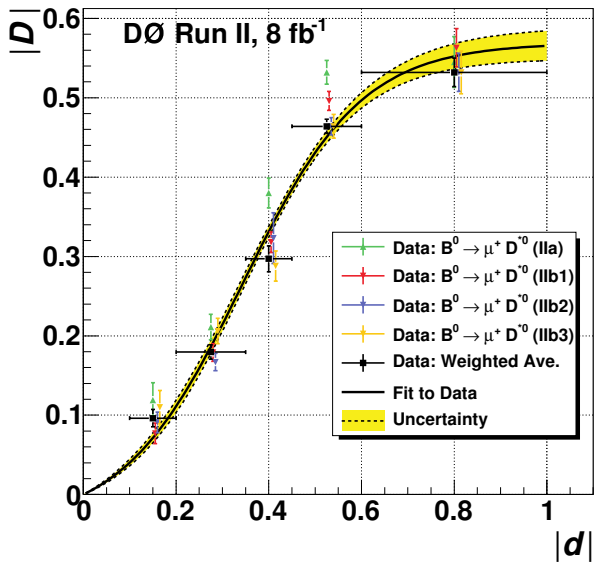}
\caption{(color online). Parametrization of the dilution $|{\cal D}|$
as a function of the tagging parameter $|d|$ for the combined 
opposite-side tagger.
The curve is the result of the weighted fit to four self-tagging
control data samples (see text).
}
\label{fig:ostall}
\end{center}
\end{figure}

For each sample we  perform an
analysis of the $B_d^0-\overline{B}_d^0$ oscillations described in Ref.~\cite{Abazov:2006dm}.
We divide the samples in five ranges of the tagging parameter $|d|$,
and for each range we obtain a mean value of the dilution ${\cal |D|}$.
The mixing frequency $\Delta M_d$ is fitted simultaneously and  is found
to be stable and consistent with the world average value.
The measured values of the tagging dilution  $\cal |D|$  for 
the four data samples above, 
in different ranges of  $|d|$, are 
shown in Fig.~\ref{fig:ostall}.
The dependence of the dilution on $|d|$  is parametrized as
\begin{equation}
\label{dileq}
{\cal |D|} = \frac{p_0}{(1 + \exp((p_1 - |d|)/p_2))} - \frac{p_0}{(1+\exp(p_1/p_2))}.  
\end{equation}
and the function is fitted to the data.
All four measurements are in good agreement and hence a weighted average is taken.


\newcommand{\Ao}{|A_0(0)|}
\newcommand{\All}{|A_\parallel(0)|}
\newcommand{\Ap}{|A_\perp(0)|}
\newcommand{\DG}{\Delta \Gamma_s}
\newcommand{\Phis}{\phi_s^{J/\psi \phi}}
\newcommand{\DeltaUno}{\delta_1}
\newcommand{\DeltaDos}{\delta_2}
\newcommand{\TAU}{\overline \tau_s}
\newcommand{\DeltaM}{\Delta M_s}


\section{\label{sec:fitting} Maximum Likelihood Fit}

We perform a six-dimensional (6D) unbinned maximum likelihood fit to the proper decay time
and its uncertainty,
three decay angles characterizing the final state, and the mass of the $B_s^0$ candidate. 
We use events for which the invariant mass of the $K^+K^-$ pair is
within the range 1.01 -- 1.03 GeV.
There are 104683 events in the BDT-based sample and 66455 events in the
Square-cuts sample.
We adopt the formulae and notation of  Ref.~\cite{formulae}.  
The normalized  functional form
of the differential decay rate 
includes an $\cal S$-wave $KK$ contribution in addition to the dominant
$\cal P$-wave $\phi \rightarrow K^+K^-$ decay. 
To model  the distributions of the signal and background we use 
the software library {\sc RooFit}~\cite{refroofit}.

\subsection{Signal model}

\begin{figure}[htbp]
 \begin{center}
 \includegraphics*[width=0.46\textwidth]{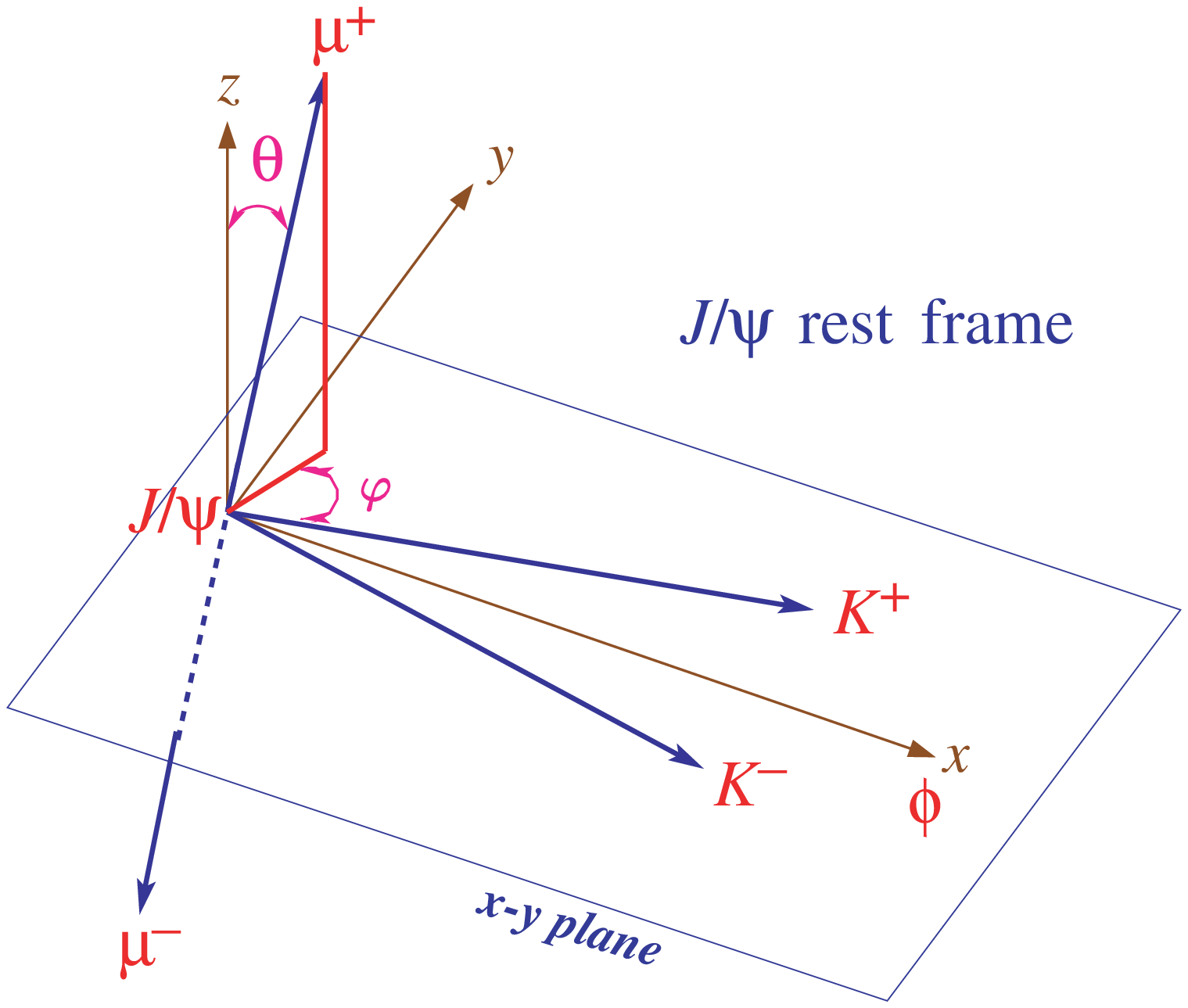}
\includegraphics*[width=0.40\textwidth]{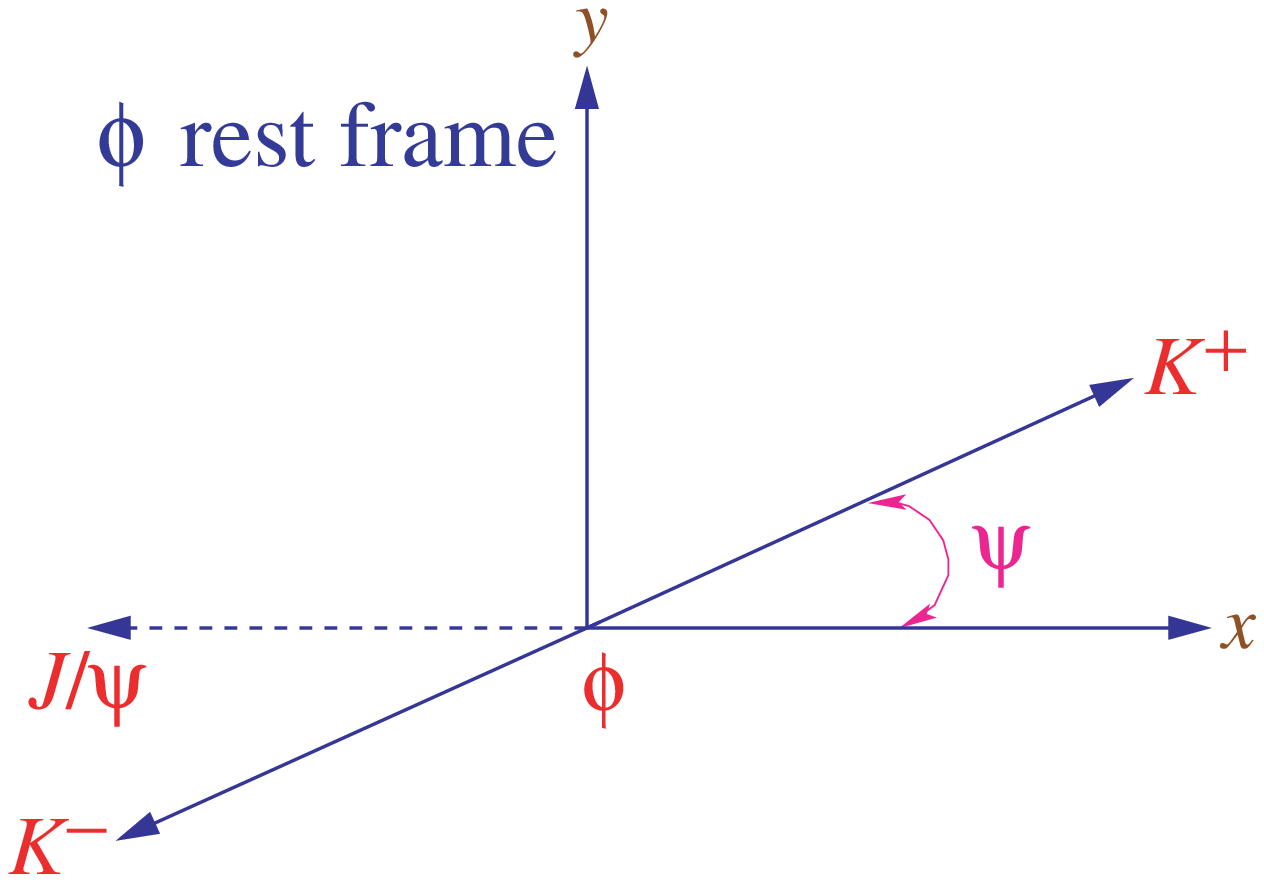}
\caption{ (color online). 
Definition of the angle $\psi$, and the transversity angles $\theta$ and $\varphi$.
}
 \label{fig:defAngles}
 \end{center}
 \end{figure}

The angular distribution of the signal is expressed in the transversity basis~\cite{Trans}.
In the coordinate system of the $J/\psi$ rest frame, 
 where the $\phi$ meson moves in the $x$ direction,
 the $z$  axis is perpendicular to 
the decay plane of $\phi \to K^+ K^-$, and $p_y(K^+)\geq 0$.
The transversity polar and azimuthal angles 
$\theta$ and  $\varphi$ describe the
direction of the positively-charged muon, while $\psi$ is 
the angle between  $\vec p(K^+)$ and  $-\vec{p}(J/\psi)$ 
 in the $\phi$ rest frame.
The angles are shown in Fig.~\ref{fig:defAngles}.

In this basis, 
the decay amplitude of the $B_s^0$ and $\overline B_s^0$ mesons 
is decomposed into three independent
components corresponding to linear polarization states of 
the vector mesons  $J/\psi$ and  $\phi$,
which are polarized either 
longitudinally (0) or transversely to their direction of motion,
and parallel ($\parallel$) or perpendicular ($\perp$) to each other.

The time dependence of amplitudes ${\cal A}_i(t)$ and ${\bar {\cal A}_i}(t)$
($i$ denotes one of  $\{||, \perp, 0\}$),
for $B_s^0$ and $\overline B_s^0$ states to reach the final state
$J/\psi$ $\phi$ is:
\begin{eqnarray}
{\cal A}_i(t) &=&    F(t) \left[
E_+(t) \pm e^{2i\beta_s} E_-(t)
\right] a_i\,,   \nonumber \\
{\bar {\cal A}_i}(t) &=& F(t) \left[
\pm E_+(t) + e^{-2i\beta_s} E_-(t)
\right] a_i\,,  
\label{eqn:finalAmp}
\end{eqnarray}
where 
\begin{eqnarray}
F(t)  &=& \frac{e^{-\Gamma_s t /2}}{\sqrt{\tau_H + \tau_L \pm \cos{2\beta_s}\left(\tau_L-\tau_H\right)}} \,,
\end{eqnarray}
and $\tau_H $ and $ \tau_L$ are the lifetimes of the heavy and light $B_s^0$ eigenstates.

In the above equations the upper sign indicates a {\sl CP}-even final state, the lower sign indicates a {\sl CP}-odd final state,
\begin{equation}
E_{\pm}(t) \equiv  \frac{1}{2}\left[e^{\left(\frac{-\Delta\Gamma_s}{4} +
i\frac{\Delta M_s}{2}\right)t} \pm e^{-\left(\frac{-\Delta\Gamma_s}{4} + i\frac{\Delta M_s}{2}\right)t}\right],
\label{eqn:functionDef}
\end{equation}
and the amplitude parameters  $a_i$ give the time-integrated decay rate to each of the polarization states,
$|a_i|^2$,   satisfying:
\begin{equation}
\sum_i {|a_i|^2} = 1 \,.
\label{eqn:aNorm}
\end{equation}

The interference terms $\apar - \aperp$ and  $A_0 - \aperp$ are proportional
to  $(e^{-\Gamma_H t}-e^{-\Gamma_Lt}) \sin\phi_{s}^{J/\psi\phi}$.
Also, if $\cos \phi_s^{J/\psi \phi}$ is significantly
different from unity, the decay rates  of the {\sl CP}-even and {\sl CP}-odd
components have two slopes each.

The  normalized probability density functions $P_B$ and $P_{\bar B}$ for $B$ and ${\bar B}$ mesons
in the variables $t$, $\cos{\psi}$, $\cos{\theta}$, and $\varphi$, 
are
\begin{eqnarray} 
P_{B}(\theta, \varphi, \psi, t) = \frac{9}{16\pi} |{\bf A}(t)\times \hat{n}|^2,  \nonumber \\
P_{\bar{B}}(\theta, \varphi, \psi, t) = \frac{9}{16\pi} |{\bf {\bar A}(t)}\times \hat{n}|^2,
\label{eqn:finalAngle}
\end{eqnarray}
where $\hat n$ is the muon momentum direction in the $J/\psi$ rest frame,
\begin{equation}
\hat{n} = \left(\sin{\theta}\cos{\varphi}, \sin{\theta}\sin{\varphi}, \cos{\theta} \right),
\label{eqn:nhat}
\end{equation}
and  ${\bf A}(t)$ and ${\bf {\bar A}}(t)$ are complex vector functions of time defined as

\begin{eqnarray}
{\bf A}(t)=\left({\mathcal A}_0(t)\cos{\psi}, -\frac{{\mathcal A}_\parallel(t)\sin{\psi}}{\sqrt{2}}, i\frac{{\mathcal A}_\perp(t)\sin{\psi}}{\sqrt{2}}\right),  \nonumber \\
{\bf{\bar {A}}}(t)=\left({\bar {\mathcal A}}_0(t)\cos{\psi}, -\frac{{\bar {\mathcal A}}_\parallel(t)\sin{\psi}}{\sqrt{2}}, i\frac{{\bar {\mathcal A}}_\perp(t)\sin{\psi}}{\sqrt{2}}\right).
\label{eqn:fixedAngle}
\end{eqnarray}

  The values of 
${\cal A}_i(t)$ at $t=0$ are denoted as $A_i$.  They are related to the parameters  $a_i$ 
by 
\begin{eqnarray}
|A_{\perp}|^2 = \frac{|a_{\perp}|^2y}{1+(y-1)|a_{\perp}|^2},   \nonumber \\
|A_{||}|^2 = \frac{|a_{||}|^2}{1+(y-1)|a_{\perp}|^2},          \nonumber \\
|A_0|^2 = \frac{|a_0|^2}{1+(y-1)|a_{\perp}|^2},             \nonumber \\   %
\end{eqnarray}
where $y \equiv (1-z)/(1+z)$ and $z\equiv \cos{2\beta_s}\Delta\Gamma_s/(2\overline \Gamma_s)$.
By convention,  the phase of $A_0$ is set to zero and the phases 
of the other two amplitudes are denoted by $\delta_{||}$ and  $\delta_\perp$. 

For a given event, the decay rate is the sum of the 
functions $P_B$ and $P_{\bar B}$  weighted  by the flavor tagging
dilution factors $(1+ {\cal D})/2$ and $(1-{\cal D})/2$, respectively.

The contribution from the decay to $J/\psi K^+K^-$ with the kaons in an $\cal S$-wave
is expressed in terms of the $\cal S$-wave fraction $F_S$ and a phase $\delta_s$.
The squared sum of the $\cal P$ and $\cal S$ waves is integrated over the $KK$ mass.
For the $\cal P$-wave, we assume the non-relativistic Breit-Wigner model

\begin{equation}
  g(M(KK)) = \sqrt{\frac {\Gamma_{\phi}/2}{\Delta M(KK)}} \cdot \frac{1}{M(KK) - M_{\phi} + i \Gamma_{\phi}/2}
\label{eqn:gDef}
\end{equation}
with the  $\phi$ meson mass $M_{\phi} = 1.019$ GeV and width $\Gamma_{\phi} =4.26$ MeV~\cite{PDG},
and with $\Delta M(KK) = 1.03 -1.01 = 0.02$ GeV.

For the $\cal S$-wave component, we assume
a uniform distribution in the range $1.01 < M(KK)<1.03$ GeV.
We constrain the oscillation frequency to $\Delta M_s = 17.77 \pm 0.12$ ps$^{-1}$, 
as measured in Ref.~\cite{dms}. 
Table~\ref{sigpar} lists all physics parameters used in the fit.

\begin{table}[htbp]
\begin{tabular}{cc}
\hline
\hline
                  Parameter           &    Definition \tabularnewline
\hline

$|A_0|^2$                       &  $\cal P$-wave longitudinal  amplitude squared, at $t=0$ \tabularnewline
$A1$        & $|A_{\|}|^2/(1-|A_0|^2)$     \tabularnewline
$\overline{\tau}_s$ (ps)        & $B^0_s$ mean lifetime  \tabularnewline
$\Delta\Gamma_s$ (ps$^{-1}$)    &  Heavy-light decay width difference   \tabularnewline
$F_S$                           &  $K^+K^-$ $\cal S$-wave fraction \tabularnewline
$\beta_s$                       &  {\sl CP}-violating phase  ( $\equiv -\phi_s^{J/\psi \phi}/2$)  \tabularnewline
$\delta_{\|}$                   &  $\arg(A_{\|}/A_0)$     \tabularnewline
$\delta_{\perp}$                &  $\arg(A_{\perp}/A_0)$  \tabularnewline
$\delta_s$                      &  $\arg(A_{s}/A_0)$      \tabularnewline
\hline
\hline
\end{tabular}
\caption {Definition of nine real measurables for the decay \bsdec\ used in the
Maximum Likelihood fitting.
}
\label{sigpar}
\end{table}

For the signal mass distribution we use a Gaussian function with
a free  mean value, width, and normalization.
The function describing the signal rate in the 6D
space is invariant under 
 the combined
 transformation 
$\beta_s \rightarrow \pi/2 - \beta_s$,
$\Delta \Gamma_s  \rightarrow  -\Delta \Gamma_s$, 
$\delta_{\|}  \rightarrow  2\pi - \delta_{\|} $, 
$\delta_{\perp}  \rightarrow  \pi - \delta_{\perp} $, and
$\delta_s  \rightarrow  \pi - \delta_s$.
In addition, with a limited flavor-tagging power,
there is an approximate symmetry around $\beta_s =0$ 
for a given sign of $\Delta \Gamma_s$.

We correct the signal decay rate by 
a detector acceptance factor $\epsilon(\psi, \theta, \varphi)$ 
parametrized by  coefficients 
of expansion in Legendre polynomials
$P_k(\psi)$ and real harmonics $Y_{lm}(\theta,\varphi)$. The coefficients
are obtained from Monte Carlo simulated samples, as described in Appendix~\ref{appmcdata}.

\begin{figure}[htbp]
 \begin{center}
 \includegraphics*[width=0.4\textwidth]{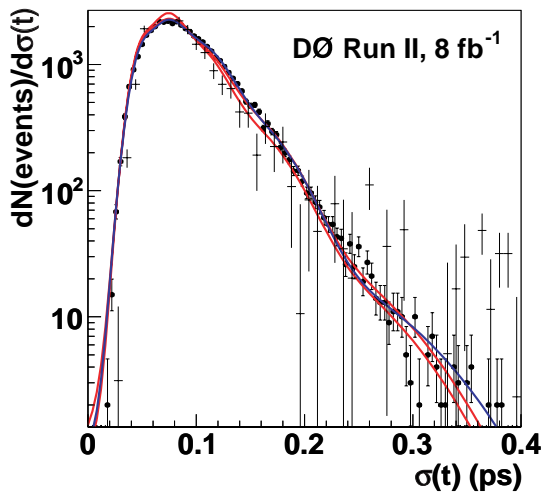}
\caption{ (color online). The distribution of the uncertainty in the decay time for the signal,
MC (squares) and background-subtracted data (crosses).
The blue curve is the sum of five Gaussian functions fitted to the MC distribution. 
The two red lines are variations of the default function
used in the studies of systematic effects.}
 \label{fig:et_signal_BDT20}
 \end{center}
 \end{figure}

The signal decay time resolution is given by a Gaussian centered at zero and width given
by the product of a global scale factor and the event-by-event uncertainty in the decay time
measurement. The distribution of the uncertainty in the decay time measurement in the MC simulation
is modeled by a superposition of five Gaussian functions.
The background-subtracted signal distribution agrees well with the MC model,
as seen in Fig.~\ref{fig:et_signal_BDT20}.
Variations of the parameters within one sigma of the best fit are used
to define two additional functions, also shown in the figure, that are 
used in alternative fits to estimate the systematic effect due to time resolution.

\subsection{Background model}

The proper decay time distribution  of the  background is described by a sum of
a prompt component, modeled as the same resolution function used in the signal 
decay time, and a non-prompt component.
The non-prompt component  is modeled as a superposition of one 
exponential decay for $t<0$ 
and two exponential decays for $t>0$, with free slopes and normalizations.
The lifetime resolution is modeled by an exponential convoluted with 
a Gaussian function,
with two separate parameters for prompt and non-prompt background.
To allow for the possibility of the lifetime uncertainty to be 
systematically
underestimated, we introduce a free scale factor.

The mass distributions of the  two components of background
are parametrized by low-order polynomials: a linear function for the
prompt background and a quadratic function for the non-prompt background.
The angular distribution of background is parametrized
by Legendre and real harmonics expansion coefficients. A separate set of 
expansion coefficients $c^k_{lm}$ and  $c^k_{lm}$, with
$k=0$ or $2$ and $l=0,1,2$, is used for the prompt and non-prompt background.
A preliminary fit is first performed with all  17 parameters ${c^{(P)}}^k_{lm}$ for
prompt and 17 parameters ${c^{(NP)}}^k_{lm}$ for non-prompt allowed to vary. 
In subsequent fits
those that converge at values  within two standard deviations of zero are set to zero.
Nine free parameters remain, five for non-prompt background:
${c^{(NP)}}^0_{1-1}$,   ${c^{(NP)}}^0_{20}$,  ${c^{(NP)}}^0_{22}$, ${c^{(NP)}}^2_{00}$, and
${c^{(NP)}}^2_{22}$, 
and four for prompt background:
${c^{(P)}}^0_{1-1}$,   ${c^{(P)}}^0_{20}$,  ${c^{(P)}}^0_{22}$,  and  ${c^{(P)}}^2_{2-1}$.
All background parameters described above are varied simultaneously with
physics  parameters. 
In total, there are 36 parameters used in the fit. In addition
to the  nine physics parameters defined
in Table~\ref{sigpar}, they are: signal yield, mean mass and width, 
non-prompt background contribution, six non-prompt background lifetime
parameters, four background time resolution parameters, one 
time resolution scale factor, three 
background mass distribution parameters, and nine parameters describing
background angular distributions.

\subsection{Fit results}

The maximum likelihood fit results for the nominal fit (Default), for two 
alternative time resolution functions, $\sigma_A(t)$ and $\sigma_B(t)$ shown
in Fig. \ref{fig:et_signal_BDT20}, and for an alternative $M(KK)$ dependence of 
the $\phi(1020) \rightarrow K^+ K^-$ decay with the decay width increased 
by a factor of two are shown in Table \ref{roofitresults-cut10} and 
Table \ref{roofitresults-sq}. These alternative fits are used to estimate the
systematic uncertainties.
The fit assigns $5598 \pm 113$  ($5050\pm105$) events to the signal
for the BDT (Square-cuts) sample.
Only the parameters whose values do not suffer from multi-modal
effects are shown. A single fit does not provide  meaningful 
point estimates and uncertainties for the four phase parameters.
Their estimates are obtained using the MCMC 
technique.
Figures~\ref{fig:cut10bkg}~--~\ref{fig:a2withfit} 
illustrate the quality of the fit for
the background, for all data,  and for the signal-enhanced sub-samples.

An independent measurement of the $\cal S$-wave fraction is described
 in Appendix~\ref{appfs} and the result is in agreement with  $F_S$
 determined from the  maximum likelihood fit.

 \begin{widetext} 

\begin{table}[htbp]
\begin{tabular}{ccccc}
\hline
\hline
     Parameter                        &  Default  & $\sigma_A(t)$ & $\sigma_B(t)$  &  $\Gamma_{\phi} = 8.52$ MeV    \tabularnewline
\hline

$|A_0|^2$                       & $0.553 \pm 0.016$  & ~~$0.553 \pm 0.016$ & ~~$0.552 \pm 0.016$ &  ~~$0.553 \pm 0.016$  \tabularnewline
$|A_{\|}|^2/(1-|A_0|^2)$        & $0.487 \pm 0.043$  & ~~$0.483 \pm 0.043$ & ~~$0.485 \pm 0.043$ & ~~$0.487 \pm 0.043$ \tabularnewline
$\overline{\tau}_s$ (ps)        & $1.417 \pm 0.038$    & ~~$1.420 \pm 0.037$ &   ~~$1.417 \pm 0.037$  & ~~$1.408 \pm 0.434$  \tabularnewline
$\Delta\Gamma_s$ (ps$^{-1}$)    & $0.151 \pm 0.058$   & ~~$0.136 \pm 0.056$ &  ~~$0.145 \pm 0.057$   & ~~$0.170 \pm 0.067$   \tabularnewline
$F_S$                           & $0.147 \pm 0.035$   & ~~$0.149 \pm 0.034$  & ~~$0.147 \pm 0.035$  & ~~$0.147 \pm 0.035$   \tabularnewline
\hline
\hline
\end{tabular}
\caption {
Maximum likelihood fit results for the BDT selection.
The uncertainties are statistical.
}
\label{roofitresults-cut10}
\end{table}


\begin{table}[htbp]
\begin{tabular}{ccccc}
\hline
\hline
     Parameter                  &  Default           & ~~$\sigma_{A}(t)$   & ~~$\sigma_{B}(t)$   & ~~$\Gamma_{\phi} = 8.52$ MeV    \tabularnewline
\hline
$|A_0|^2$                       & $0.566 \pm 0.017$  & ~~$0.564 \pm 0.017$ & ~~$0.567 \pm 0.017$ & ~~$0.566 \pm 0.017$  \tabularnewline
$|A_{\|}|^2/(1-|A_0|^2)$        & $0.579 \pm 0.048$  & ~~$0.579 \pm 0.048$ & ~~$0.577 \pm 0.048$ & ~~$0.579 \pm 0.048$ \tabularnewline
$\overline{\tau}_s$ (ps)        & $1.439 \pm 0.039$  & ~~$1.450 \pm 0.038$ & ~~$1.457 \pm 0.037$ & ~~$1.438 \pm 0.042$  \tabularnewline
$\Delta\Gamma_s$ (ps$^{-1}$)    & $0.199 \pm 0.058$  & ~~$0.194 \pm 0.057$ & ~~$0.185 \pm 0.056$ & ~~$0.202 \pm 0.060$   \tabularnewline
$F_S$                           & $0.175 \pm 0.035$  & ~~$0.169 \pm 0.035$ & ~~$0.171 \pm 0.035$ & ~~$0.175 \pm 0.035$   \tabularnewline
\hline
\hline
\end{tabular}

\caption {Maximum likelihood fit results for the `Square-cuts' sample.
}
\label{roofitresults-sq}
\end{table}

\end{widetext}

\begin{widetext}

 \begin{figure}[htbp]
 \begin{center}
\centering
 \includegraphics*[width=0.27\textwidth]{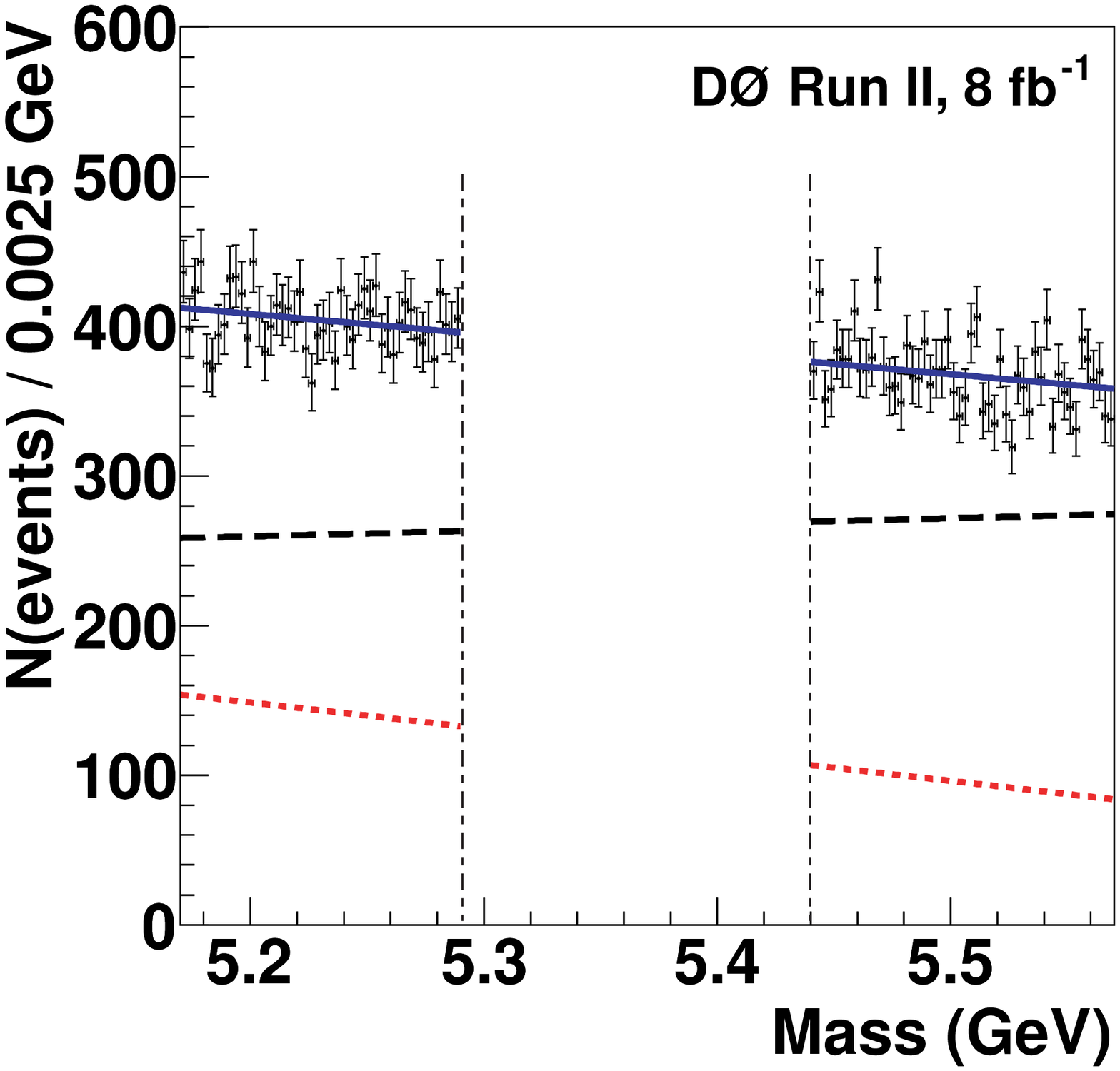}
 \includegraphics*[width=0.27\textwidth]{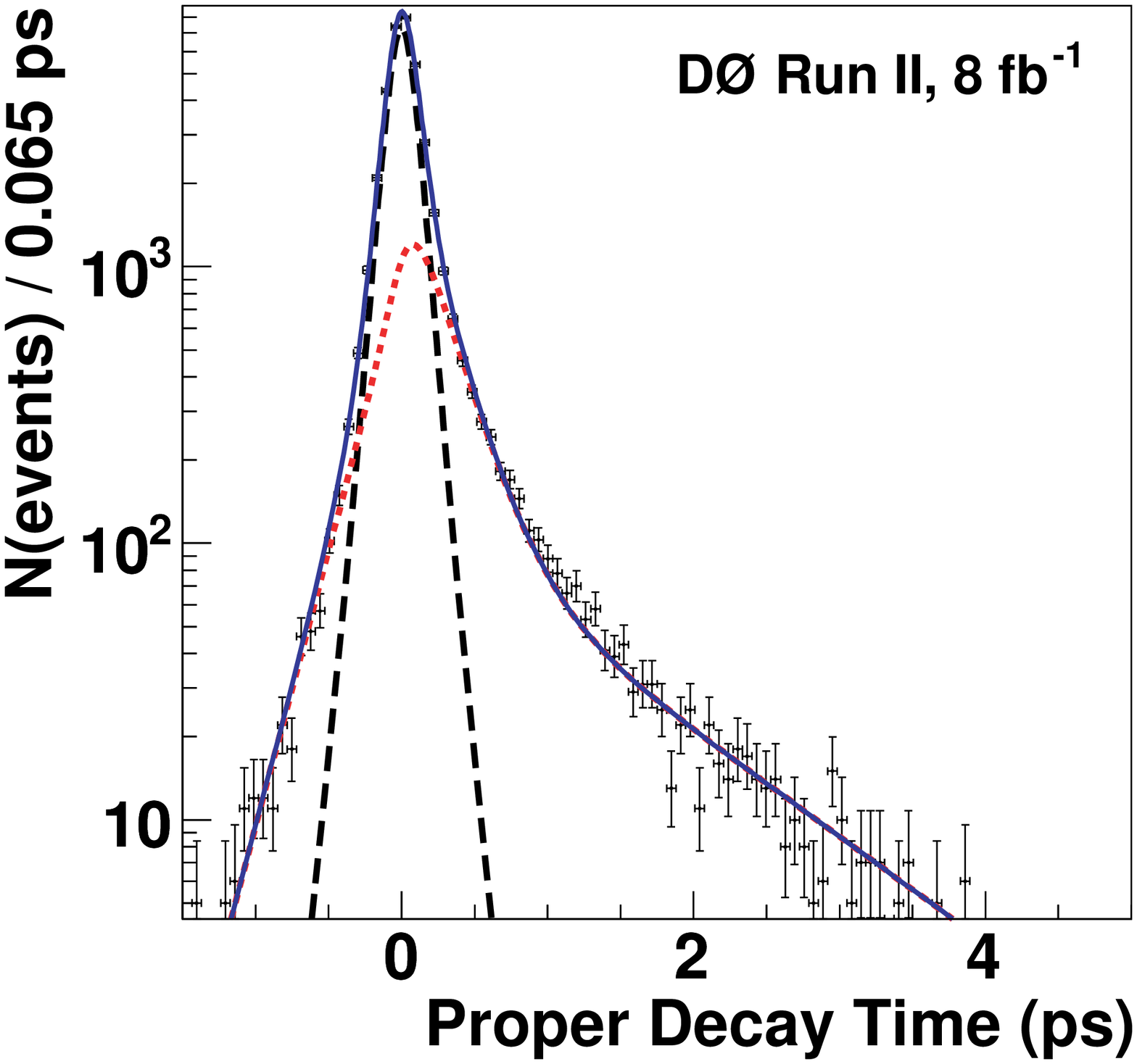}
\includegraphics*[width=0.27\textwidth]{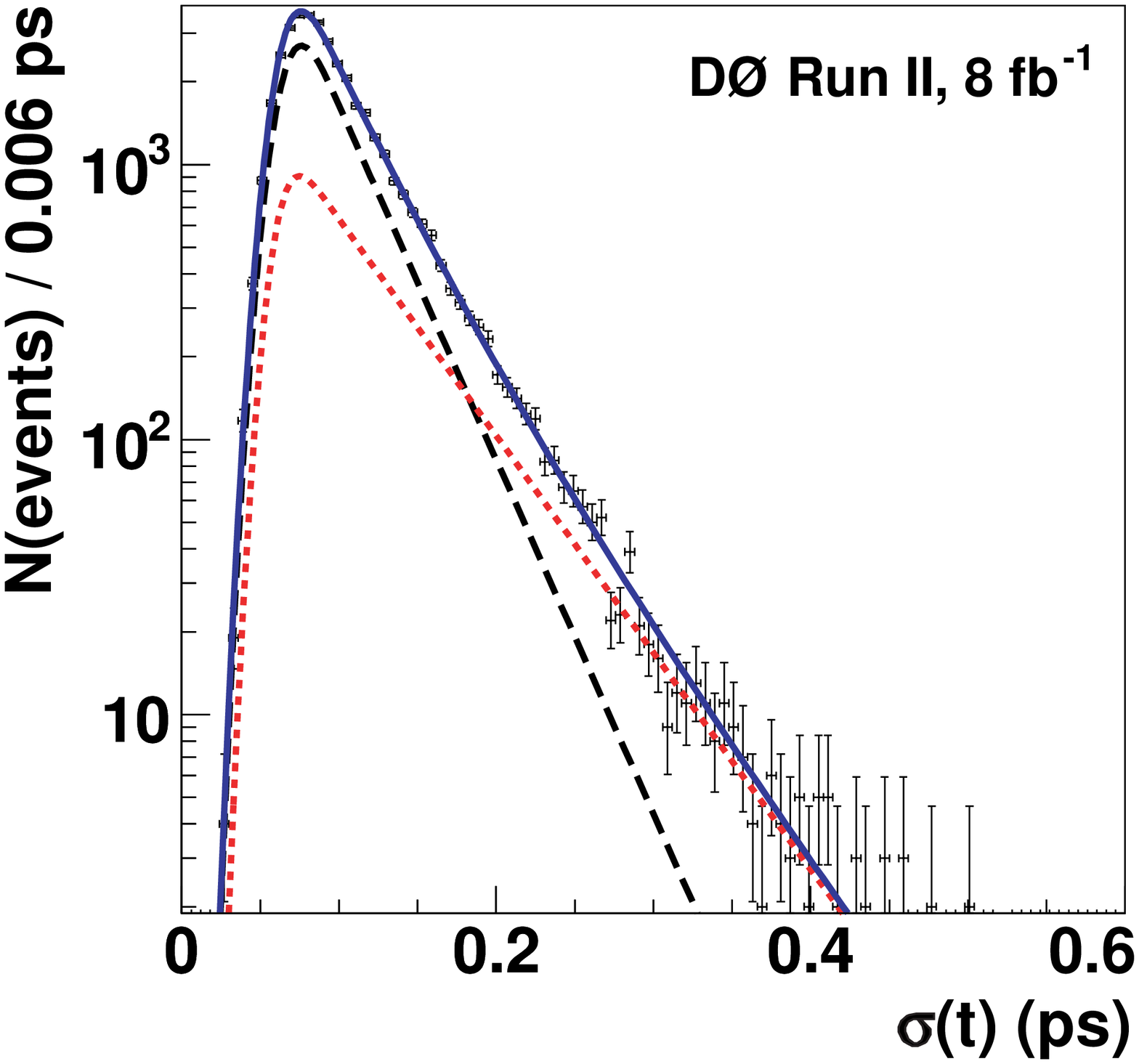}
 \includegraphics*[width=0.27\textwidth]{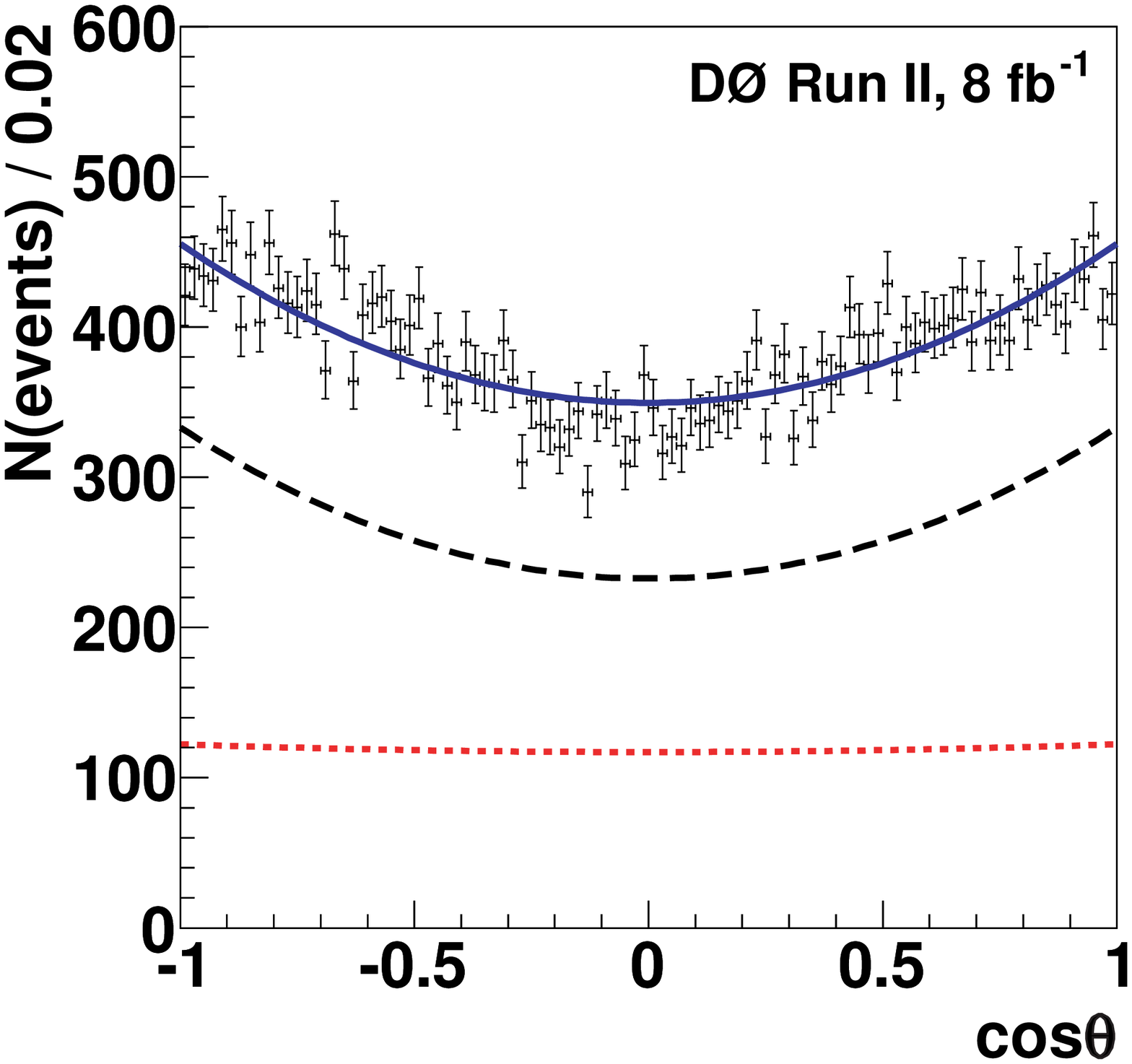}
\includegraphics*[width=0.27\textwidth]{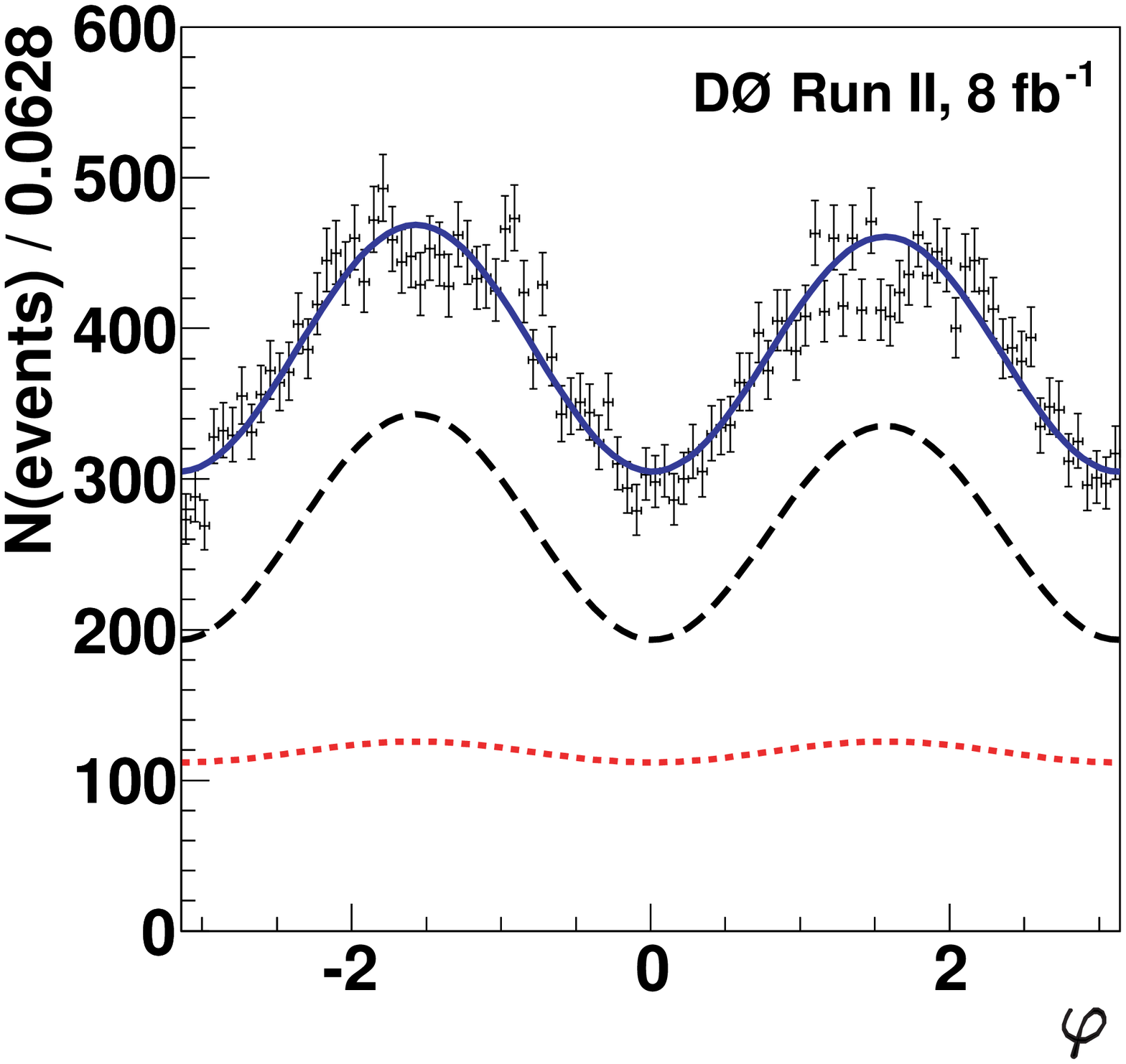}
\includegraphics*[width=0.27\textwidth]{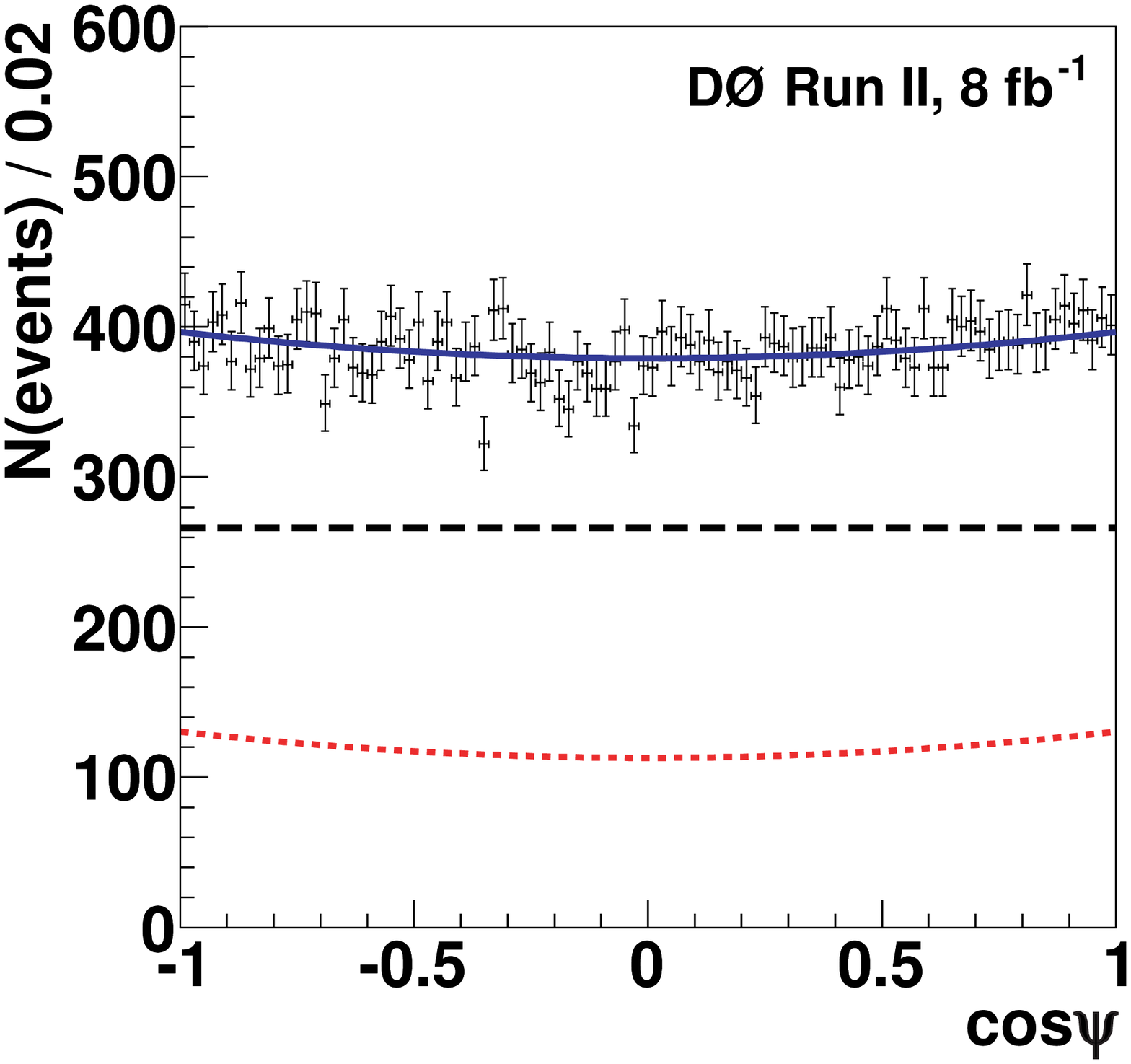}
\caption{(color online). The distributions in the background ($B_s^0$ mass sidebands) region of candidate mass, 
proper decay time, decay time uncertainty, transversity polar and azimuthal angles, and $\cos \psi$
for the BDT sample.
The curves show the prompt (black dashed) and non-prompt (red dotted)  components, and their sum (blue solid).  }
 \label{fig:cut10bkg}
 \end{center}
 \end{figure}



 \begin{figure}[htbp]
 \begin{center}
               \includegraphics[width=0.27\textwidth]{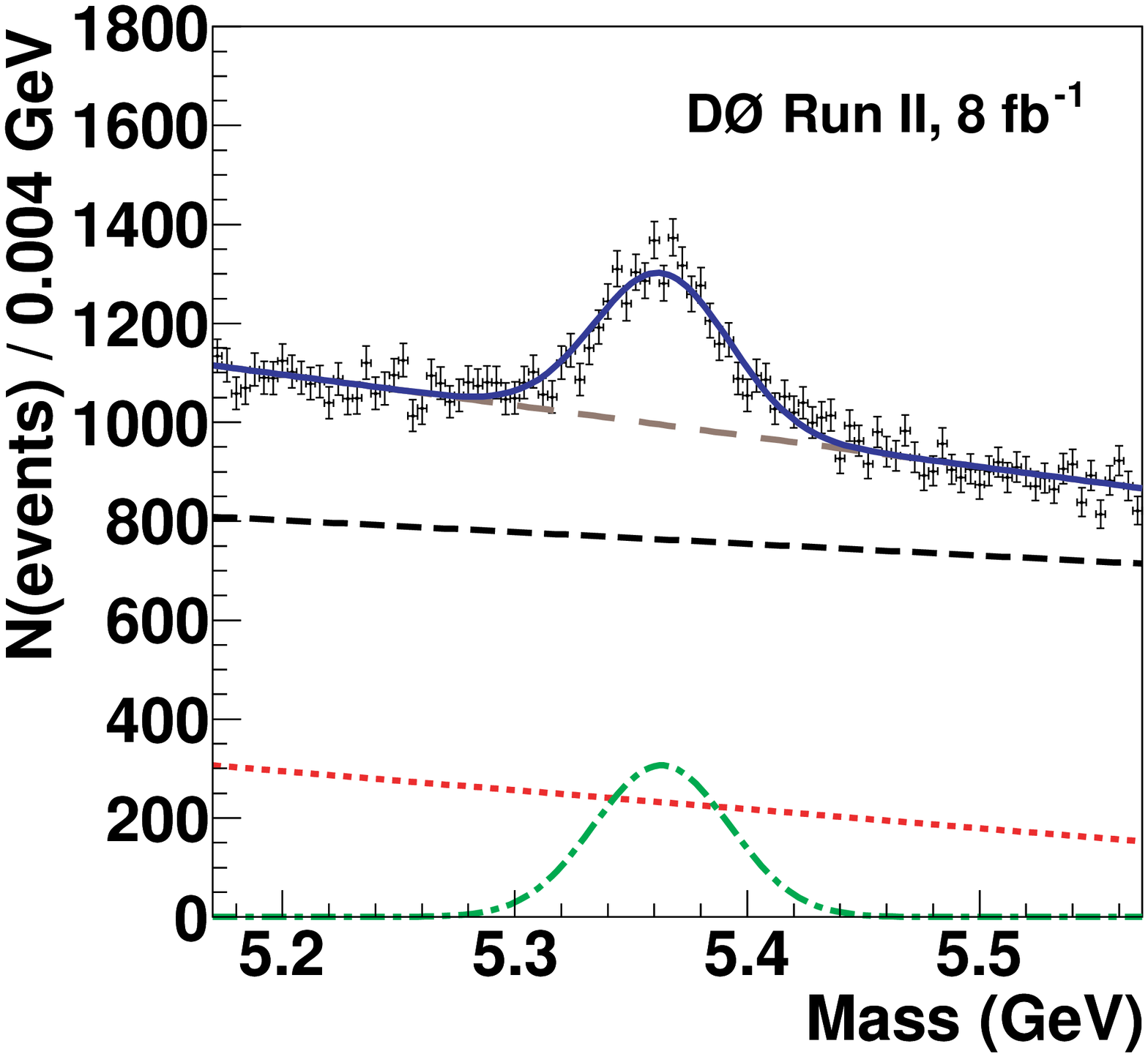}
               \includegraphics[width=0.27\textwidth]{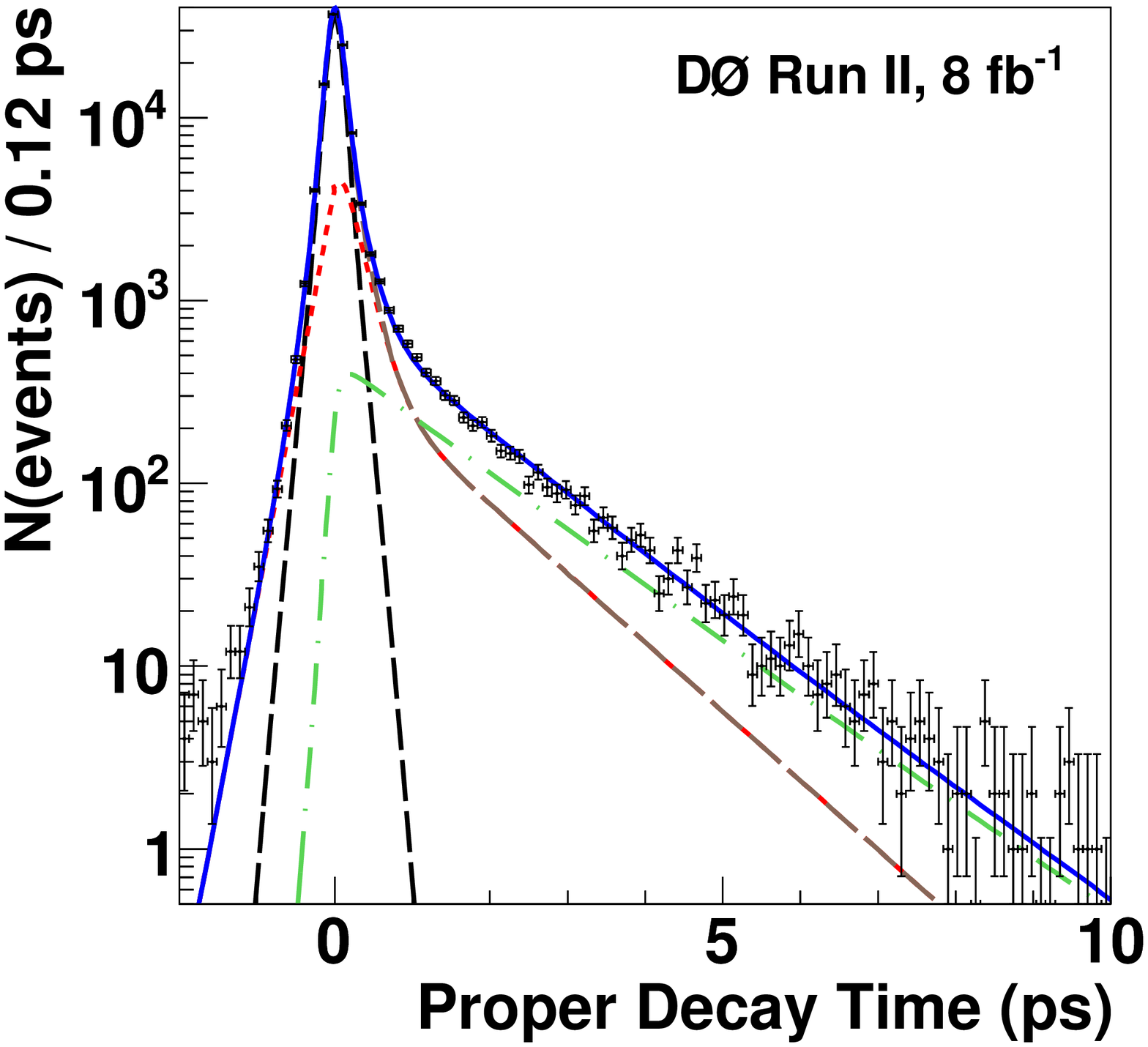}  
                \includegraphics[width=0.27\textwidth]{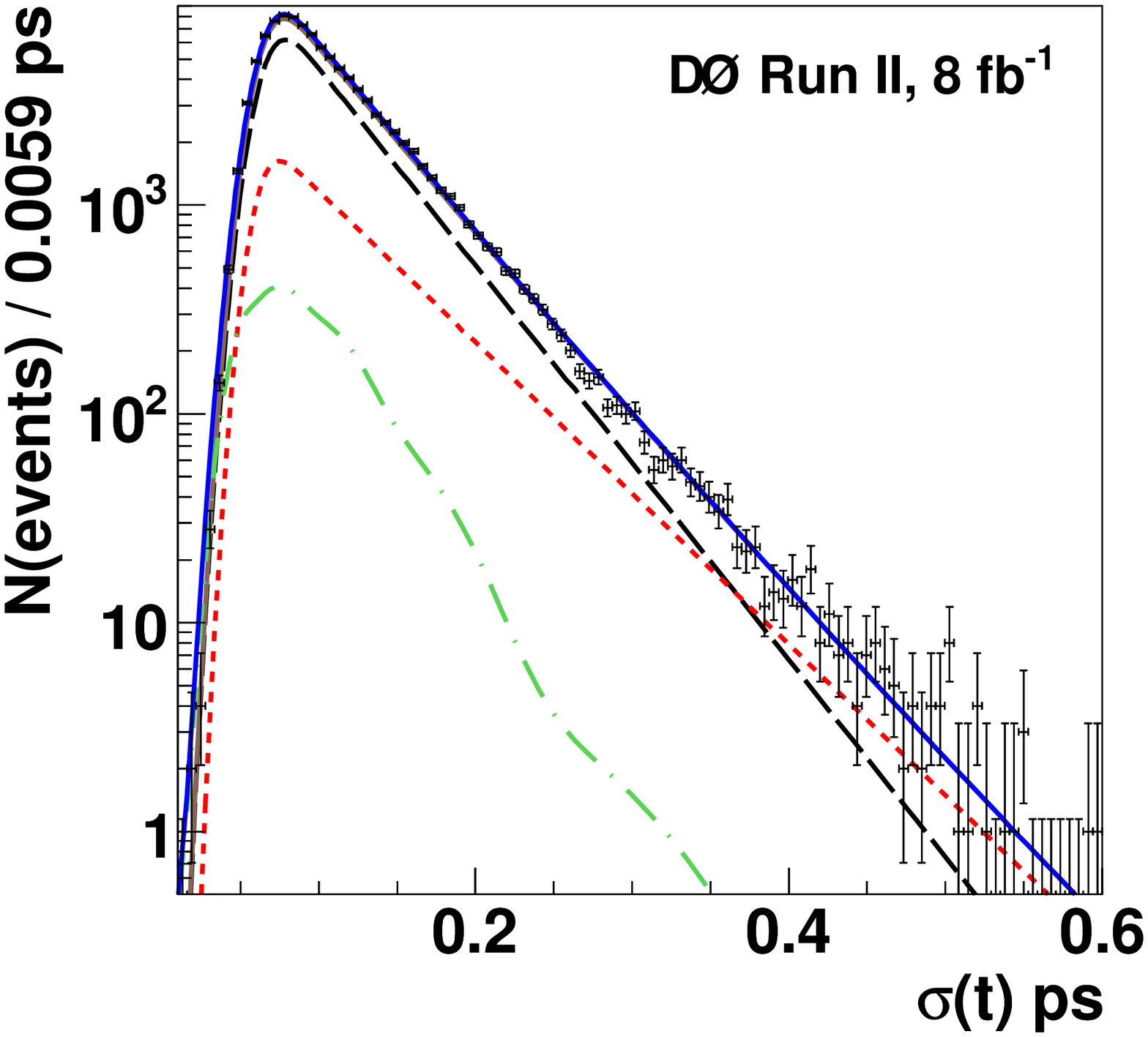}
               \includegraphics[width=0.27\textwidth]{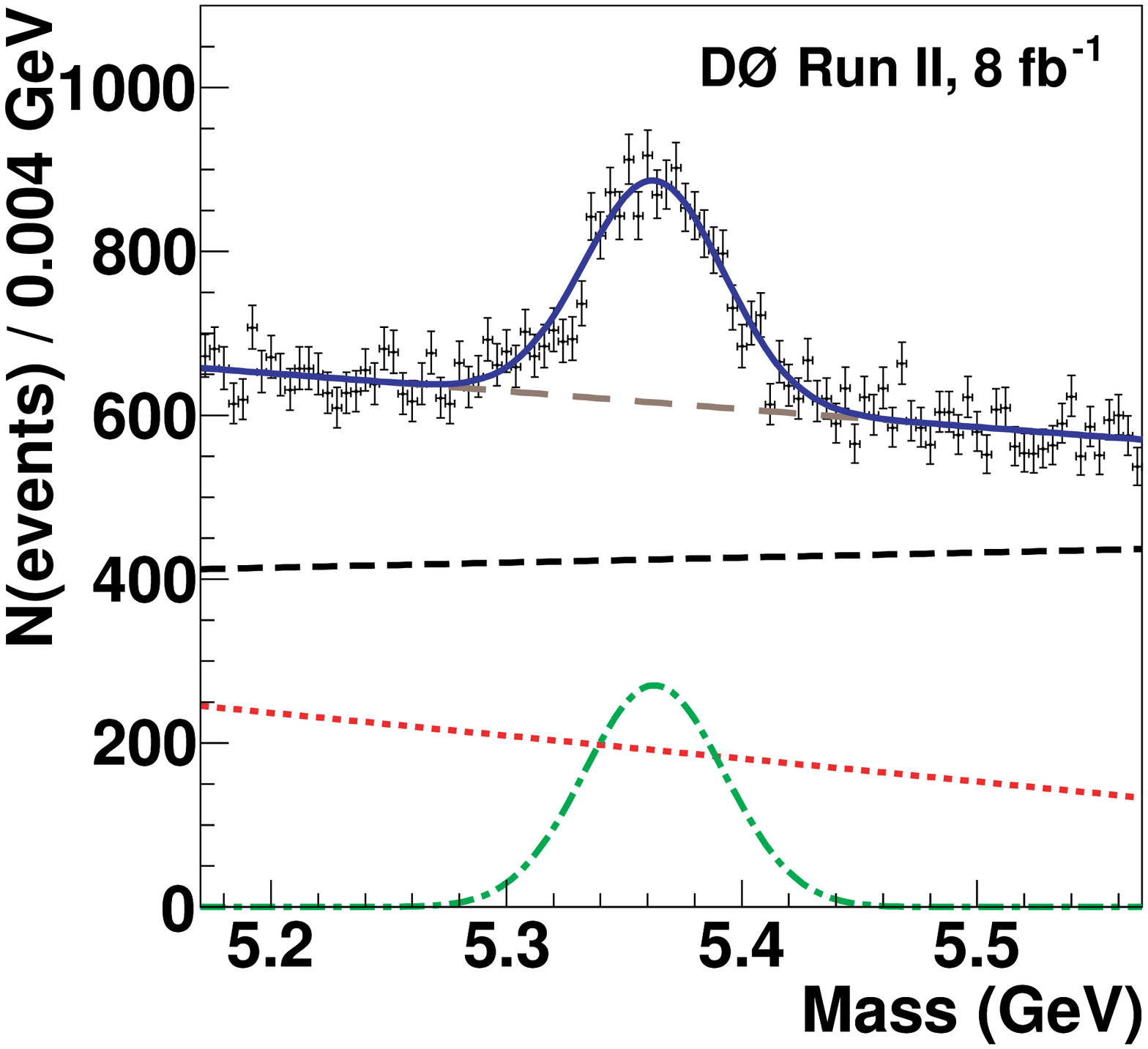}
               \includegraphics[width=0.27\textwidth]{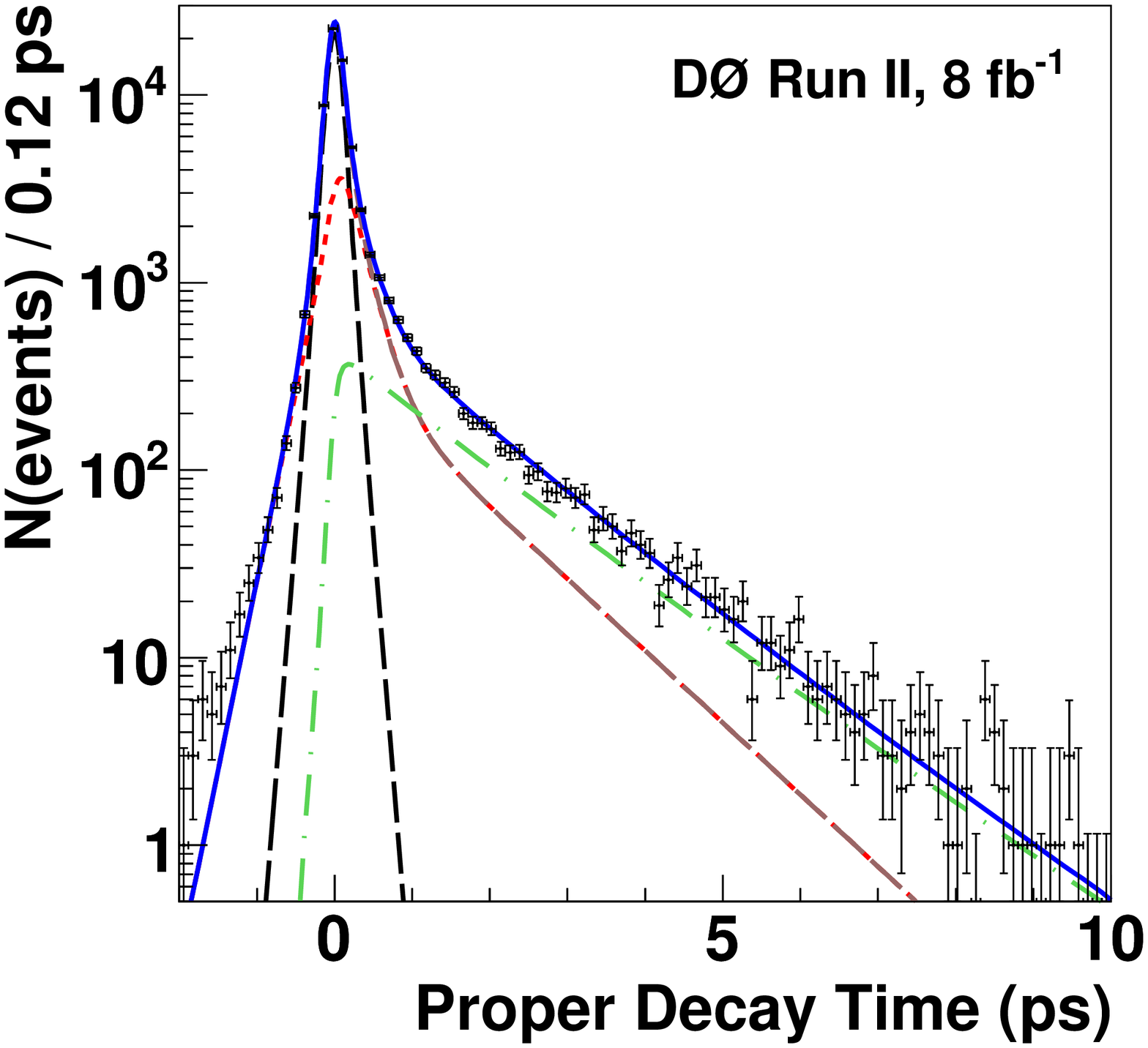}  
                \includegraphics[width=0.27\textwidth]{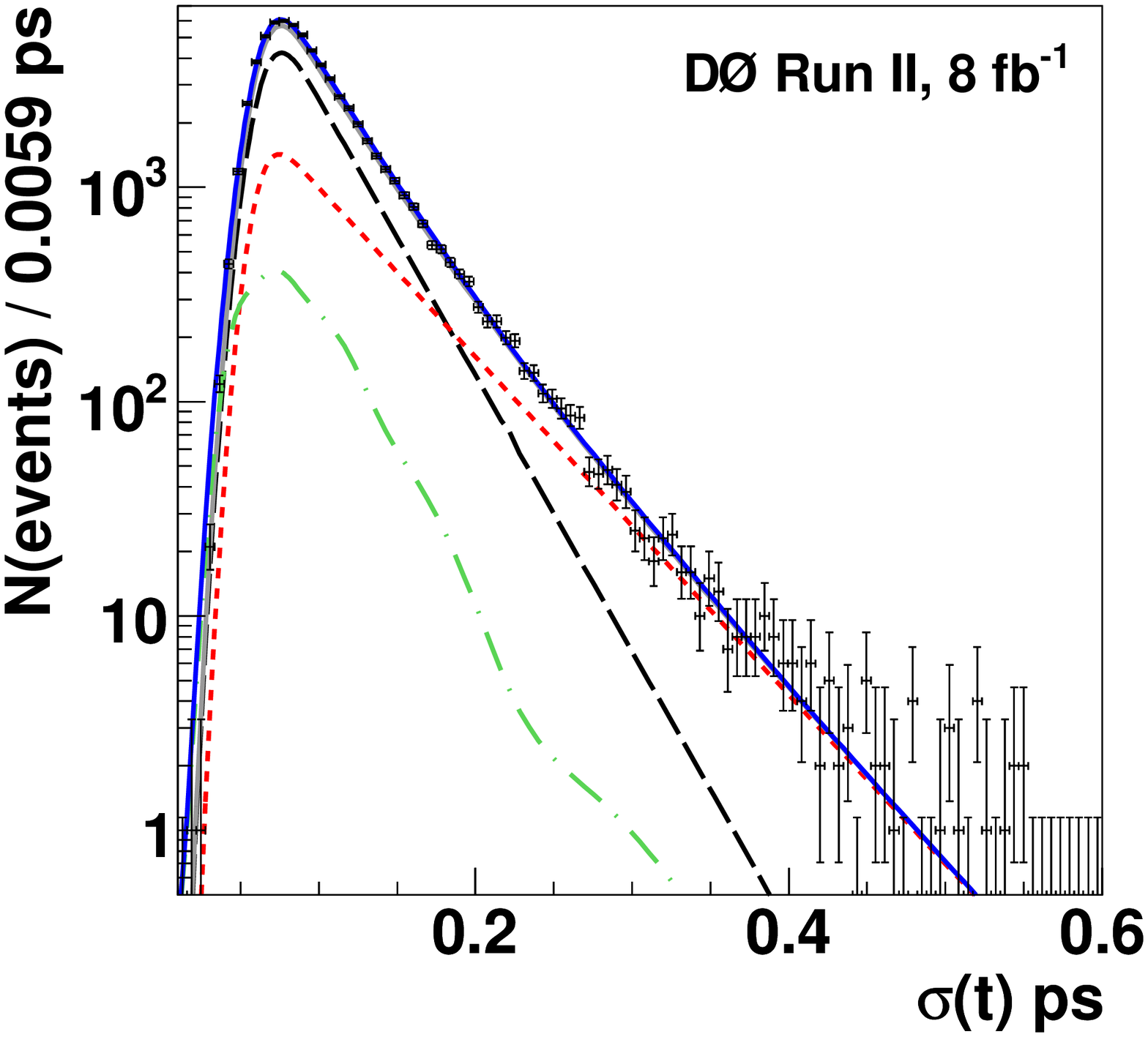}
\caption{(color online). Invariant mass,  proper decay time,  and
proper decay time uncertainty 
distributions for \bs\ candidates in the (top)  BDT sample
 and (bottom)  Square-cuts sample.
The curves are projections of the maximum likelihood fit. 
Shown are the signal (green dashed-dotted curve), prompt background (black dashed curve), non-prompt
background (red dotted curve), total background (brown long-dashed curve), and the sum of signal and
total background (solid blue curve).
}
 \label{fig:mwithfit}
 \end{center}
 \end{figure}


 \begin{figure}[htbp]
 \begin{center}
               \includegraphics[width=0.30\textwidth]{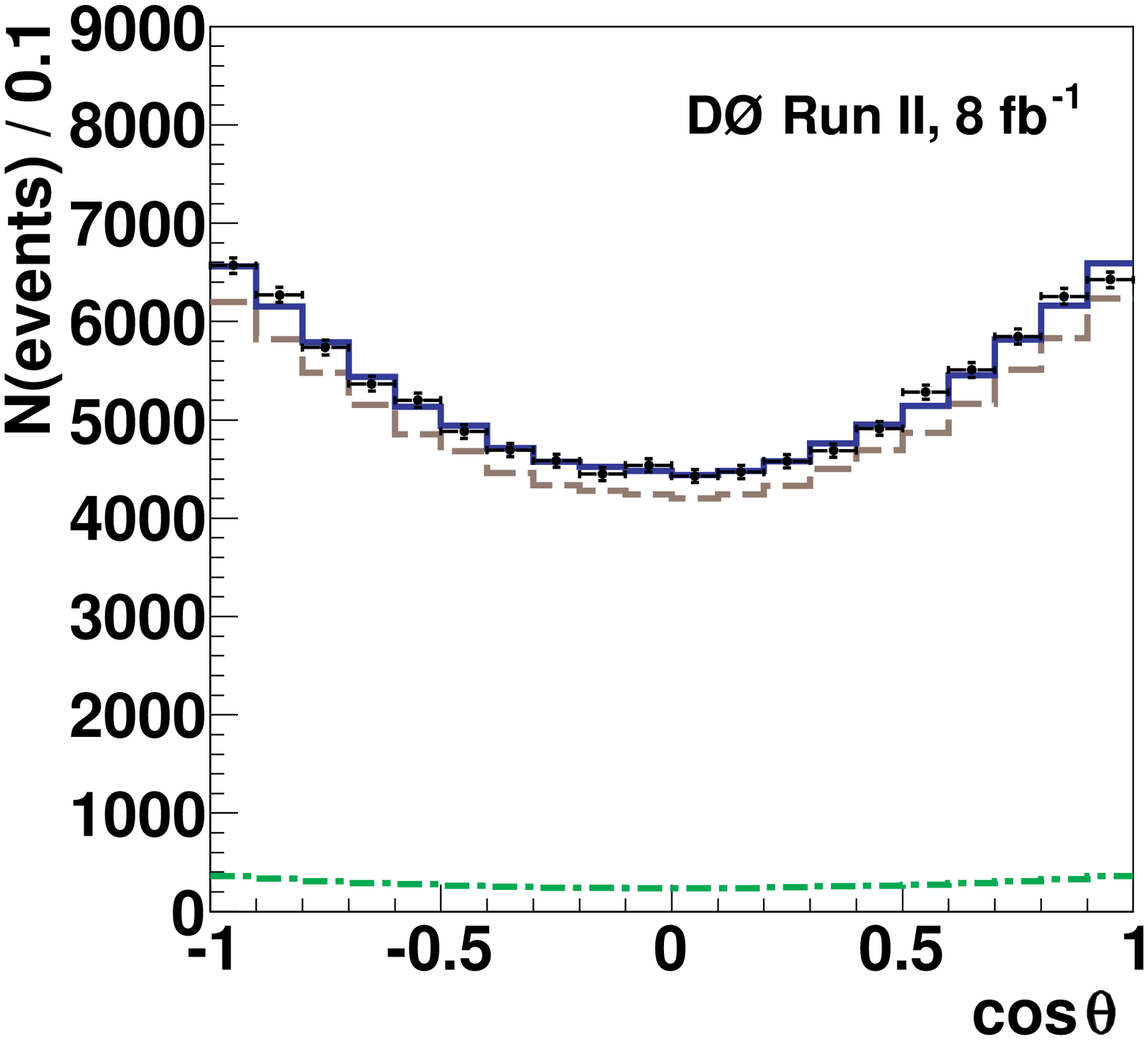}
               \includegraphics[width=0.30\textwidth]{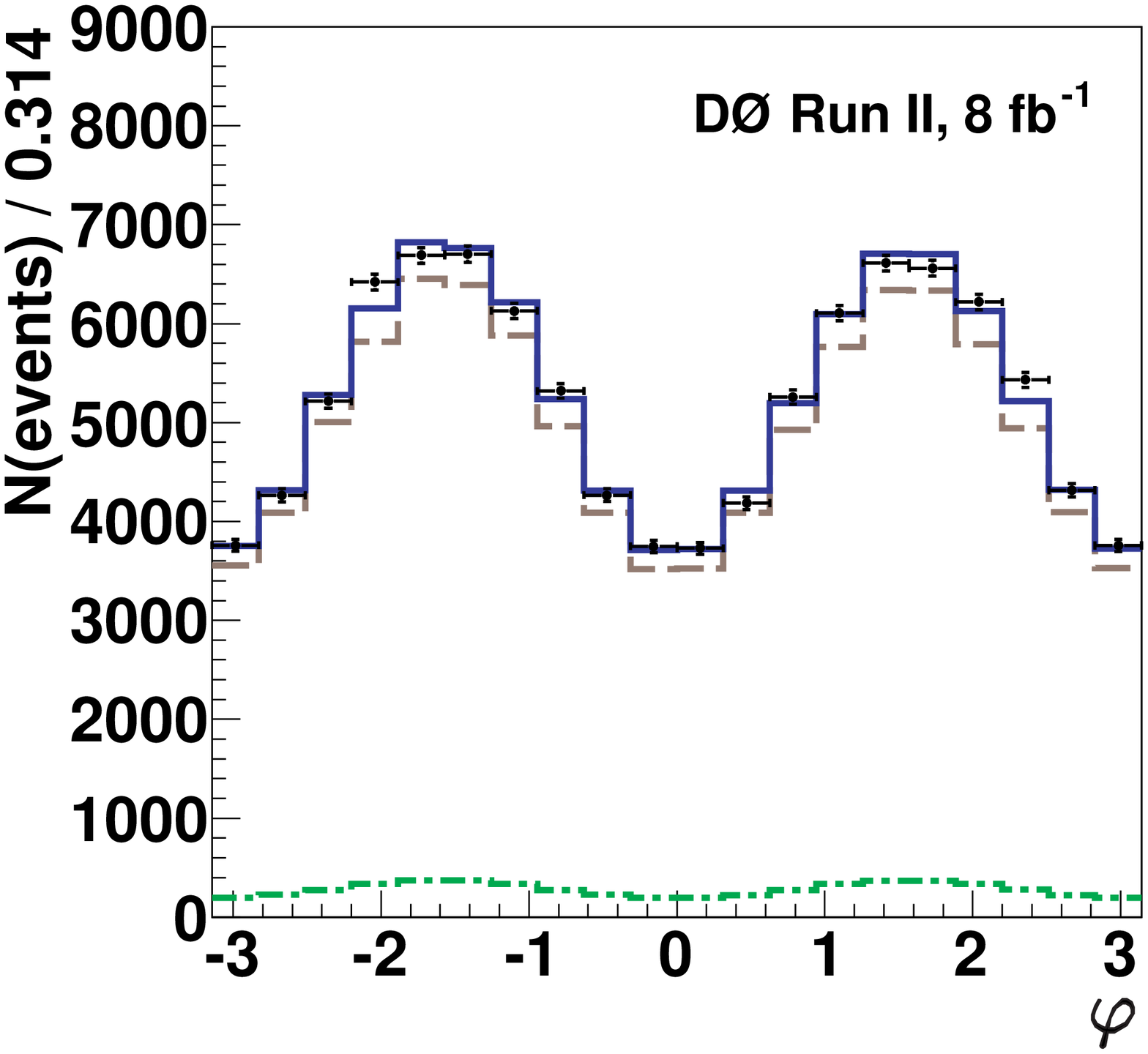}  
                \includegraphics[width=0.30\textwidth]{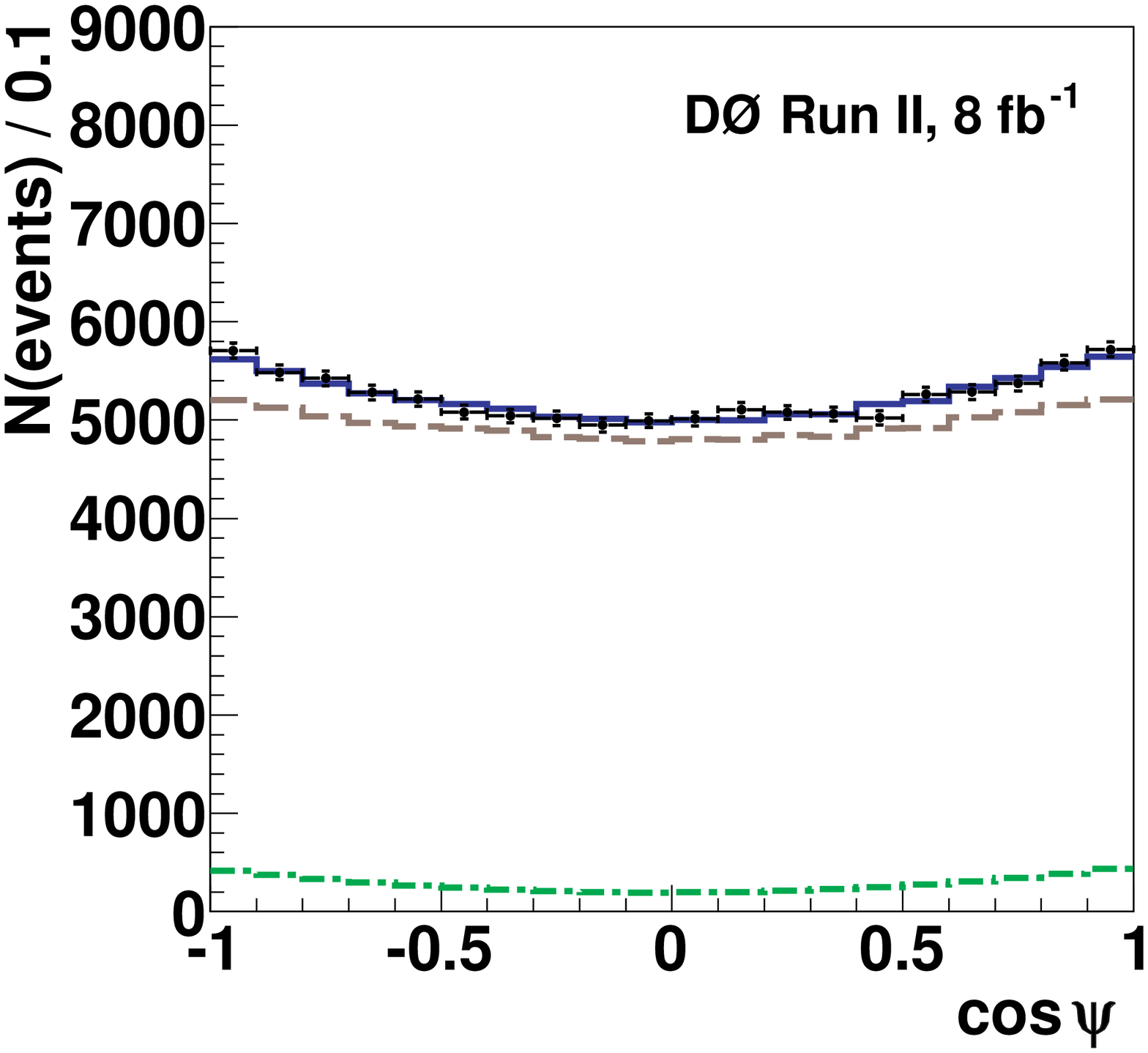}
               \includegraphics[width=0.30\textwidth]{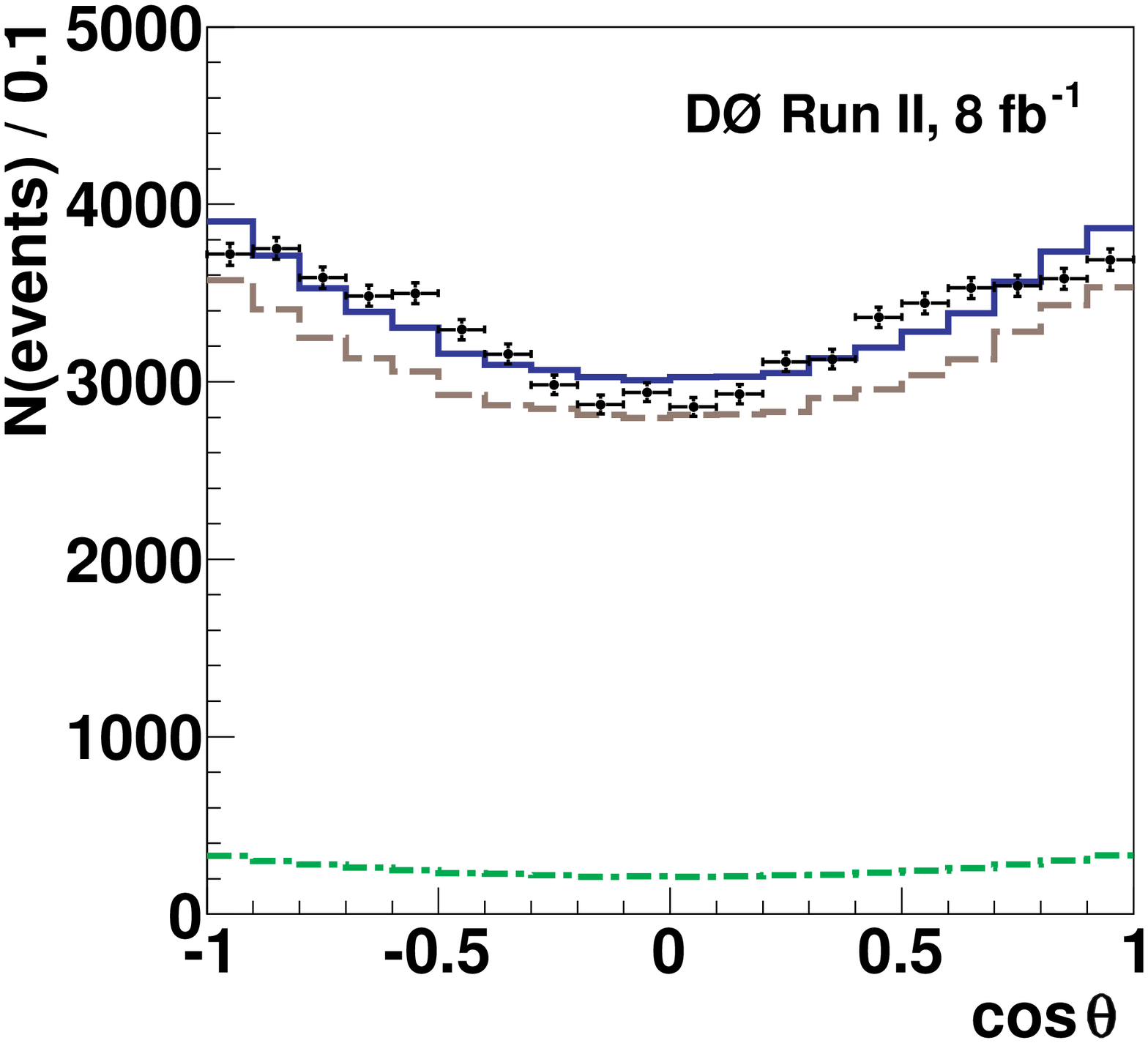}
               \includegraphics[width=0.30\textwidth]{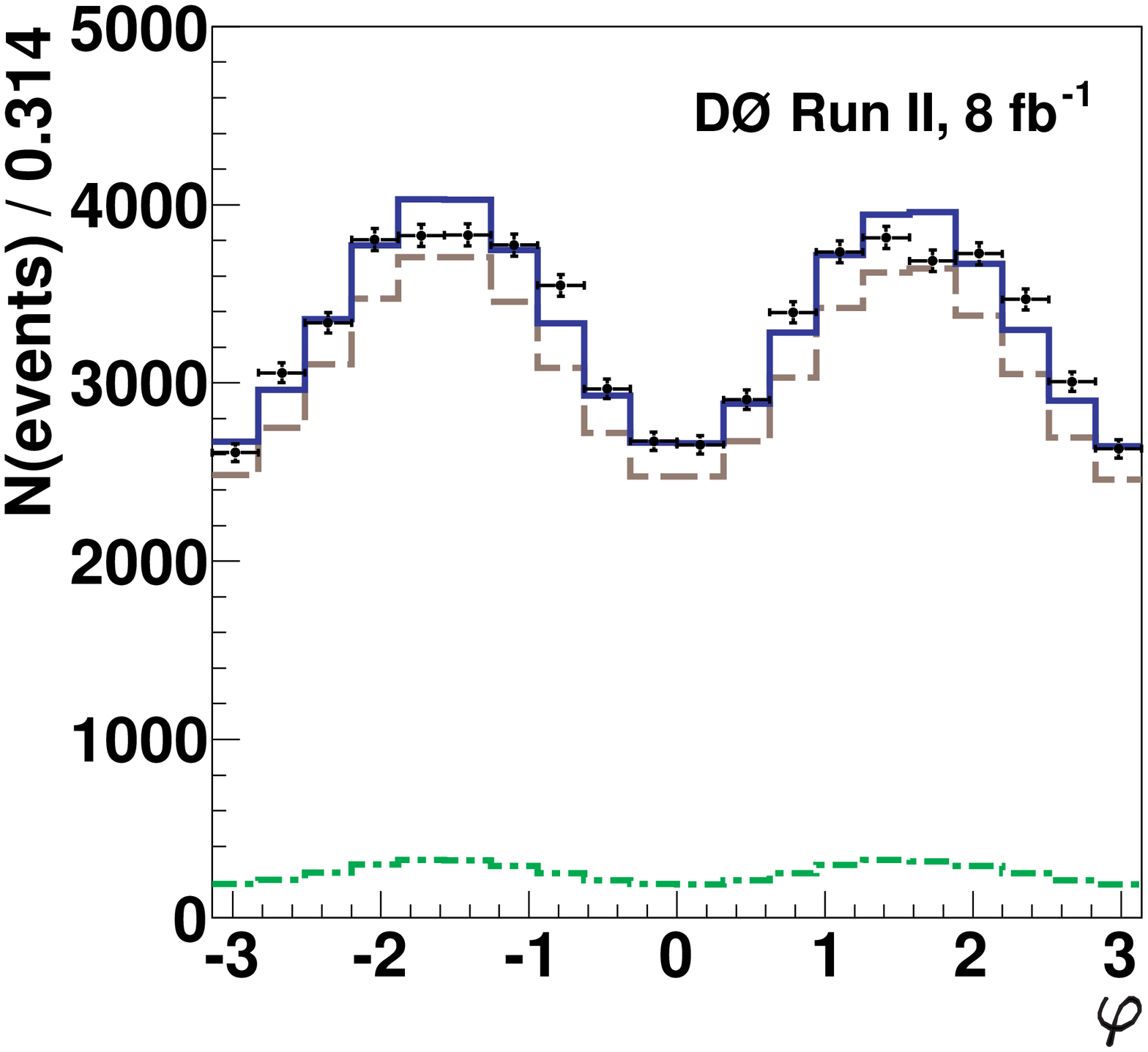}  
                \includegraphics[width=0.30\textwidth]{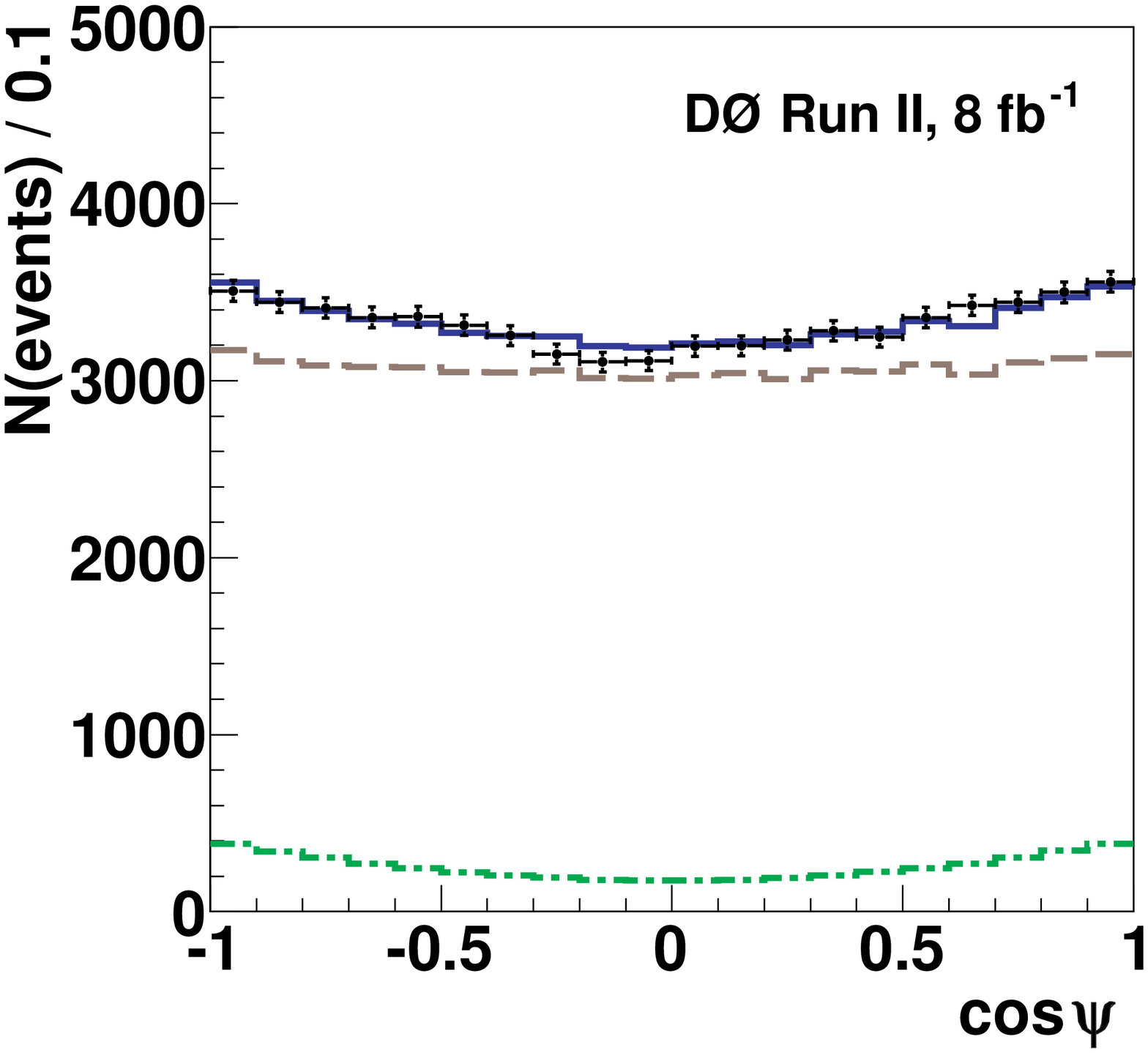}
\caption{(color online).  Distributions of transversity polar and azimuthal angles and $\cos \psi$
 for \bs\ candidates in the BDT sample (top)
 and  Square-cuts sample (bottom).
The curves are projections of the maximum likelihood fit. 
Shown are the signal (green dashed-dotted curve), total background (brown long-dashed curve)
and the sum of signal and total background (blue solid curve).
}
 \label{fig:awithfit}
 \end{center}
 \end{figure}



 \begin{figure}[htbp]
 \begin{center}
               \includegraphics[width=0.30\textwidth]{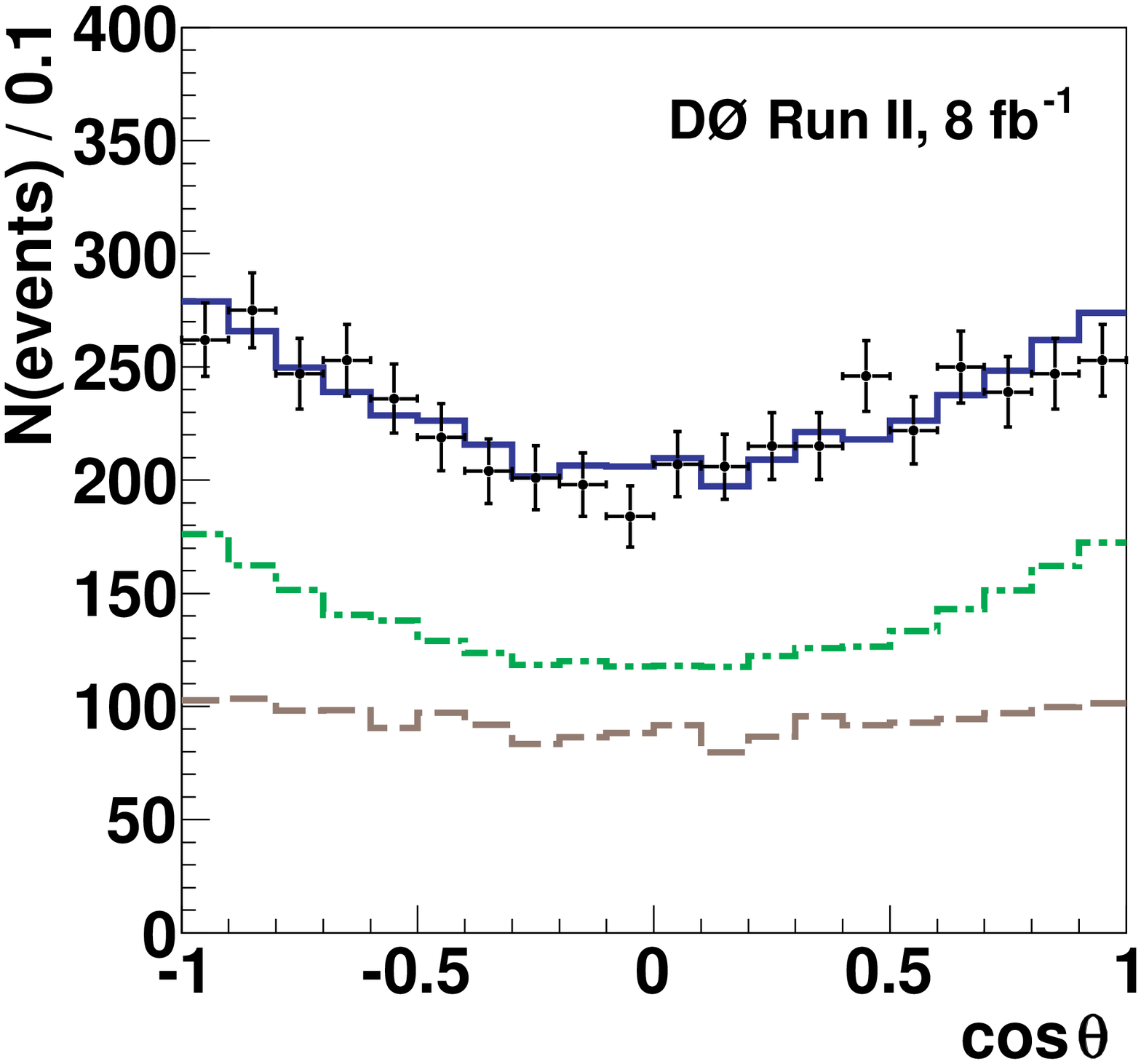}
               \includegraphics[width=0.30\textwidth]{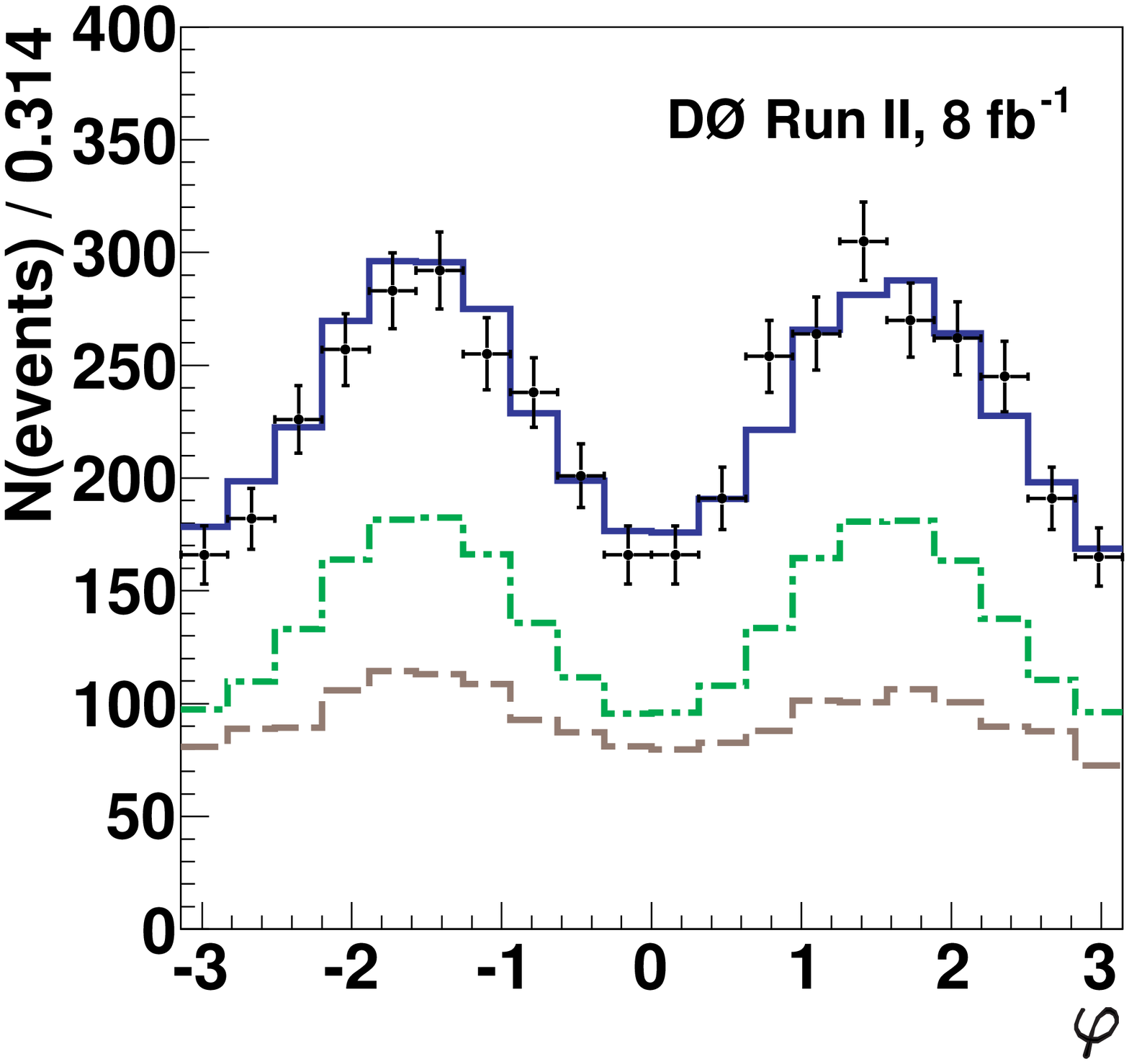}  
                \includegraphics[width=0.30\textwidth]{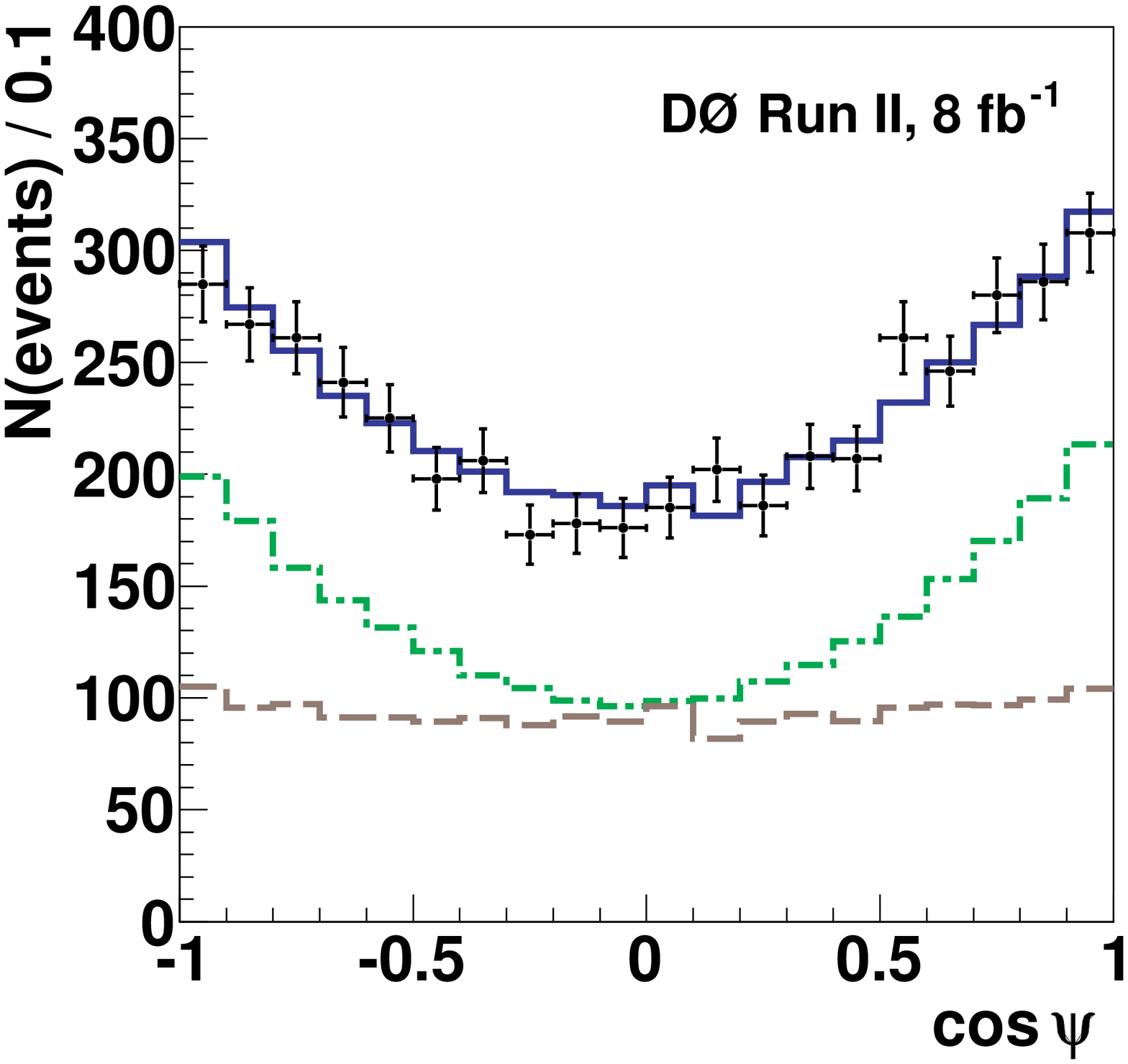}
               \includegraphics[width=0.30\textwidth]{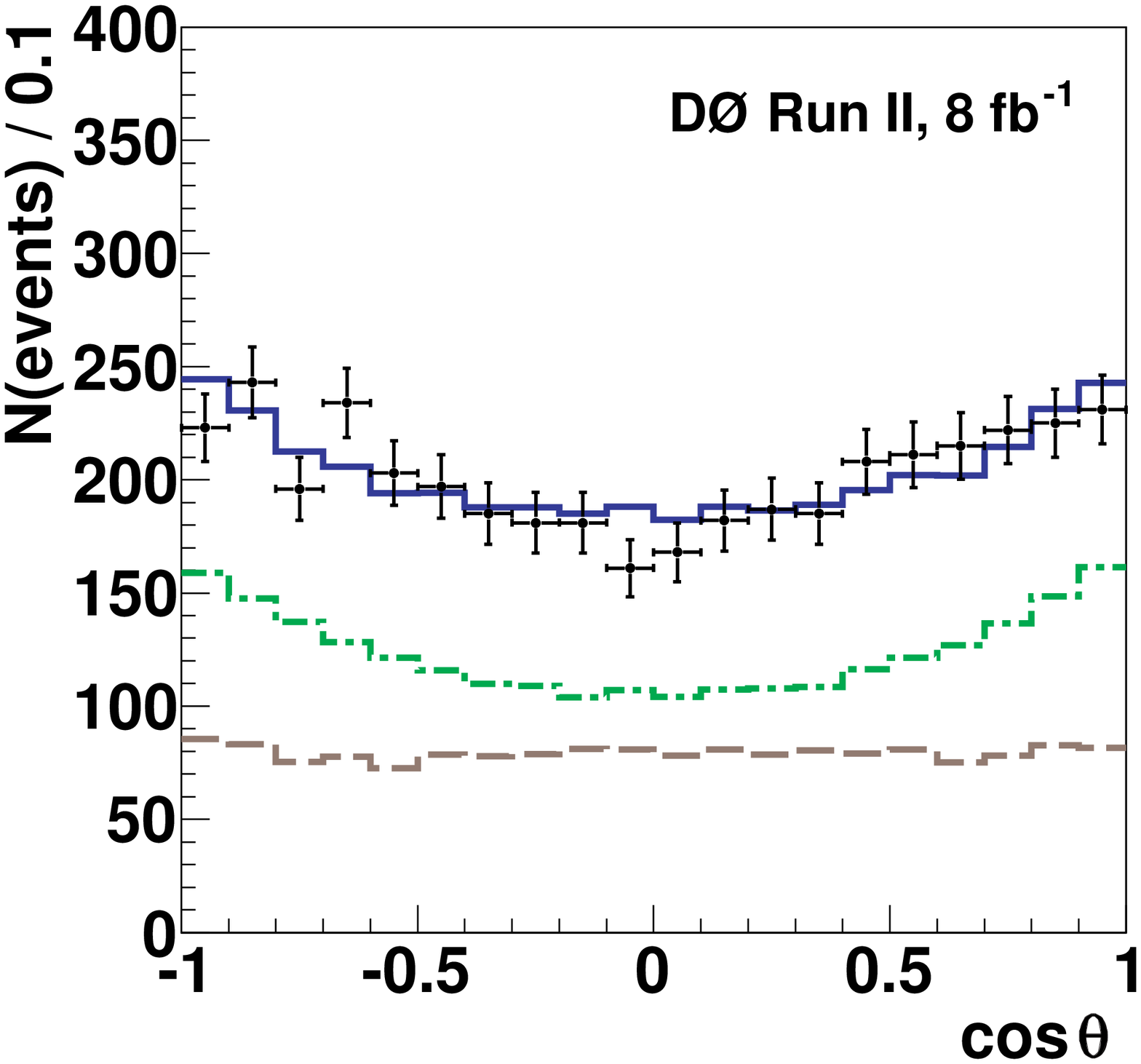}
               \includegraphics[width=0.30\textwidth]{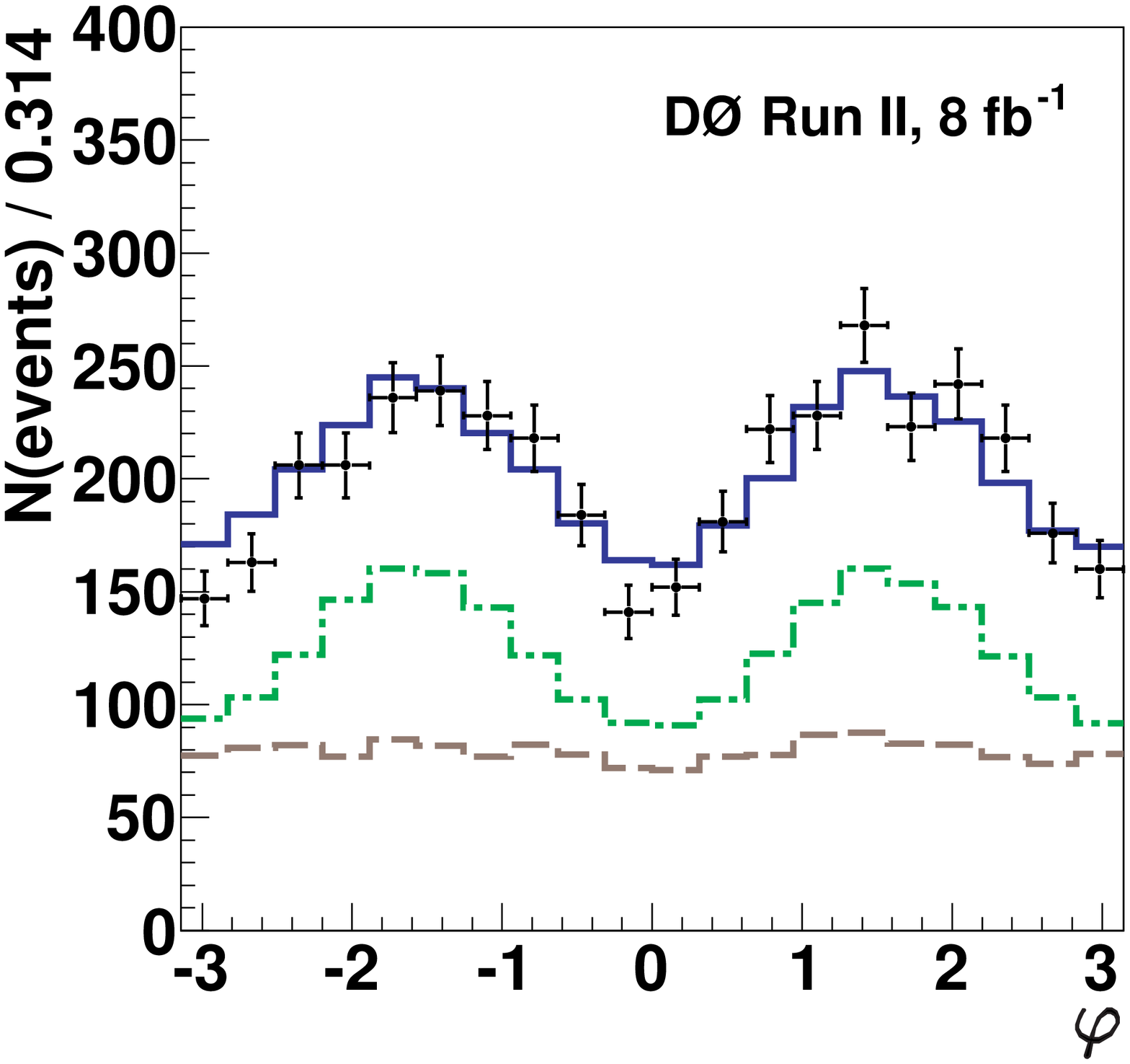}  
                \includegraphics[width=0.30\textwidth]{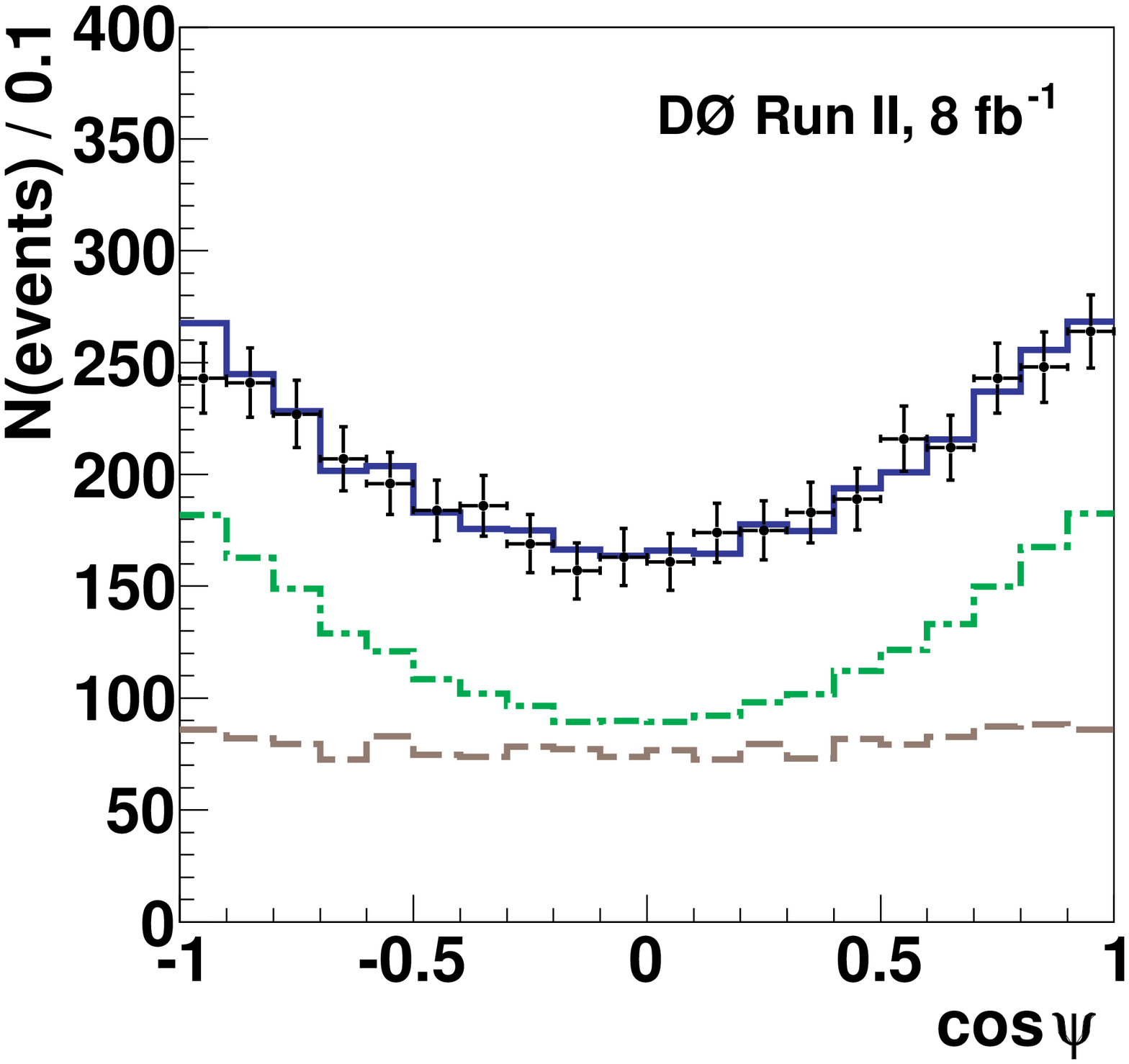}
\caption{(color online).
Distributions of transversity polar and azimuthal angles and $\cos \psi$
for \bs\ candidates in the BDT sample (top) and  Square-cuts sample (bottom).
The signal contribution is enhanced, relative to the distributions
shown in Fig.~\ref{fig:awithfit}, by additional requirements on the reconstructed
mass of the $B^0_s$ candidates ($5.31 < M(B^0_s) < 5.43 $ GeV)
and on the proper time $t > 1.0$ ps.
The curves are projections of the maximum likelihood fit. 
Shown are the signal (green dashed-dotted curve), total background (brown long-dashed curve)
and the sum of signal and total background (blue solid curve).
}
 \label{fig:a2withfit}
 \end{center}
 \end{figure}

\end{widetext}

\subsection{\label{sec:syst} Systematic uncertainties}

There are several possible sources of systematic uncertainty in
the measurements. 
These uncertainties are estimated as described below.

\begin{itemize}

\item {\bf Flavor tagging}:
The measured flavor mistag fraction suffers from uncertainties
due to the limited number of events in  the data samples for the decay 
$B_d^0\rightarrow \mu \nu D^{(*)\pm}$. The nominal calibration of the
flavor tagging dilution is determined as a weighted average of four
samples separated by the running period. As an alternative,
we  use two separate calibration parameters, one for the data collected 
in running periods
IIa and  IIb1, and one for the IIb2 and IIb3 data.
 We also
alter the nominal parameters by their uncertainties.
We find the effects of the changes to the flavor mistag variation
negligible. 

\item {\bf Proper decay time resolution}:
Fit results can be affected by the uncertainty of 
the assumed proper decay time resolution function. To assess the effect,
we have used two alternative parametrizations obtained by random
sampling of the resolution function.

\item {\bf Detector acceptance}:
The effects of imperfect modeling of the
detector acceptance and of the
selection requirements are estimated by investigating the
consistency of the fit results
for the sample based on the BDT selection and on the Square-cuts selection.
Although the overlap between the two samples is
70\%, and some statistical differences are
expected, we interpret the differences in the results
as a measure of systematic effects.

The two event selection approaches have different merits.
The BDT-based approach uses more information on each event,
and hence it allows a higher signal yield at lower background.
However, it accepts signal events of lower quality
(large vertex $\chi^2$ or proper decay time uncertainty)
that are rejected by the square cuts.
Also, the BDT-based approach uses the $M(KK)$ distribution
as a discriminant 
in the event selection, affecting the results
for the parameters entering the $\cal S - \cal P$
interference term, particularly the ${\cal S}$-wave fraction
$F_S$ and the phase parameters.

The main difference between the two samples is in the kinematic ranges
of final-state kaons, and so the
angular acceptance functions and MC weights (see Appendix~\ref{appmcdata})
are different
for the two samples.
Imperfections in the modelling of the $B_s^0$ decay kinematics and
estimated acceptances, and in the
treatment of the MC weighting, are reflected in differences
between fit results. The differences are used
as an estimate of this class of systematic uncertainty.

\item {\bf {\boldmath $M(KK)$} resolution:}
The limited $M(KK)$ resolution may affect the results
of the analysis, especially the phases and the $\cal S$-wave fraction $F_S$, 
through the dependence of the $\cal S - \cal P$ interference term
on the $\cal P$-wave mass model. 
In principle, the function of Eq.~(\ref{eqn:gDef}) should be replaced by a Breit-Wigner
function convoluted with a Gaussian. We avoid this complication by 
approximating the smeared $\cal{P}$-wave amplitude by a Breit-Wigner
function where the width $\Gamma_{\phi}$ of Eq.~(\ref{eqn:gDef}) is set to
twice the world average value to account for the detector resolution effects.
A MC simulation-based estimate of the scale factor for the event selection criteria 
used in this analysis yields a value in the range 1.5 -- 1.7. The  resulting 
complex integral of the  $\cal  S- \cal P$ interference 
has an absolute value behavior closer to the data, but
a distorted ratio of the
real and imaginary parts compared to Eq.~(\ref{eqn:gDef}).
 We repeat the fits using this altered 
$\phi(1020)$ propagator as a measure of the sensitivity to
the $M(KK)$ resolution.

\end{itemize}

Tables  \ref{roofitresults-cut10} and  \ref{roofitresults-sq}
compare results for the default fit and the alternative fits
discussed above. The differences between the best-fit values
provide a measure of systematic effects.
For the best estimate of the credible intervals for all the measured physics 
quantities, 
we conduct MCMC studies
described in the next section.

Other sources of systematic uncertainties like the functional model
of the background mass, lifetime and angle distributions were 
studied and give a negligible contribution.

\section{\label{sec:mcmc} Bayesian credibility intervals from MCMC studies}

The maximum likelihood fit provides the best values of all free parameters,
including the signal observables and background model parameters,
their statistical uncertainties and  their full correlation matrix. 

In addition to the free parameters determined in the fit,
the model depends on a number of external constants whose inherent
uncertainties are not taken into account in a given fit.
Ideally, effects of uncertainties of external constants,
such as time resolution parameters, flavor tagging dilution
calibration, or detector acceptance, should be included in the model
by introducing the appropriate parametrized probability density functions
and allowing the parameters to vary. Such a procedure of maximizing the
likelihood function over the external parameter space would greatly increase
the number of free parameters and would be prohibitive.
Therefore, as a trade-off, we apply a random sampling
of external parameter values within their uncertainties,
we perform the analysis for thus created ``alternative universes'',
and we  average the results. To do the averaging in the multidimensional
space, taking into account non-Gaussian parameter distributions
and correlations, we use the MCMC technique.

\subsection{The method }

The MCMC  technique uses the
Metropolis-Hastings algorithm \cite{MetHast} to generate a sample 
representative to a given probability distribution.
The algorithm generates  a sequence of  ``states'', a Markov chain,
 in which each state depends
only on the previous state. 

To generate a Markov chain for a given data sample, we start from the best-fit point $\vec{x}$. 
We randomly generate a point $\vec{x}'$ in the parameter space according to the multivariate normal  distribution
$\exp(-(\vec{x}'-\vec{x})\cdot \Sigma \cdot (\vec{x}'-\vec{x})/2)$, where
$\Sigma$ is the covariance matrix between the best fit current point $\vec{x}$ in the chain and next random point $\vec{x}'$.
The best-fit point and the covariance matrix are obtained
from a maximum likelihood fit over the same data sample. The new point
is accepted if ${\cal L}(x')/{\cal L}(x)>1$,
otherwise it is accepted with the probability 
${\cal L}(x')/{\cal L}(x)$.
The process is continued
until a desired number of states is achieved.
To avoid a bias due to the choice of the initial state,
we discard the early states which may ``remember'' the initial state.
Our studies show  that the initial state is ``forgotten'' after approximately
50 steps. We discard the first 100 states in each chain.

\subsection{General properties of MCMC chains for the BDT-selection  and Square-cuts samples}

We generate 8 MCMC chains, each containing one million states:
a nominal and three alternative chains each for the BDT-selection and Square-cuts samples,
according to the fit results  presented in 
Tables~\ref{roofitresults-cut10} and  \ref{roofitresults-sq}.

\begin{widetext}

 \begin{figure}H]
 \begin{center}
\subfigure[]
{ \includegraphics*[width=0.30\textwidth]{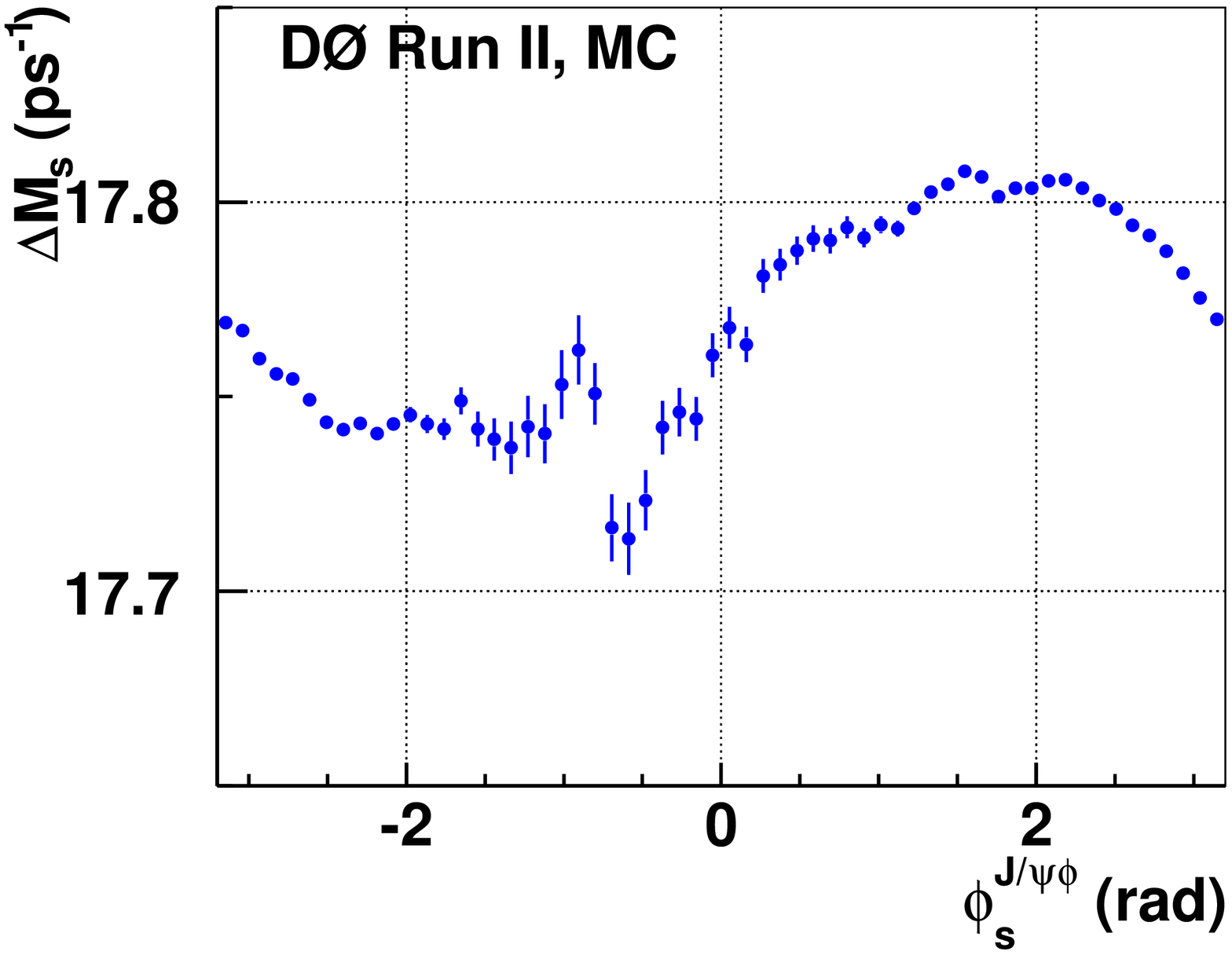}}
\subfigure[]
{\includegraphics*[width=0.30\textwidth]{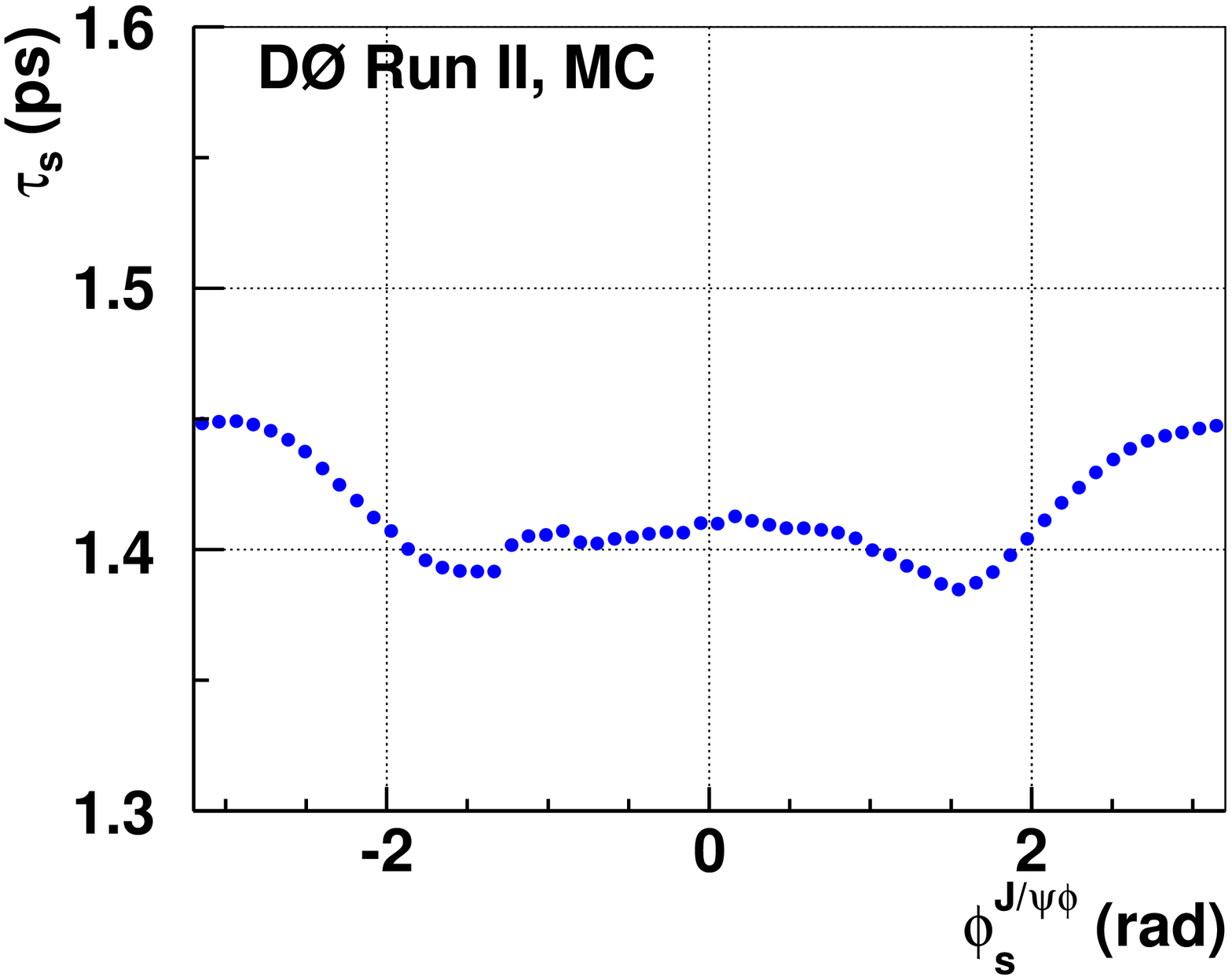}}
\subfigure[]
{\includegraphics*[width=0.30\textwidth]{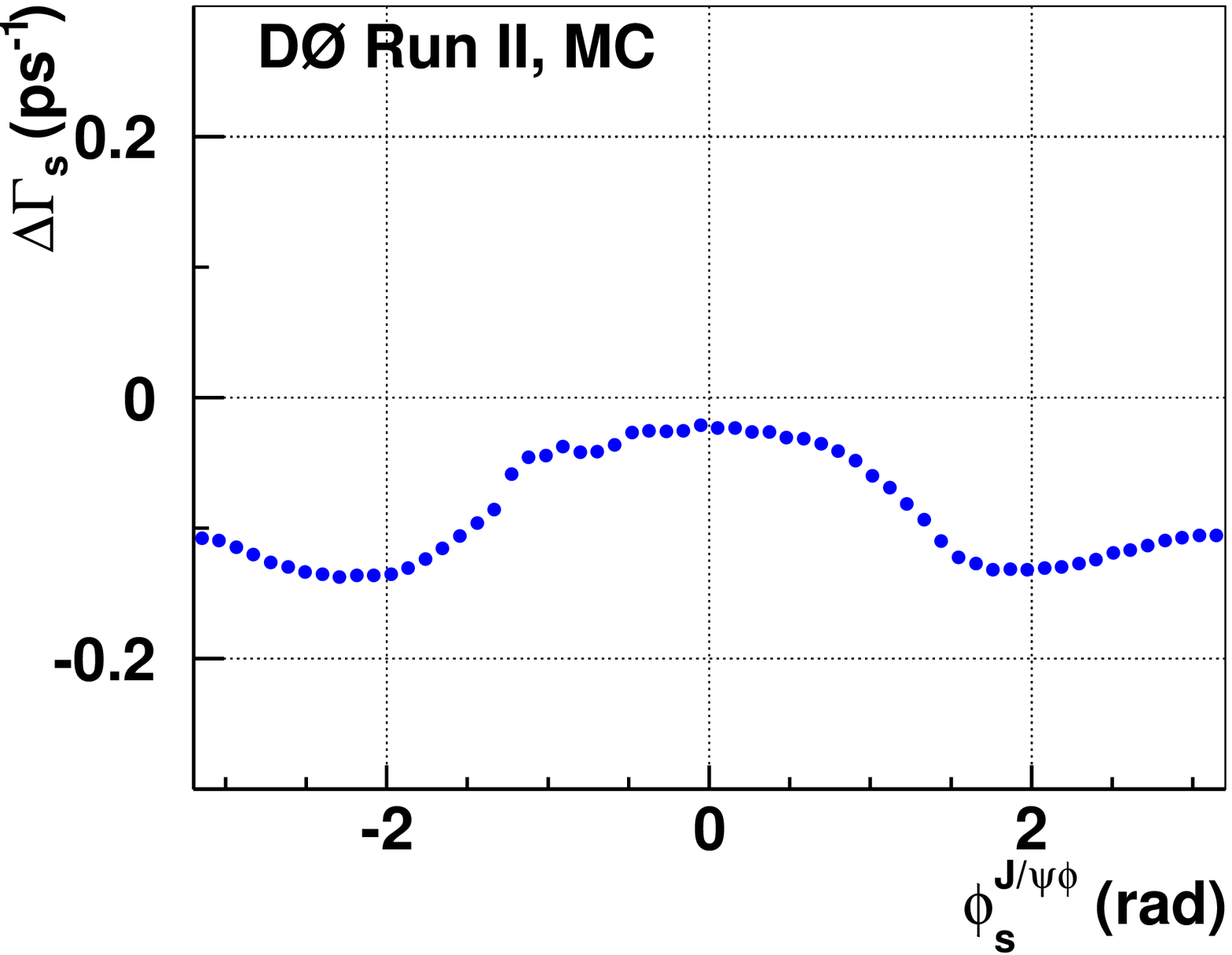}}
\subfigure[]
{\includegraphics*[width=0.30\textwidth]{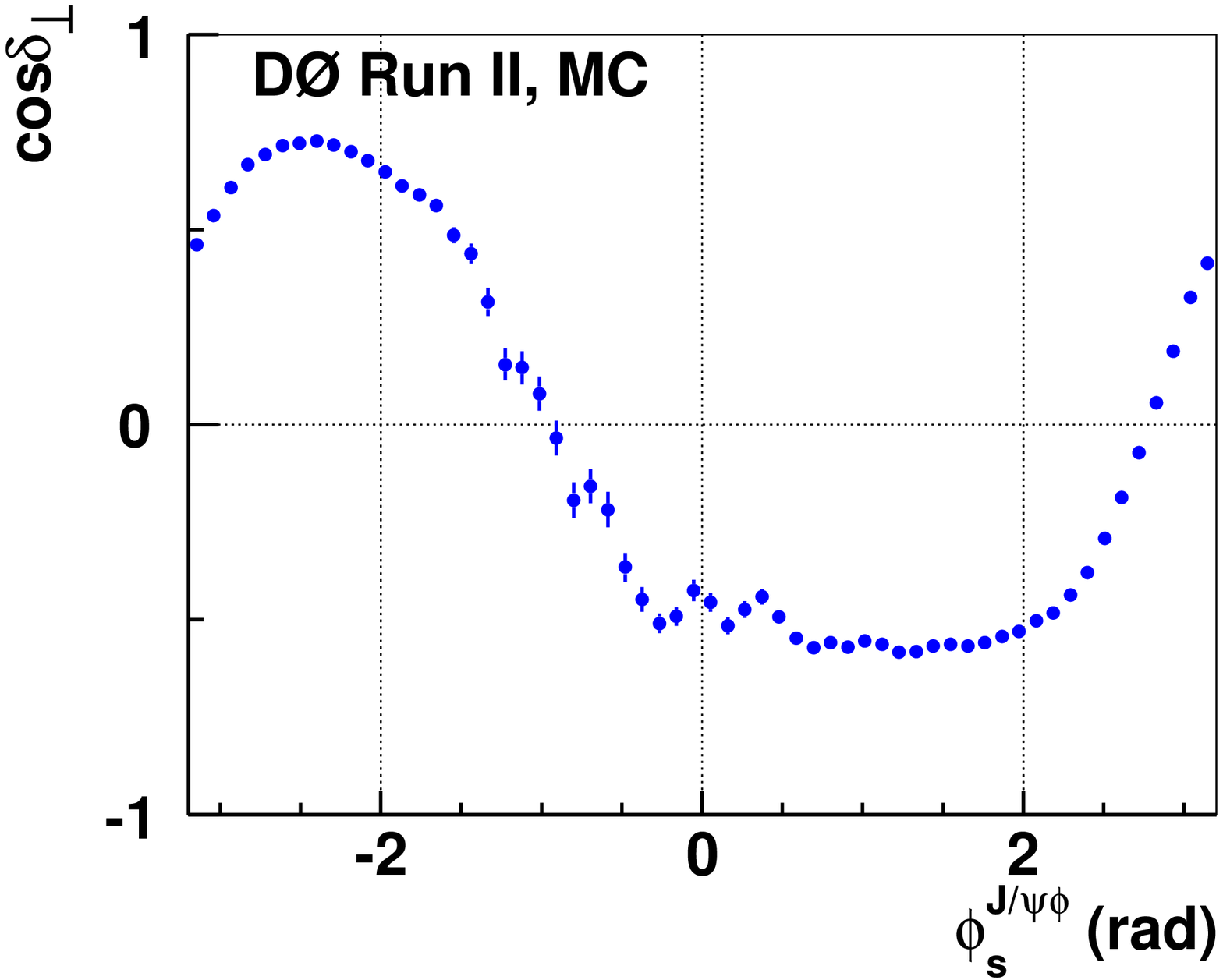}}
\subfigure[]
{ \includegraphics*[width=0.30\textwidth]{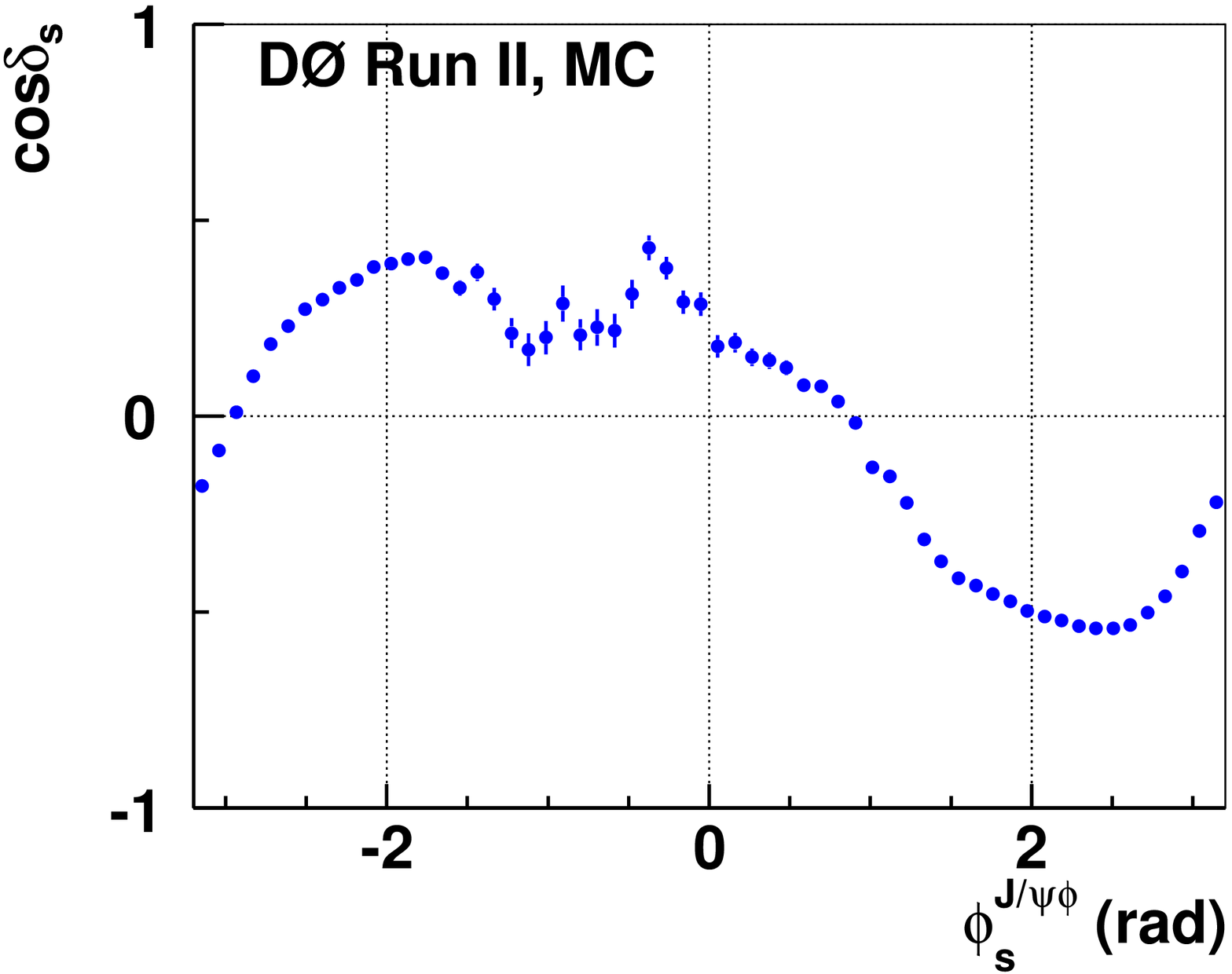}}
\subfigure[]
{\includegraphics*[width=0.30\textwidth]{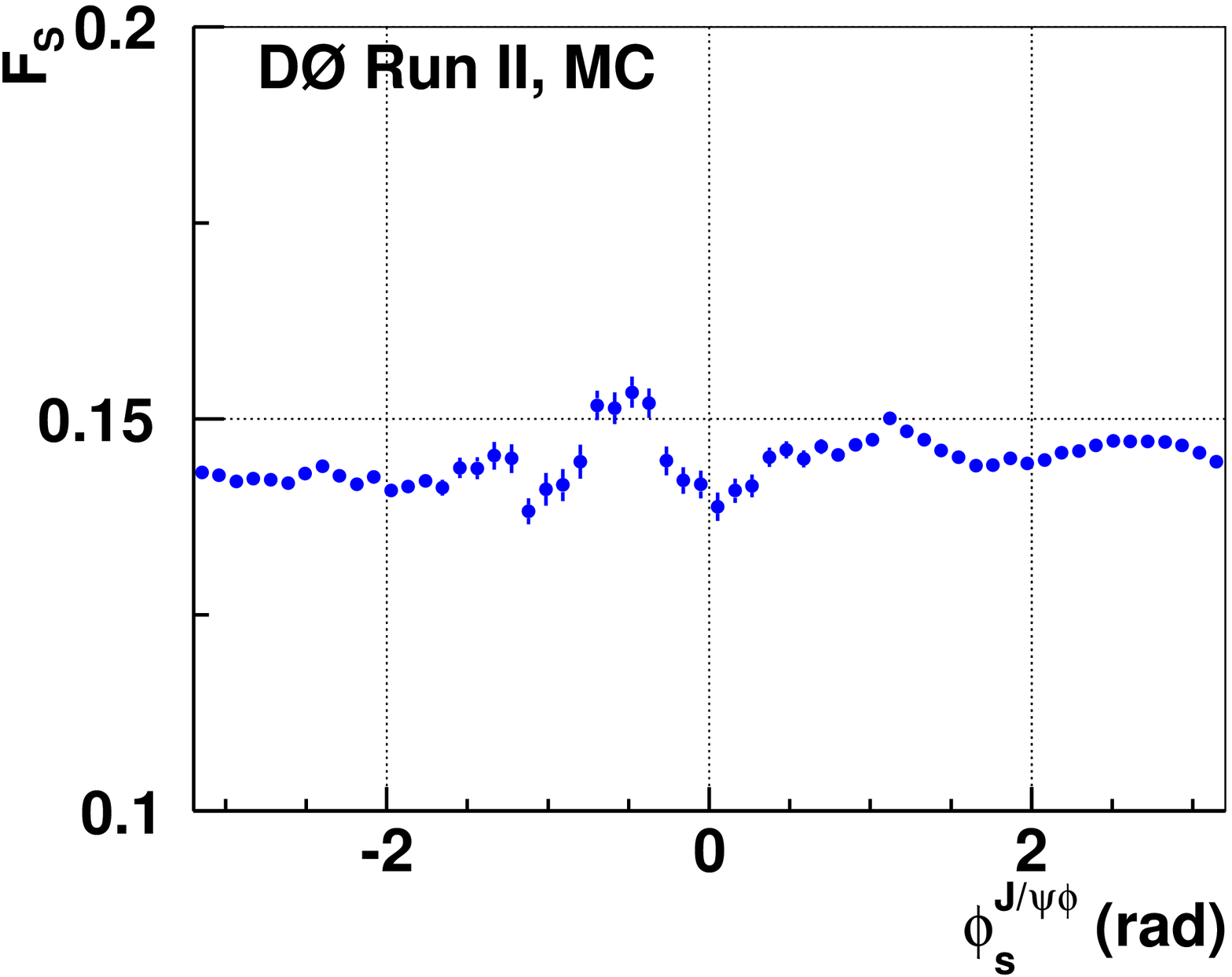}}
\caption{Profiles of $\Delta M_s$,  $\overline \tau_s$, $\Delta \Gamma_s$,
$\cos\delta_\perp$, $\cos\delta_s$, and  $F_S$, for  $\Delta \Gamma_s<0$,
 versus   \phis\
from the MCMC simulation for the BDT selection data sample.
}
 \label{fig:profDmvsphis_cut10dgm}
 \end{center}
 \end{figure}

\end{widetext}

 Figures~\ref{fig:profDmvsphis_cut10dgp}  and~\ref{fig:profDmvsphis_cut10dgm}  
illustrate the dependence of  $\phi_s^{J/\psi \phi}$
on other physics parameters, in particular on 
$\cos\delta_\perp$ and  $\cos\delta_s$. Each point shows the Markov Chain 
representation of the likelihood function integrated over all parameters
except the parameter of interest in a slice of \phis. For clarity,
the profiles are shown for $\Delta \Gamma_s>0$ and  $\Delta \Gamma_s<0$
separately.
The distributions for the Square-cuts sample are similar. 
We note the following salient features of these correlations
for  $\Delta \Gamma_s>0$:

\begin{widetext}

 \begin{figure}[H]
 \begin{center}
\subfigure[]
 {\includegraphics*[width=0.30\textwidth]{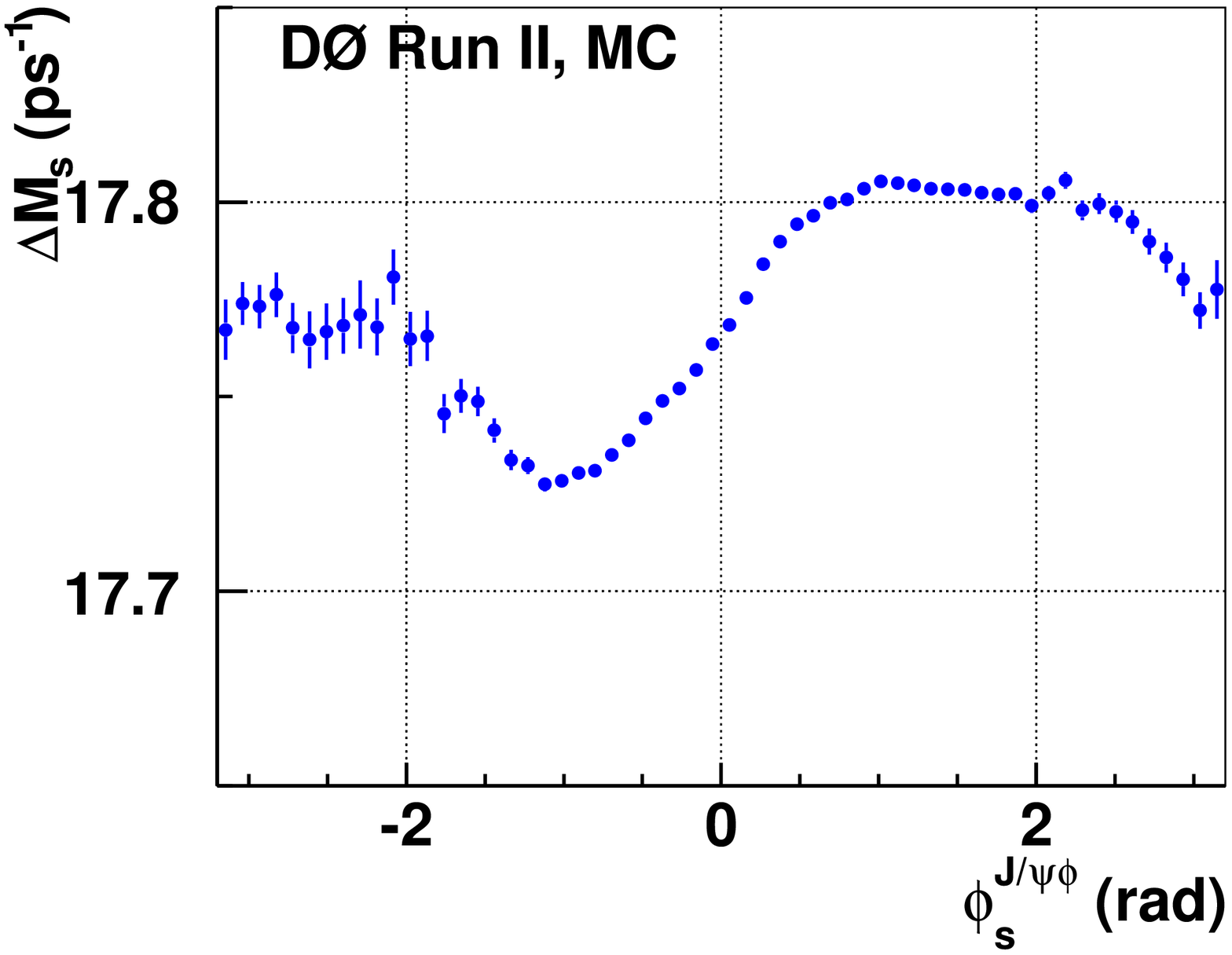}}
\subfigure[]
{\includegraphics*[width=0.30\textwidth]{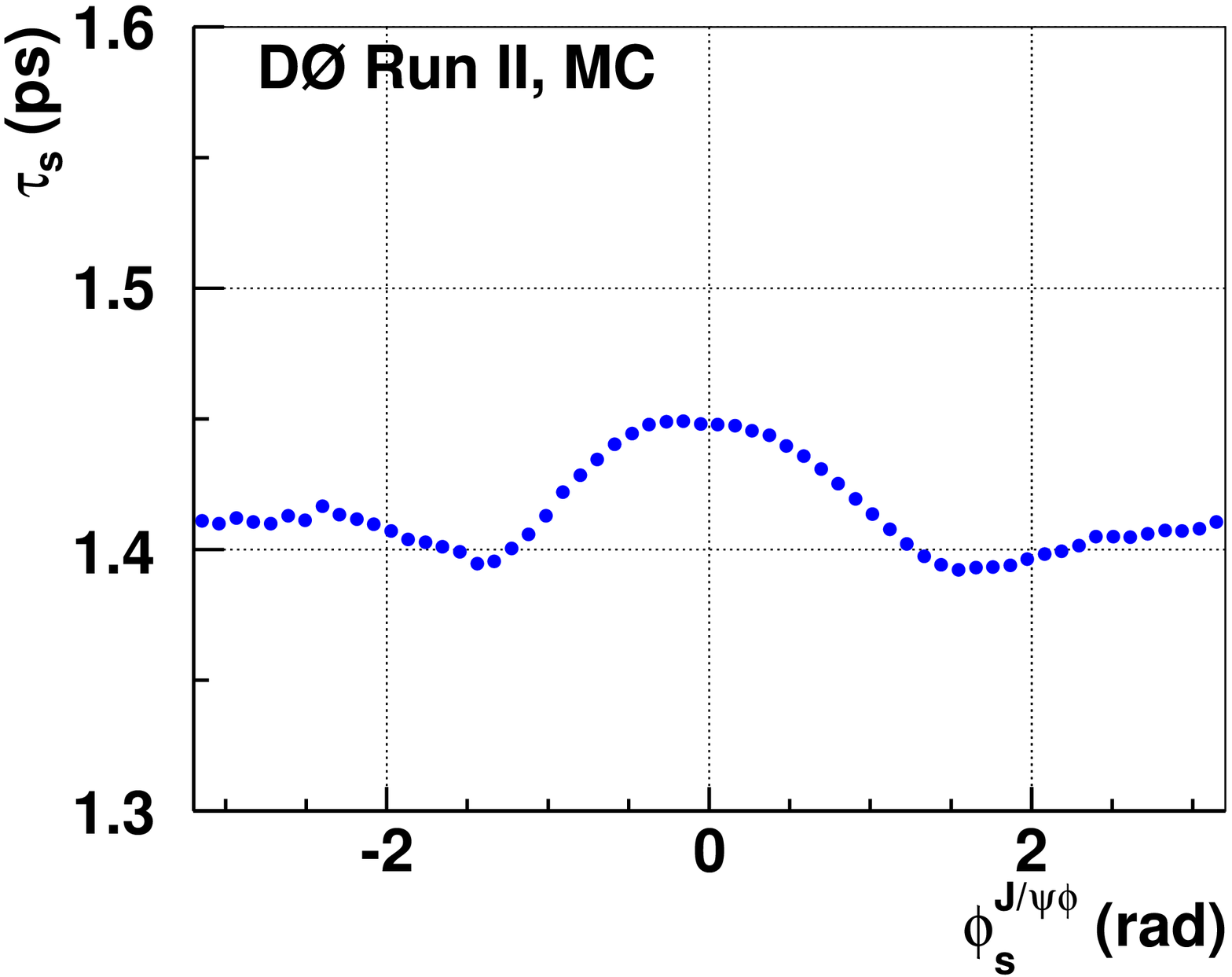}}
\subfigure[]
{\includegraphics*[width=0.30\textwidth]{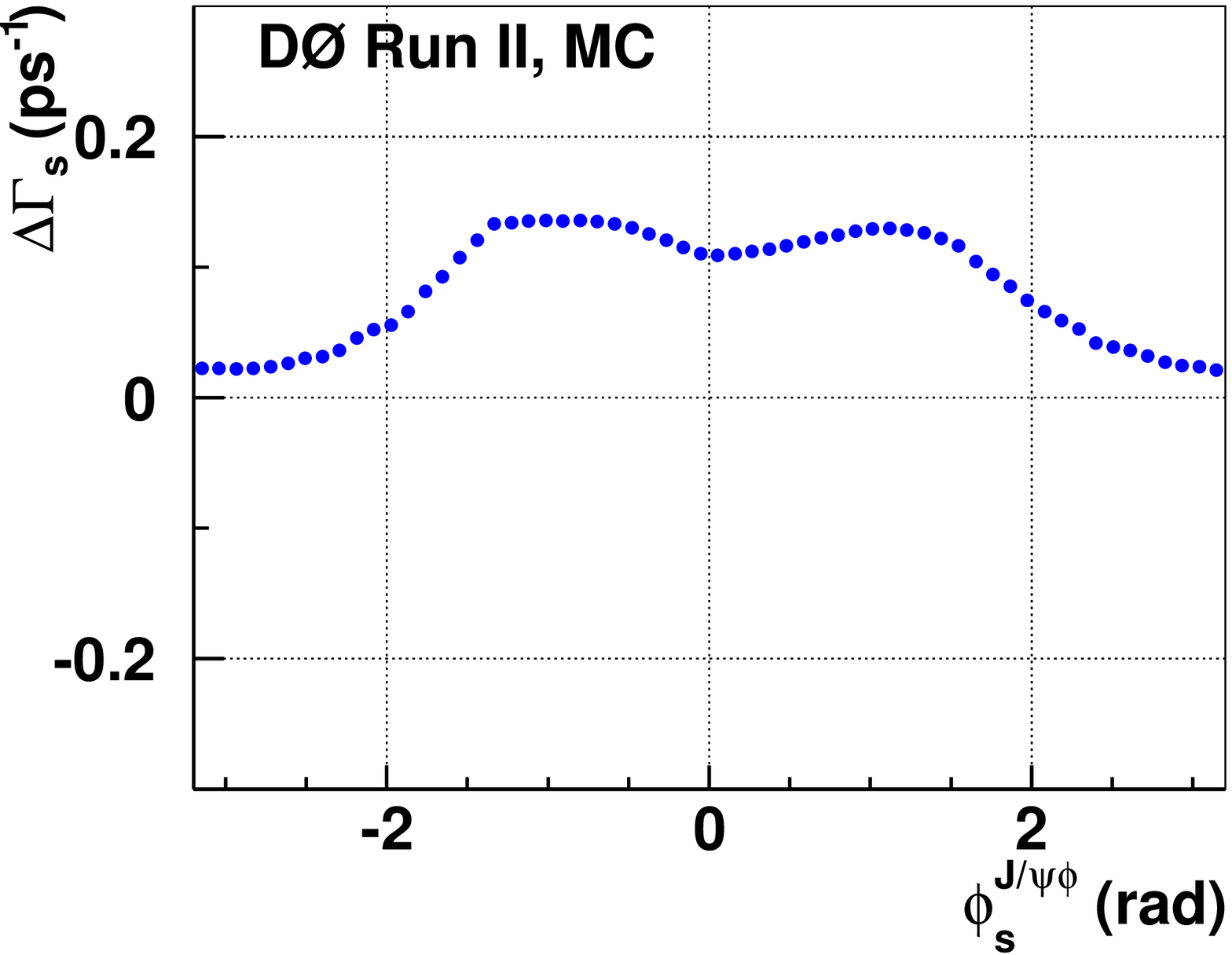}}
\subfigure[]
{\includegraphics*[width=0.30\textwidth]{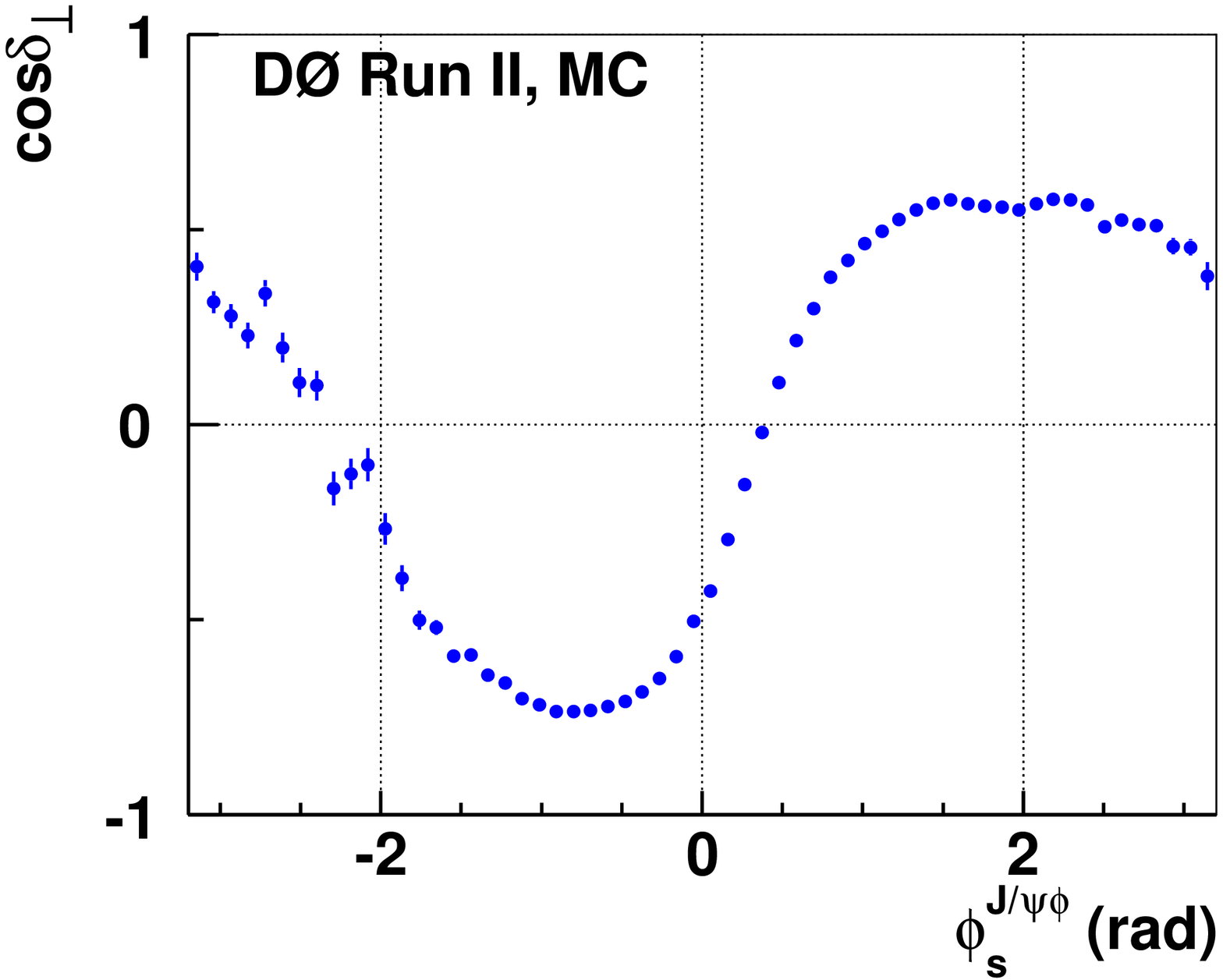}}
\subfigure[]
 {\includegraphics*[width=0.30\textwidth]{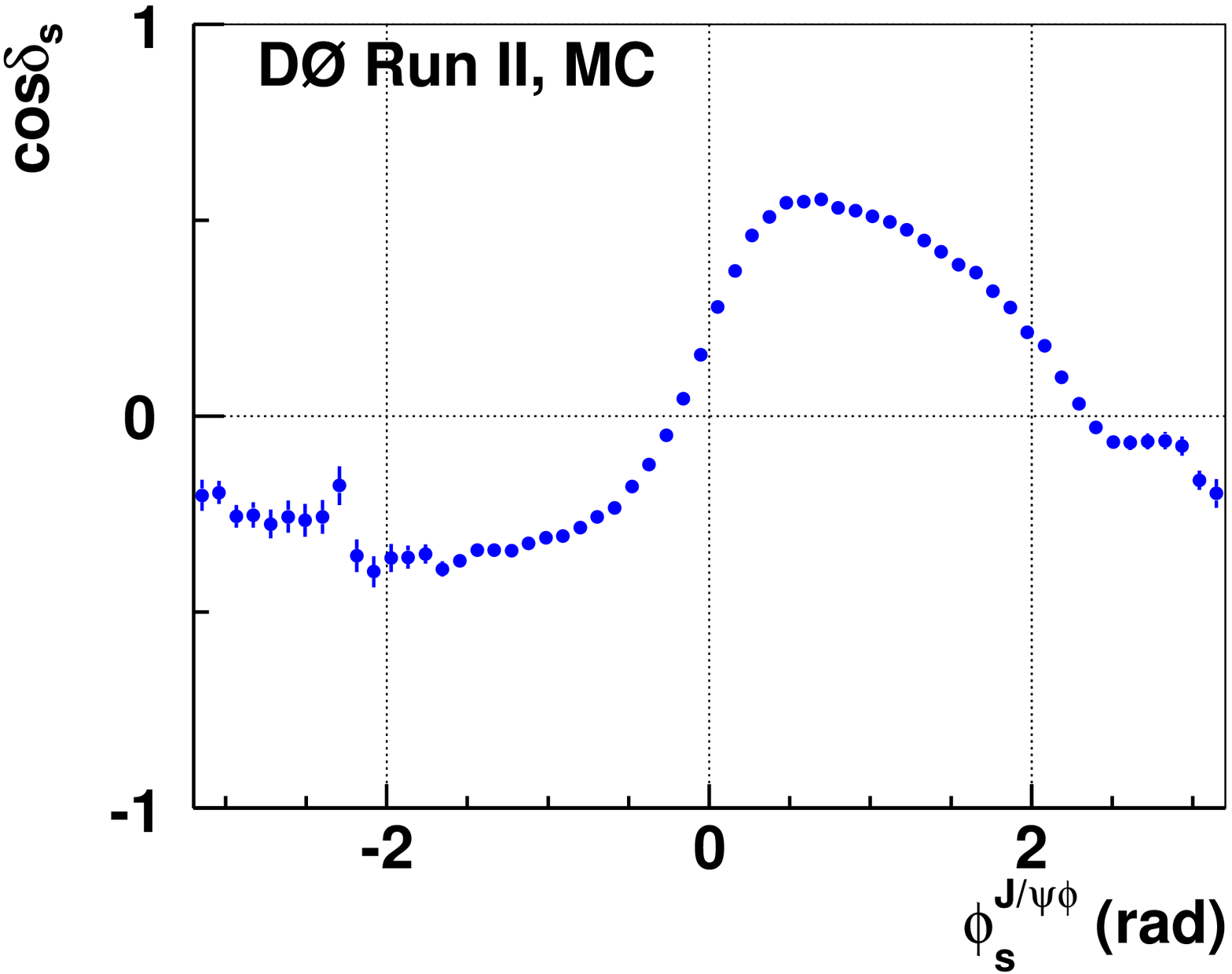}}
\subfigure[]
{\includegraphics*[width=0.30\textwidth]{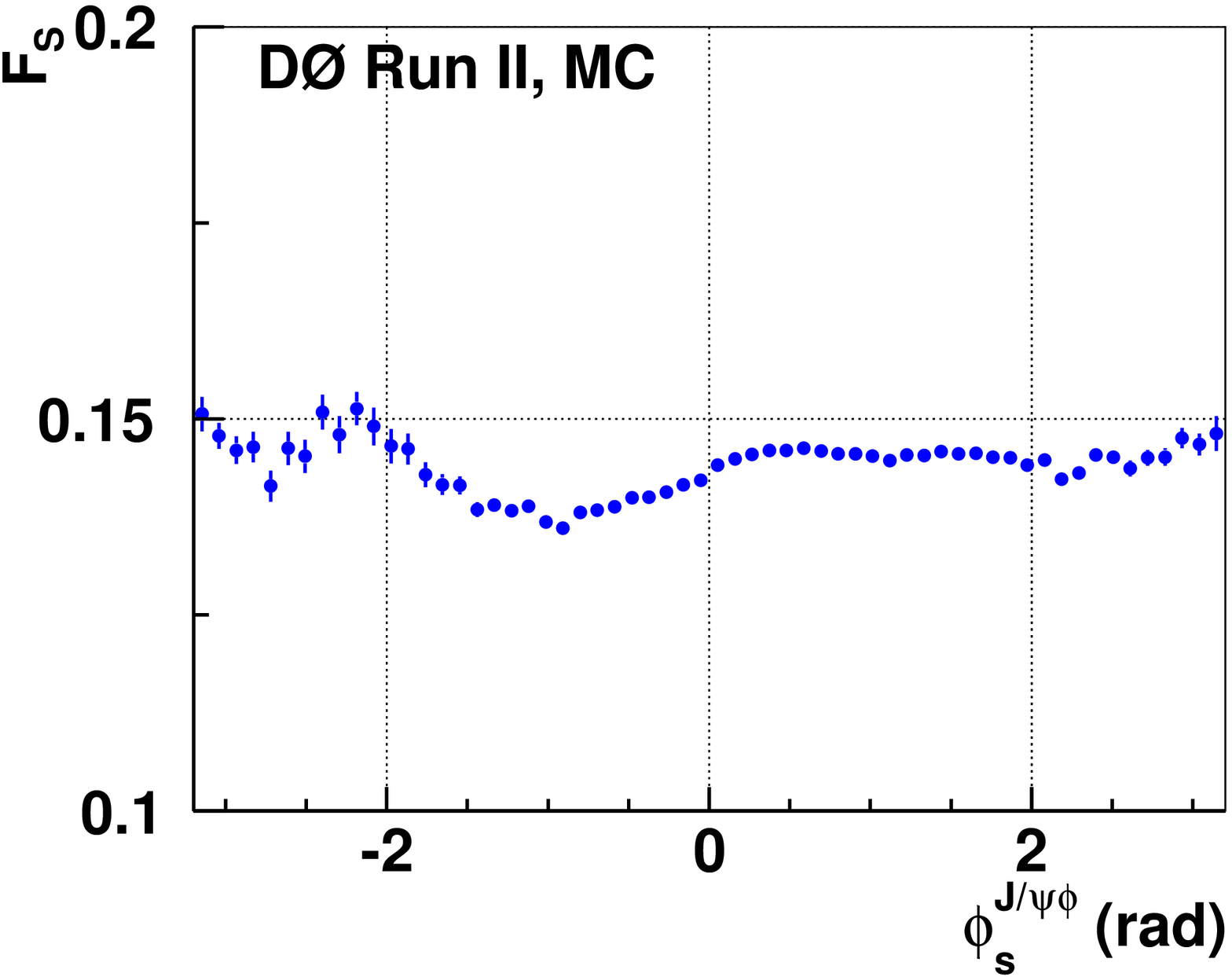}}
\caption{Profiles of $\Delta M_s$,  $\overline \tau_s$, $\Delta \Gamma_s$,
$\cos\delta_\perp$, $\cos\delta_s$, and  $F_S$, for  $\Delta \Gamma_s>0$,
 versus   \phis\
from the MCMC simulation for the BDT selection data sample.
}
 \label{fig:profDmvsphis_cut10dgp}
 \end{center}
 \end{figure}

\end{widetext}

\begin{itemize}

\item [a)]A positive correlation between  \phis\ and  $\Delta M_s$,
with the best fit of \phis\ changing sign as $\Delta M_s$ increases
(see also Fig.~\ref{fig:bs_dms_fourier} in Appendix~\ref{apposcill}).

\item [b)] A  correlation between   $|\phi_s^{J/\psi \phi}|$   
and $\overline \tau_s$, with  the highest $\overline \tau_s$
occuring at $\phi_s^{J/\psi \phi}=0$.

\item [c)] For \phis\ near zero,  $|\Delta \Gamma_s|$ 
increases with $|\phi_s^{J/\psi \phi}|$.

\item [d)] A strong positive correlation between  \phis\  and
 $\cos\delta_\perp$ near  $\phi_s^{J/\psi \phi}=0$, with  \phis\  changing sign as the average  $\cos\delta_\perp$
increases between $-0.8$ and $+0.8$. 
For the related decay \bddec\ the measured 
value is 
 $\cos\delta_\perp = -0.97$. This indicates  that a constraint
of  $\cos\delta_\perp$   to the \bddec\ value would
result in $\phi_s^{J/\psi \phi}<0$ with a smaller uncertainty.

\item  [e)] A strong positive correlation between  \phis\  and
 $\cos\delta_s$  near  $\phi_s^{J/\psi \phi}=0$,  with  \phis\  changing sign as the average  $\cos\delta_s$
increases between $-0.4$ and $+0.4$.

\item [f)] A weak correlation between  \phis\  and $F_S$,
with $F_S$ a few percent lower for  $\phi_s^{J/\psi \phi}<0$.

\end{itemize}

While we do not use any external numerical constraints on the polarization
amplitudes, we note that the best-fit values of their magnitudes
and phases
are consistent with  those measured in the $U(3)$-flavor related 
decay \bddec\ \cite{PDG}, up to the
sign ambiguities. Ref.~\cite{gr} predicts
that the phases of the polarization amplitudes in the two decay processes 
should agree within approximately 0.17 radians.
For $\delta_{\perp}$, our measurement gives equivalent solutions
near $\pi$ and near zero, with only the former being in agreement with
the value of $2.91 \pm 0.06$ measured for \bddec\ by $B$ factories.
Therefore, in the following we limit the range of  $\delta_{\perp}$
to $\cos \delta_{\perp}<0$.

To obtain the credible intervals for physics parameters, taking into account
non-Gaussian tails and  systematic effects, we combine 
the MCMC chains for the nominal and alternative fits.
This is equivalent to an effective averaging of the resulting
probability density functions from the fits.
First, we combine the four MCMC chains for each sample.
We then combine all eight chains, to produce the final result.


\subsection{Results  }

\begin{figure}[htbp]
\begin{center}
\subfigure[BDT selection]
{\includegraphics*[width=0.35\textwidth]{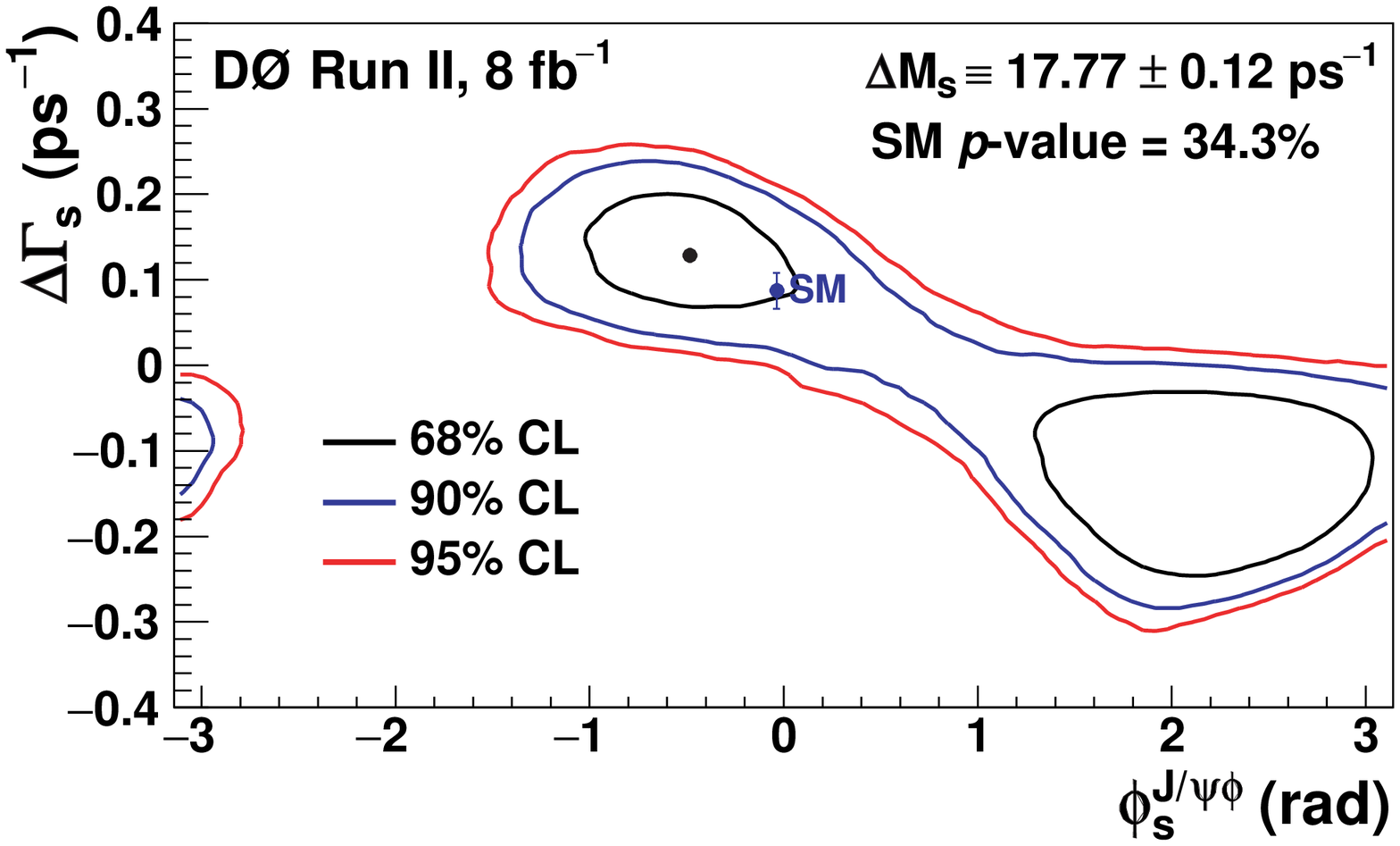}}\hfill
\subfigure[Square-cuts selection]
{\includegraphics*[width=0.35\textwidth]{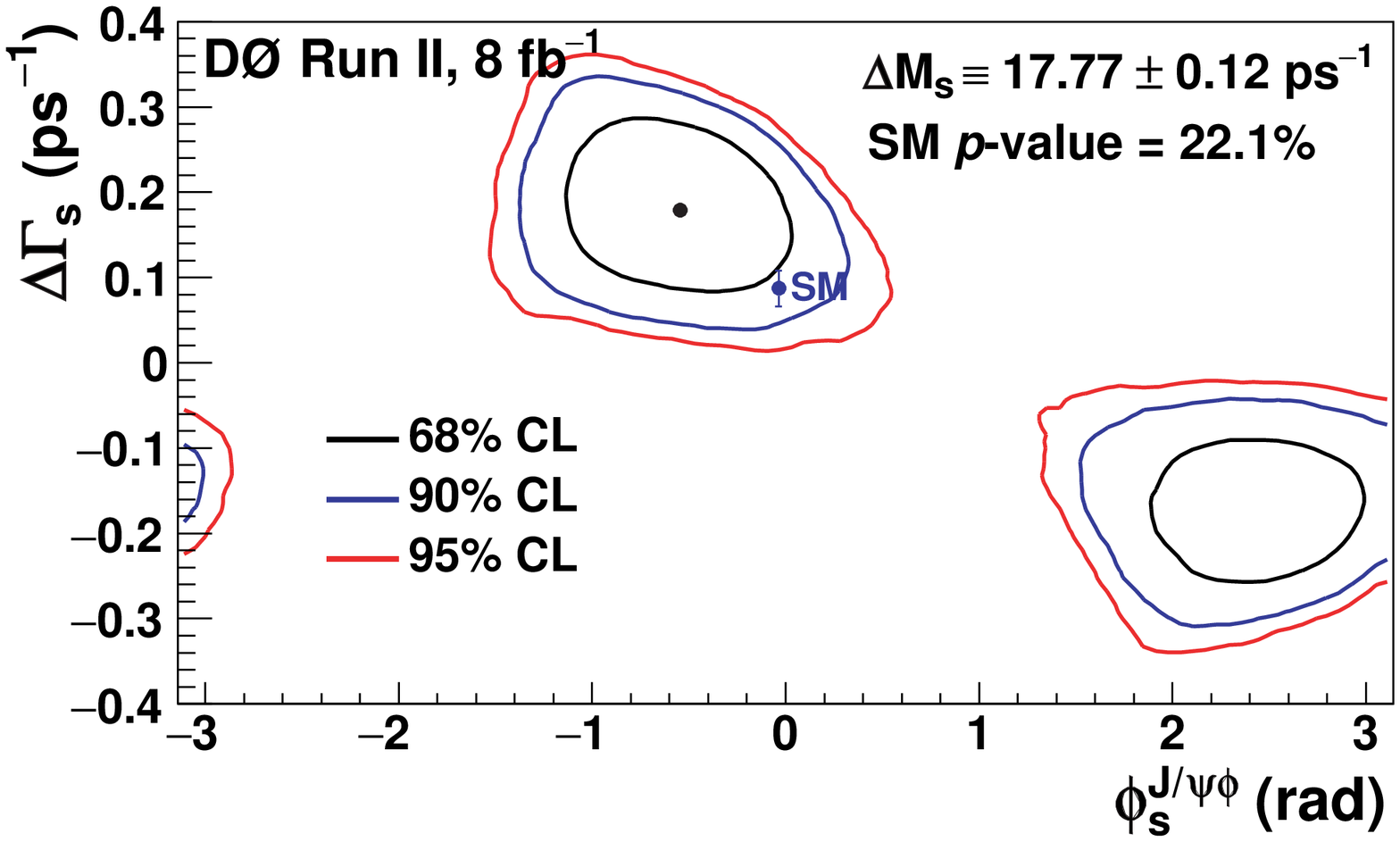}}
\caption{(color online). Two-dimensional 68\%, 90\% and and 95\% credible regions
for (a) the BDT selection and (b)  the Square-cuts sample.
The standard model expectation is indicated as a point with an error. }
\label{fig:contour_bdtsq}
\end{center}
\end{figure}

Figure~\ref{fig:contour_bdtsq} shows 68\%, 90\% and 95\% credible regions
in the   (\phis,$\Delta \Gamma_s$) plane for the BDT-based  
and for the Square-cuts samples.
The point estimates of physics parameters are obtained from
one-dimensional projections.
The minimal range containing 68\% 
of the area of the probability density function defines
the one standard deviation credible interval for each parameter,
while the most probable value defines the central value.

The large correlation coefficient (0.85) between the two phases, 
$\delta_{\perp}$ and $\delta_s$, prevents us from making separate point
estimates. Their individual errors are much larger than the uncertainty on
their difference. For the BDT selection, the measured $\cal S$-wave fraction
$F_S({\text {eff}})$ is an effective fraction of the $K^+K^-$  $\cal S$-wave
in the accepted sample, in the mass range $1.01 <M(K^+K^-)<1.03$ GeV.
It includes the effect of the diminished acceptance for
the $\cal S$-wave with respect to the $\cal P$-wave in the event selection.

This procedure gives the following results for the BDT-based sample:

\begin{eqnarray}
\overline \tau_s   & = & 
 1.426 ^{+0.035} _{-0.032} \rm~{ps},  \nonumber \\ 
\Delta \Gamma_s & = &
 0.129 ^{+0.076} _{-0.053} \rm~{ps}^{-1},\nonumber \\ 
\phi_s^{J/\psi \phi}  & = &
 -0.49 ^{+0.48} _{-0.40},\nonumber \\ 
|A_0|^2  & = &
 0.552 ^{+0.016} _{-0.017},\nonumber \\
|A_{\parallel}|^2  & = &
 0.219 ^{+0.020} _{-0.021},\nonumber \\
\delta_{\|}  & = & 
3.15 \pm 0.27, \nonumber \\ 
\cos(\delta_{\perp} -\delta_s)   & = & 
-0.06 \pm 0.24\nonumber, \\ 
F_S({\text {eff} }) & = &
 0.146 \pm 0.035. \nonumber \\ 
\nonumber 
\end{eqnarray}

$F_S({\text{eff}})$ in this case refers to the ``effective'' $F_S$ since it is 
not a physical parameter: the BDT cut on the phi mass leads to the measurement
of $F_S$ in this case to depend on the efficiency of the selection to 
non-resonant $B^0_s \rightarrow J/\psi K^+K^-$.

The one-dimensional estimates of physics parameters  for the Square-cuts sample are:

\begin{eqnarray}
\overline \tau_s   & = & 
 1.444 ^{+0.041} _{-0.033}  \rm~{ps},\nonumber \\ 
\Delta \Gamma_s & = &
 0.179 ^{+0.059} _{-0.060} \rm~{ps}^{-1},\nonumber \\  
\phi_s^{J/\psi \phi}  & = &
 -0.56 ^{+0.36} _{-0.32}, \nonumber \\  
|A_0|^2  & = &
 0.565 \pm 0.017,\nonumber \\ 
|A_{\parallel}|^2  & = &
 0.249 ^{+0.021} _{-0.022},\nonumber \\ 
\delta_{\|}  & = & 
3.15 \pm 0.19,\nonumber \\ 
\cos(\delta_{\perp} -\delta_s)   & = & 
-0.20 ^{+0.26}_{-0.27},\nonumber \\ 
F_S & = &
 0.173 \pm 0.036. \nonumber \\ 
\nonumber 
\end{eqnarray}


 \begin{figure}[htbp]
 \begin{center}
 \includegraphics*[width=0.49\textwidth]{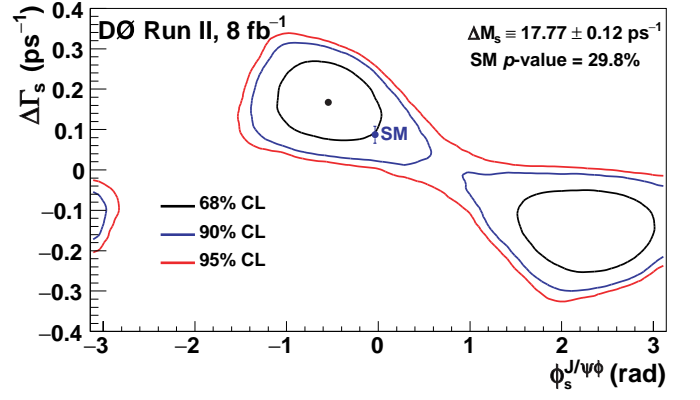}
\caption{(color online). Two-dimensional 68\%, 90\% and  95\% credible regions
including systematic uncertainties.
The standard model expectation is indicated as a point with an error. }
\label{fig:contour_final}
\end{center}
 \end{figure}



To obtain the final credible intervals for physics parameters, 
 we combine all eight  MCMC chains, effectively averaging the
probability density functions of the results of the fits to the
BDT- and Square-cuts samples.
Figure~\ref{fig:contour_final} shows 68\%, 90\% and 95\% credible regions
in the   (\phis,$\Delta \Gamma_s$) plane.
The $p$-value for the SM point~\cite{ln2011}
 ($\phi_s^{J/\psi \phi}, \Delta \Gamma_s)  = (-0.038, 0.087$ ps$^{-1}$)
is 29.8\%.
The one-dimensional 68\% credible intervals are listed in Section~\ref{sec:conclusions} below.



\section{\label{sec:conclusions}Summary and Discussion}

 We have presented  a time-dependent  angular
analysis of the decay process \bsdec.
We measure
$B_s^0$  mixing parameters,  average lifetime,
and decay amplitudes.
In addition, we  measure the amplitudes and
phases of the polarization amplitudes.
We also measure the level of the  $KK$ $\cal S$-wave
contamination in the mass range ($1.01$ -- $1.03$) GeV,
$F_S$.
The measured values and the 68\% credible intervals, including
systematic uncertainties, with the
oscillation frequency constrained to  $\Delta M_s = 17.77 \pm 0.12$ ps$^{-1}$,
 are:

\begin{eqnarray}
\overline \tau_s   & = & 
 1.443 ^{+0.038}  _{-0.035}  \rm~{ps},  \nonumber \\ 
\Delta \Gamma_s & = &
 0.163  ^{+0.065} _{-0.064} \rm~{ps}^{-1},  \nonumber \\  
\phi_s^{J/\psi \phi}  & = &
 -0.55 ^{+0.38} _{-0.36},       \nonumber \\  
|A_0|^2  & = &
 0.558 ^{+0.017} _{-0.019},        \nonumber \\ 
|A_{\parallel}|^2  & = &
 0.231  ^{+0.024} _{-0.030},        \nonumber   \\ 
\delta_{\|}  & = & 
3.15 \pm 0.22, \nonumber \\ 
\cos(\delta_{\perp} -\delta_s)   & = & 
-0.11 ^{+0.27} _{-0.25}.       \nonumber \\ 
\nonumber 
F_S  & = &
 0.173 \pm 0.036,        \nonumber   \\ 
\end{eqnarray}
The $p$-value for the SM point ($\phi_s^{J/\psi \phi}, \Delta \Gamma_s)  = (-0.038, 0.087$ ps$^{-1}$)
is 29.8\%.

In the previous publication~\cite{prl08}, 
which  was based on  a subset of this  data sample,
we constrained
the strong phases to those of \bddec\ whereas this analysis
has a large enough data sample to reliably let them float.
Also, the previous publication did not have a large enough data sample to
allow for the measurement of a  significant level of  $KK$ $\cal S$-wave,
whereas it is measured together with its relative phase
in the current analysis.
 The results supersede our previous measurements.

Independently of the Maximum Likelihood analysis, we make 
an estimate of the non-resonant $K^+K^-$ in the final state
based on the $M(KK)$ distribution of the $B_s^0$ signal yield.
The result of this study (Appendix~\ref{appfs})
is consistent with the result of the
Maximum Likelihood fit shown above.


%
We thank the staffs at Fermilab and collaborating institutions,
and acknowledge support from the
DOE and NSF (USA);
CEA and CNRS/IN2P3 (France);
FASI, Rosatom and RFBR (Russia);
CNPq, FAPERJ, FAPESP and FUNDUNESP (Brazil);
DAE and DST (India);
Colciencias (Colombia);
CONACyT (Mexico);
KRF and KOSEF (Korea);
CONICET and UBACyT (Argentina);
FOM (The Netherlands);
STFC and the Royal Society (United Kingdom);
MSMT and GACR (Czech Republic);
CRC Program and NSERC (Canada);
BMBF and DFG (Germany);
SFI (Ireland);
The Swedish Research Council (Sweden);
and
CAS and CNSF (China).
We thank J. Boudreau who has suggested and developed the use of the MCMC method for this study.

\appendix

\section{BDT Discriminants} 
\label{appbdt}

Two BDT discriminants are used to reject background. One is trained to remove the prompt background
(the ``prompt BDT''), and the other is trained to remove inclusive $B$ decays (the ``inclusive BDT''). 
The prompt BDT uses 33 variables, listed in 
Table \ref{tProVars}. The inclusive BDT uses 35 variables, listed in Table
\ref{tIncVars}. In these tables, $\Delta R$ is defined as $\Delta R = \sqrt{(\Delta \eta)^2 + 
(\Delta \phi)^2}$, where $\eta$ is the pseudorapidity and $\phi$ is the azimuthal angle.
The term ``uncorrected'' refers to the correction due to the $J/\psi$ mass constraint. 
``Leading'' (``trailing'')  muon or kaon refers to the particle with larger (smaller)
$p_T$, and $dE/dx$ is the energy loss per unit path length of a charged  particle as it
traverses the silicon detector.
Isolation is defined as $p(B)/\sum_{<\Delta R}p$ where $p(B)$ is the 
sum of the momenta of the four daughter particles of the $B_s^0$ candidate, 
and the sum is over  
all  particles within a cone defined by $\Delta R$, including the 
decay products of the $B_s^0$ candidate. 
The tables also show the importance and separation for each 
variable. 
The separation $\langle S^2 \rangle$ of a classifier $y$ is defined as
\begin{equation}
\langle S^2 \rangle = 
\frac{1}{2}\int\frac{(\hat{y}_S(y)-\hat{y}_B(y))^2}{\hat{y}_S(y)+\hat{y}_B(y)}dy,
\end{equation}
where $y_S$ is the output of the discriminant function for signal events and $y_B$ is the
discriminant function for background. 
The importance  of each  BDT input variable is derived by counting in the training how often the variable is
used to
split decision tree nodes and by weighting each split occurrence by its separation gain 
squared and by the number of events in the node. 

The  distributions for the  six most important variables in 
training on prompt $J/\psi$ decays are shown in
Fig. \ref{fvp1}.  The distributions for 
the  six  most important variables in 
the training on inclusive $B \rightarrow J/\psi X$ 
decays are shown in Fig. \ref{fvi1}.
 
Figure \ref{eff} compares the shapes of the distributions of the three angular variables and the lifetime,
before and after  the BDT requirements. The figures show that the BDT requirements do not affect these 
differential distributions significantly.

\begin{widetext}

\begin{table}[htbp!] 
\begin{tabular}{cccc} \hline 
Rank & Variable &  Importance & Separation \\\hline 
1 & $KK$ invariant mass  & 0.3655& 0.3540 \\ 
2 & Maximum $\Delta R$ between either $K$ meson and the $B_s^0$ candidate & 0.1346 & 0.4863 \\ 
3 & Isolation using the maximum $\Delta R$ between either $K$ and the $B_s^0$ & 0.0390 & 0.1784 \\ 
4 & Uncorrected $p_T$ of the $B_s^0$ & 0.0346 & 0.3626 \\ 
5 & Minimum $\Delta R$ between either $K$ and the $B_s^0$ & 0.0335 & 0.4278 \\\hline 
6 & $p_T$ of the trailing $K$ meson  & 0.0331 & 0.4854 \\ 
7 & $p_T$ of the $\phi$ meson & 0.0314 & 0.4998 \\ 
8 & $p_T$ of the leading $K$ meson & 0.0283 & 0.4884 \\ 
9 & Trailing muon momentum & 0.0252 & 0.0809 \\ 
10 & $p_T$ of the leading muon & 0.0240 & 0.1601 \\\hline 
11 & Maximum $\Delta R$ between either muon and the $B_s^0$ & 0.0223 & 0.1109 \\ 
12 & Maximum $\chi^2$ of  either $K$ meson with the $J/\psi$ vertex & 0.0217 & 0.0162 \\ 
13 & Dimuon invariant mass & 0.0215 & 0.0145 \\ 
14 & Maximum $\chi^2$ of either of the $K$ candidate track & 0.0213 & 0.021 \\ 
15 & $B_s^0$ isolation using the larger $K/B_s$ $\Delta R$ and tracks from the PV 
 & 0.0207 & 0.1739 \\\hline 
16 & $p_T$ of the $J/\psi$ meson & 0.0205 & 0.1809 \\ 
17 & Minimum $\Delta R$ between either muon and the $B_s^0$ candidate & 0.0188 & 0.1023\\
18 & Trailing $K$  momentum & 0.0105 & 0.3159 \\ 
19 & $\chi^2$ of the $B_s^0$ candidate vertex  & 0.0093 & 0.0119 \\ 
20 & $B_s^0$ isolation using $\Delta R< 0.75$ & 0.0084 & 0.0241 \\\hline 
21 & Minimum $\chi^2$ of the $J/\psi$ vertex with either $K$  & 0.0081 & 0.0069 \\ 
22 & $p_T$ of the trailing muon  & 0.0079 & 0.0922 \\ 
23 & Minimum of the $\chi^2$ of the $J/\psi$ and $\phi$ vertices & 0.0073 & 0.0057 \\ 
24 & Isolation using $\Delta R < 0.5$  & 0.0070 & 0.0405\\ 
25 & Uncorrected $B_s^0$ total momentum & 0.0068 & 0.2103 \\\hline 
26 & Minimum $\chi^2$ of either $K$ track fit & 0.0065 & 0.0266\\ 
27 & Isolation using $\Delta R <0.5$ and particles from the PV & 0.0057 & 0.0401 \\ 
28 & Leading $K$ meson momentum & 0.0051 & 0.3217 \\ 
29 & Leading muon  momentum & 0.0048 & 0.0908 \\ 
30 & $\phi$ meson momentum & 0.0048 & 0.3233 \\\hline 
31 & Maximum $\chi^2$ of the $J/\psi$ or $\phi$ vertices & 0.0044 & 0.0061 \\ 
32 & Isolation using $\Delta R<0.75$ and particles from the PV & 0.0037 & 0.0259 \\ 
33 & $J/\psi$ meson momentum & 0.0037 & 0.1004\\ \hline 
\end{tabular}\caption{Variables used to train the prompt BDT,
ranked by their importance in the training.}
\label{tProVars} 
\end{table} 



\begin{table}[htbp] 
\begin{tabular}{cccc} \hline Rank & Variable & 
Importance & Separation \\\hline 
1 & $KK$ invariant mass  & 0.2863 & 0.3603 \\ 
2 &  $B_s^0$ isolation using the larger $K/B_s$ $\Delta R$ and tracks from the PV 
& 0.1742 & 0.4511 \\ 
3 & Minimum $dE/dx$ of either $K$  & 0.0778 & 0.1076 \\ 
4 & $\chi^2$ of $B_s^0$  & 0.0757 & 0.2123 \\ 
5 & $p_T$ of the $\phi$ meson  & 0.0559 & 0.4856 \\\hline 
6 & $p_T$ of the leading $K$ meson & 0.0504 & 0.4745 \\ 
7 & Isolation using the maximum $\Delta R$ between either $K$ and the $B_s^0$ & 0.0429 & 0.4468 \\ 
8 & $p_T$ of the trailing $K$ meson  & 0.0350 & 0.4774 \\ 
9 &  Maximum $\chi^2$ of either $K$ meson with the $J/\psi$ vertex & 0.0260 & 0.2051 \\ 
10 &  Isolation using $\Delta R <0.5$ and particles from the PV  & 0.0229 & 0.1703 \\\hline 
11 & Isolation using $\Delta R<$ 0.75 and tracks from the PV  & 0.0154 & 0.2238 \\ 
12 & Minimum $\chi^2$ of of either $K$ with the $J/\psi$ vertex  & 0.0151 & 0.1308 \\ 
13 & Minimum $\Delta R$ between either $K$ meson and the $B_s^0$ candidate  & 0.0115 & 0.3104 \\
14 & Dimuon invariant mass & 0.0099 & 0.0190 \\ 
15 & Total momentum of the $\phi$ meson  & 0.0091 & 0.3307 \\\hline 
16 & $p_T$ of the $J/\psi$ meson  & 0.0089 & 0.1198 \\ 
17 & Trailing muon momentum & 0.0082 & 0.0594 \\ 
18 & Isolation using $\Delta R<0.5$  & 0.0073 & 0.1695 \\ 
19 & Maximum $\Delta R$ between either $K$ meson and the $B_s^0$ candidate  & 0.0070 & 0.3794 \\
20 & Maximum $dE/dx$ of either $K$ meson & 0.0069 & 0.0528 \\\hline
21 & Trailing $K$ meson momentum & 0.0068 & 0.3253 \\ 
22 & $J/\psi$ vertex $\chi^2$  & 0.0063 & 0.0057 \\ 
23 & Leading $K$ meson momentum & 0.0058 & 0.3277 \\ 
24 & Maximum $\chi^2$ of either $K$ candidate track & 0.0054 & 0.0267 \\ 
25 & Isolation using $\Delta R<0.75$  & 0.0046 & 0.2203 \\\hline 
26 & Minimum $\Delta R$ between either muon and the $B_s^0$ candidate  & 0.0041 & 0.0729 \\ 
27 & Minimum $\chi^2$ of either $K$ candidate track & 0.0039 & 0.0284 \\ 
28 & uncorrected $p_T$ of $B_s^0$ candidate  & 0.0036 & 0.2485 \\ 
29 & $p_T$ of the trailing muon  & 0.0029 & 0.0702 \\ 
30 & $J/\psi$  momentum & 0.0027 &  0.0645 \\\hline 
31 & Maximum $\Delta R$ between either muon and the $B_s^0$ candidate & 0.0026 & 0.0872 \\ 
32 & Vertex $\chi^2$ of the $\phi$ meson & 0.0017 & 0.0098 \\ 
33 & Uncorrected $B_s^0$ momentum & 0.0014 & 0.1675 \\ 
34 & $p_T$ of the leading muon  & 0.0011 &  0.1008 \\ 
35 & Leading muon momentum & 0.0009 & 0.0547\\ \hline
\end{tabular}\caption{Variables used to train the non-prompt BDT, 
ranked by their importance in the training.}\label{tIncVars} \end{table}


 
 \begin{figure}[htbp]
  
{\includegraphics[width=1.0\textwidth]{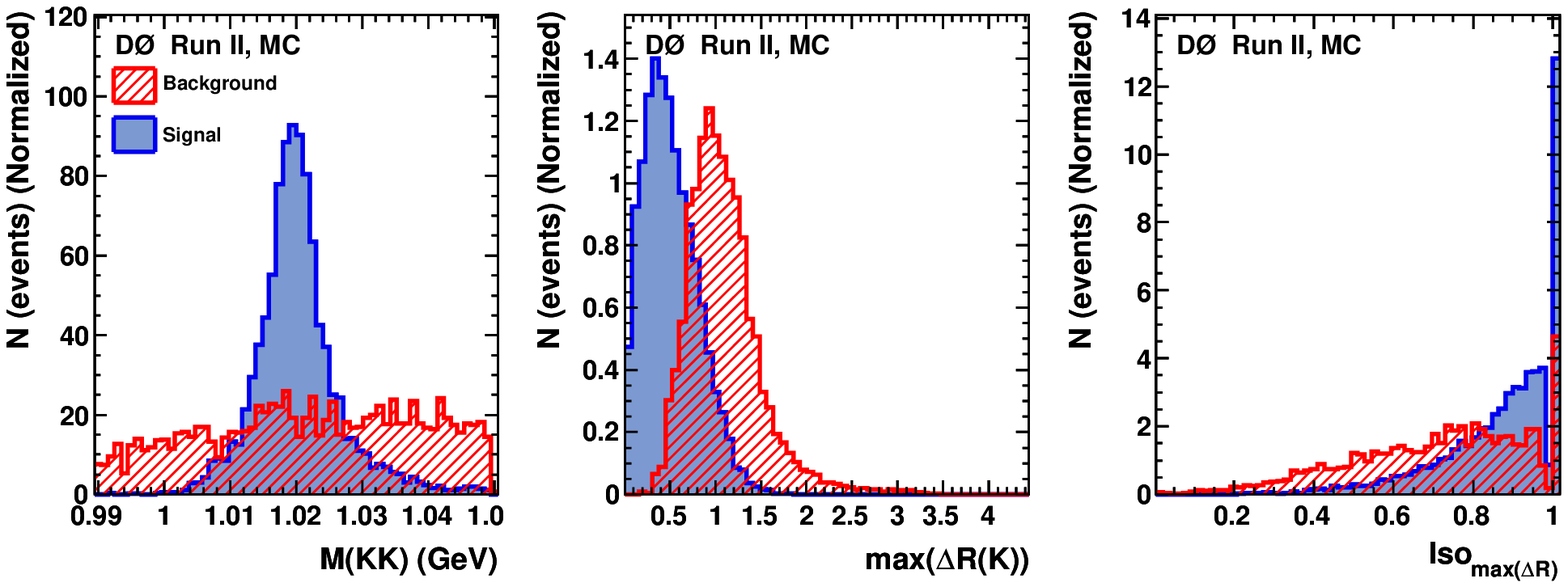}}
 
{\includegraphics[width=1.0\textwidth]{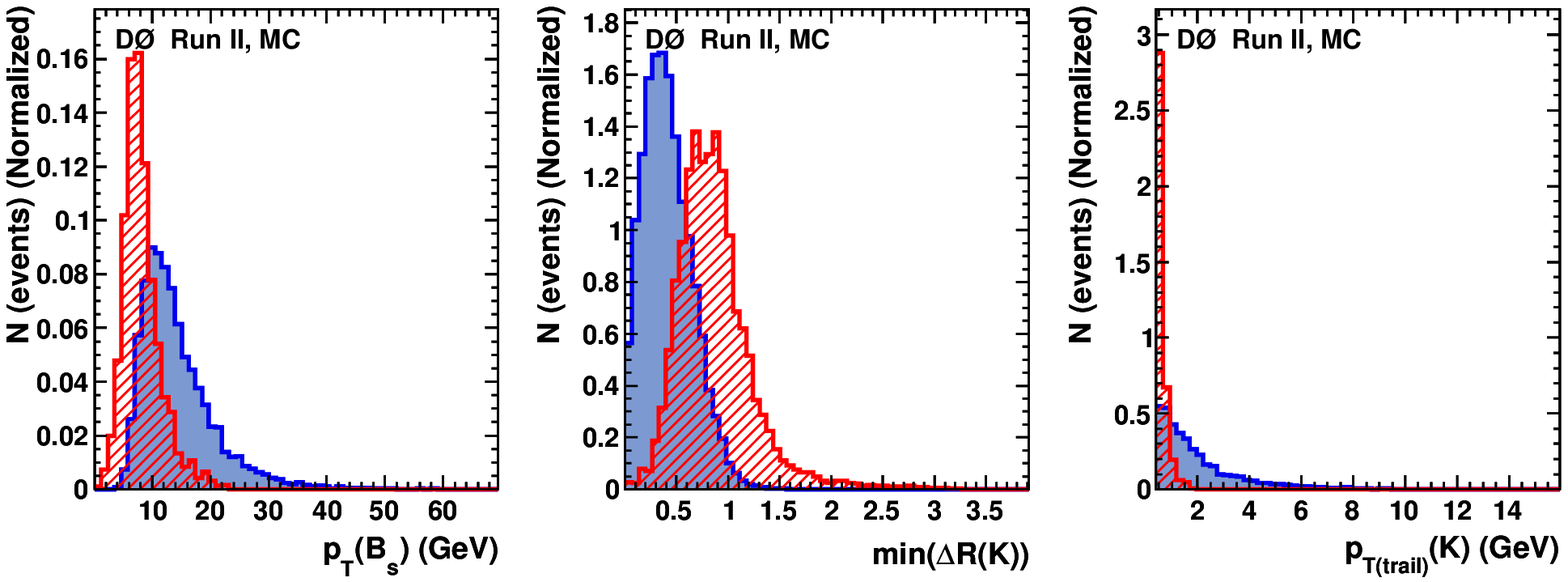}}
\caption{(color online) The distributions of  the six most important variables used in the BDT trained on prompt 
$J/\psi$ production for the \bsdec\ signal (solid blue) and prompt $J/\psi$ events (red dashed) histograms. }
\label{fvp1}
 \end{figure}


 \begin{figure}[htbp]
  
{\includegraphics[width=1.0\textwidth]{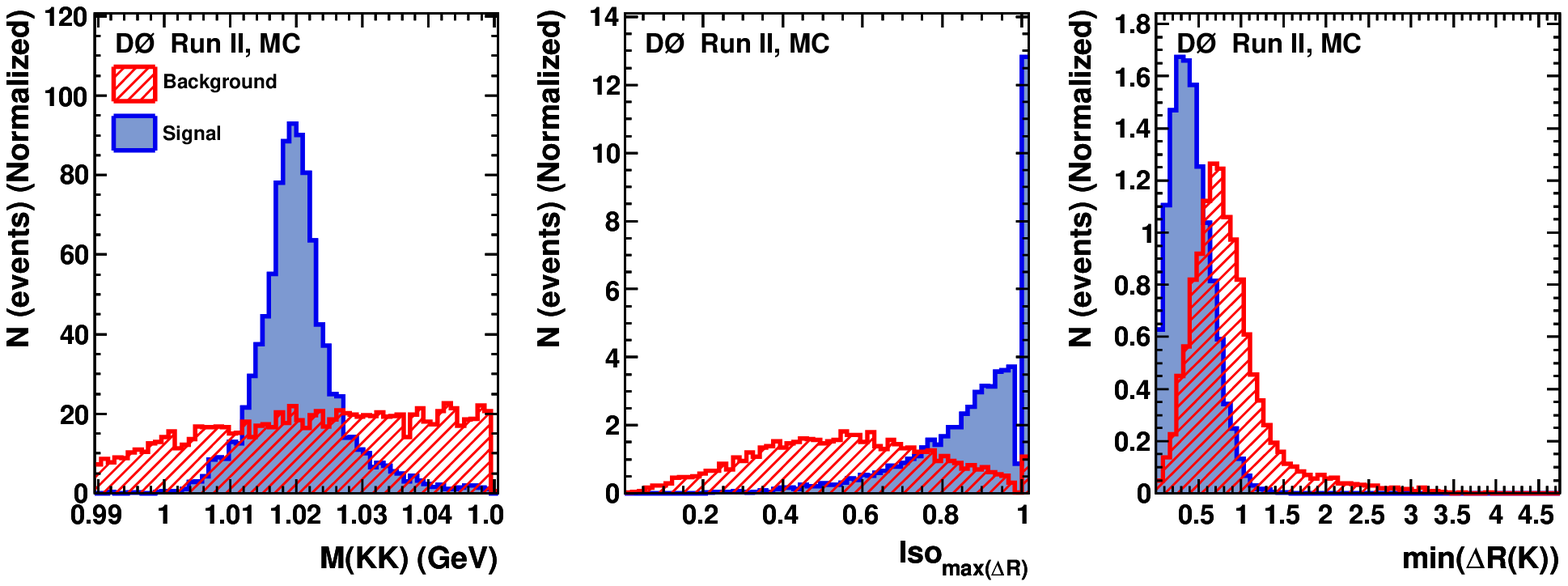}}
  
{\includegraphics[width=1.0\textwidth]{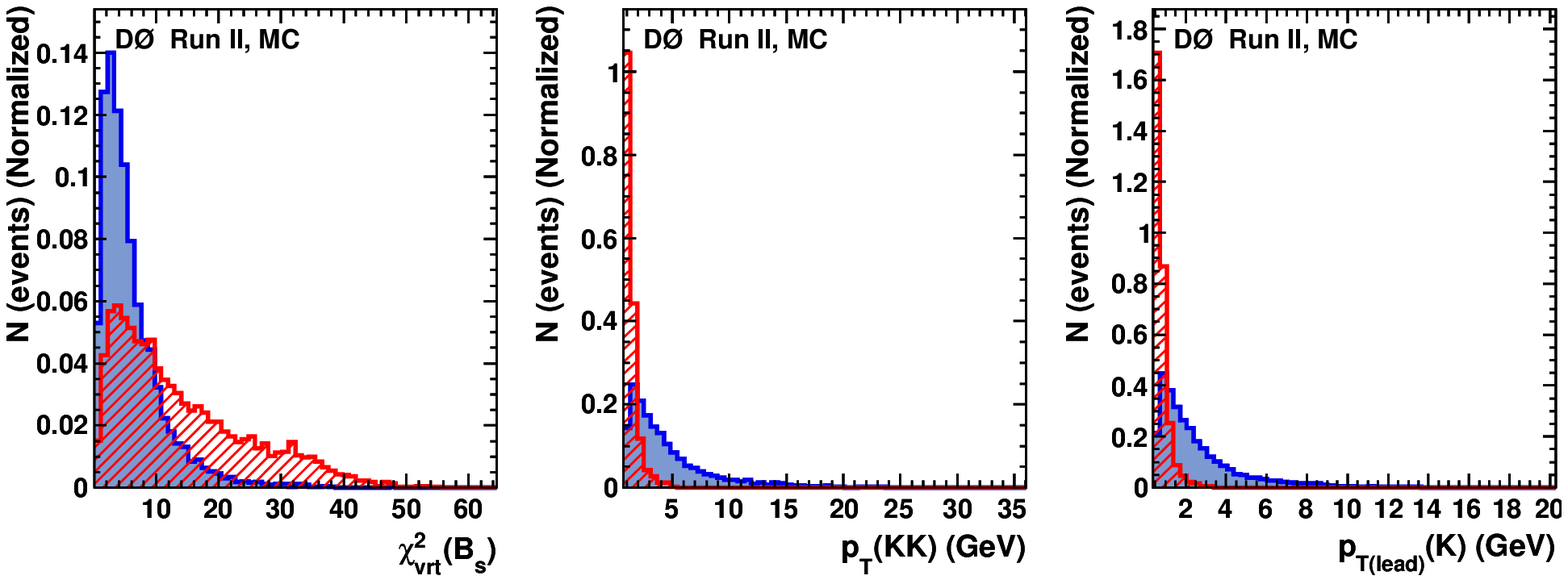}}
\caption{(color online)  The distributions of  the  six most important variables
 used in the BDT trained on inclusive 
$B \rightarrow J/\psi X$ decays  for the \bsdec\ signal (solid blue) and  inclusive $B \rightarrow J/\psi X$
 decays (red dashed) histograms. 
}
\label{fvi1}
\end{figure}



\begin{figure}[htbp]

\begin{center}
\subfigure[]
{\includegraphics*[height=5.0cm]{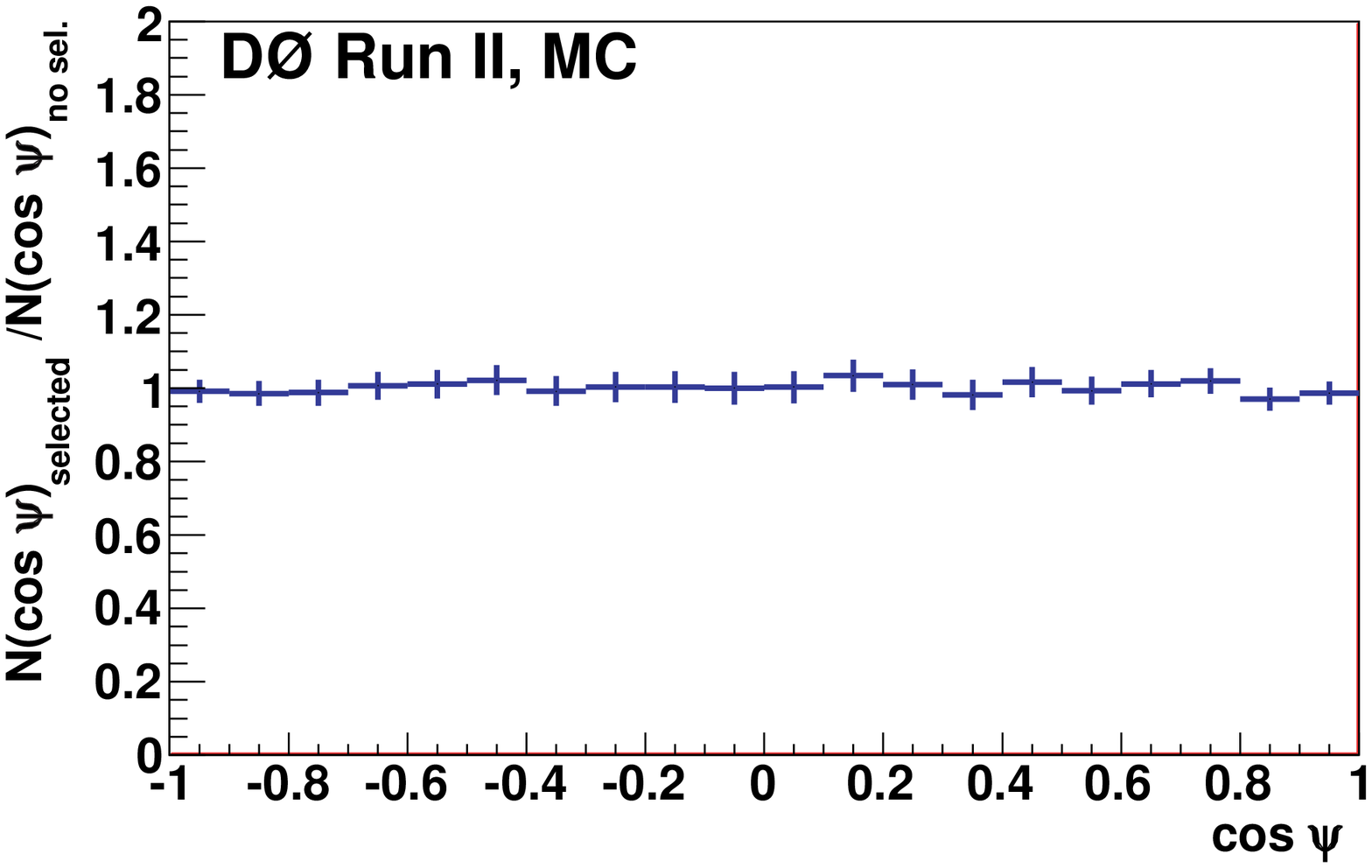}}
\subfigure[]
{\includegraphics*[height=5.0cm]{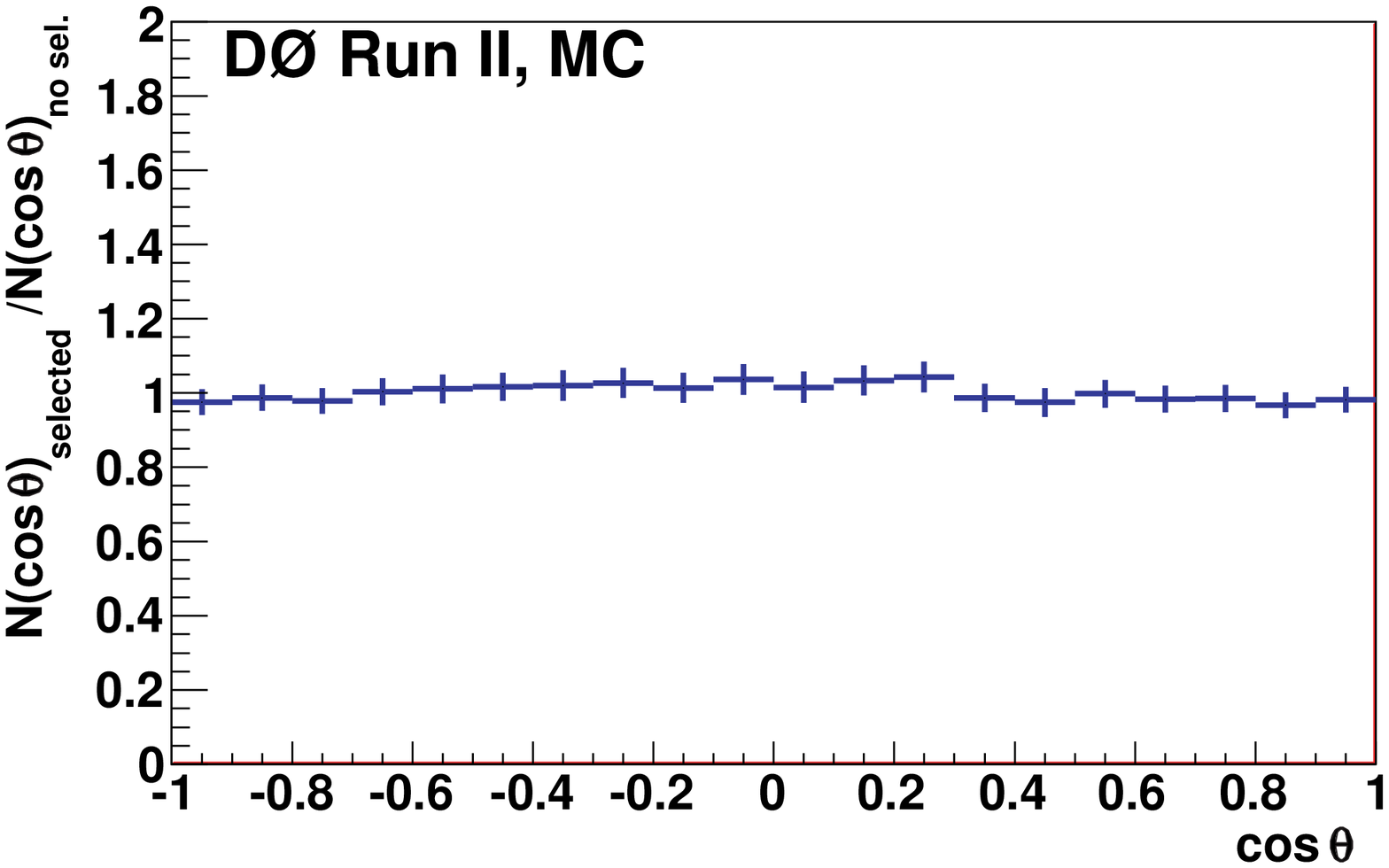}}
\subfigure[]
{\includegraphics*[height=5.0cm]{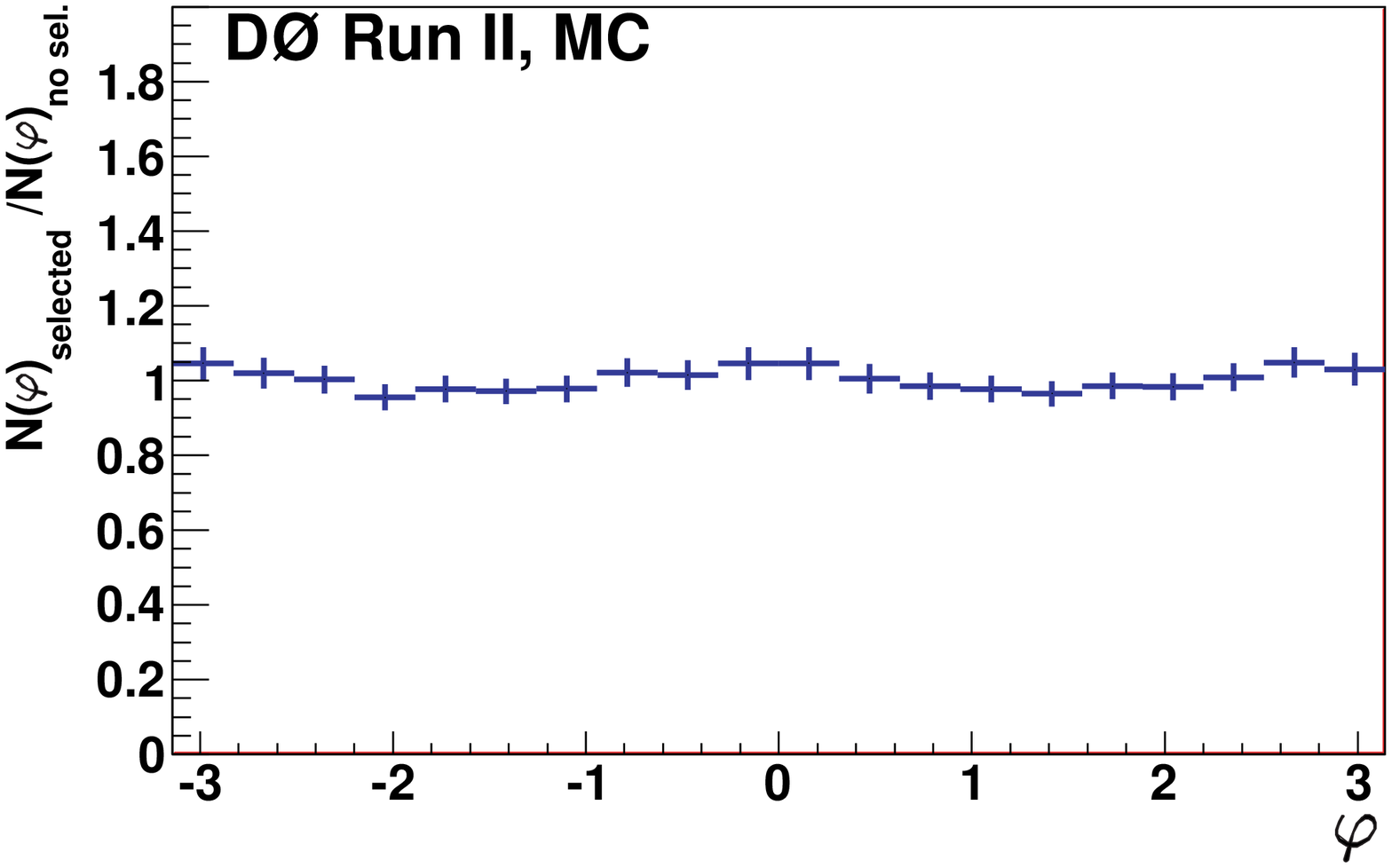}}
\subfigure[]
 {\includegraphics*[height=5.0cm]{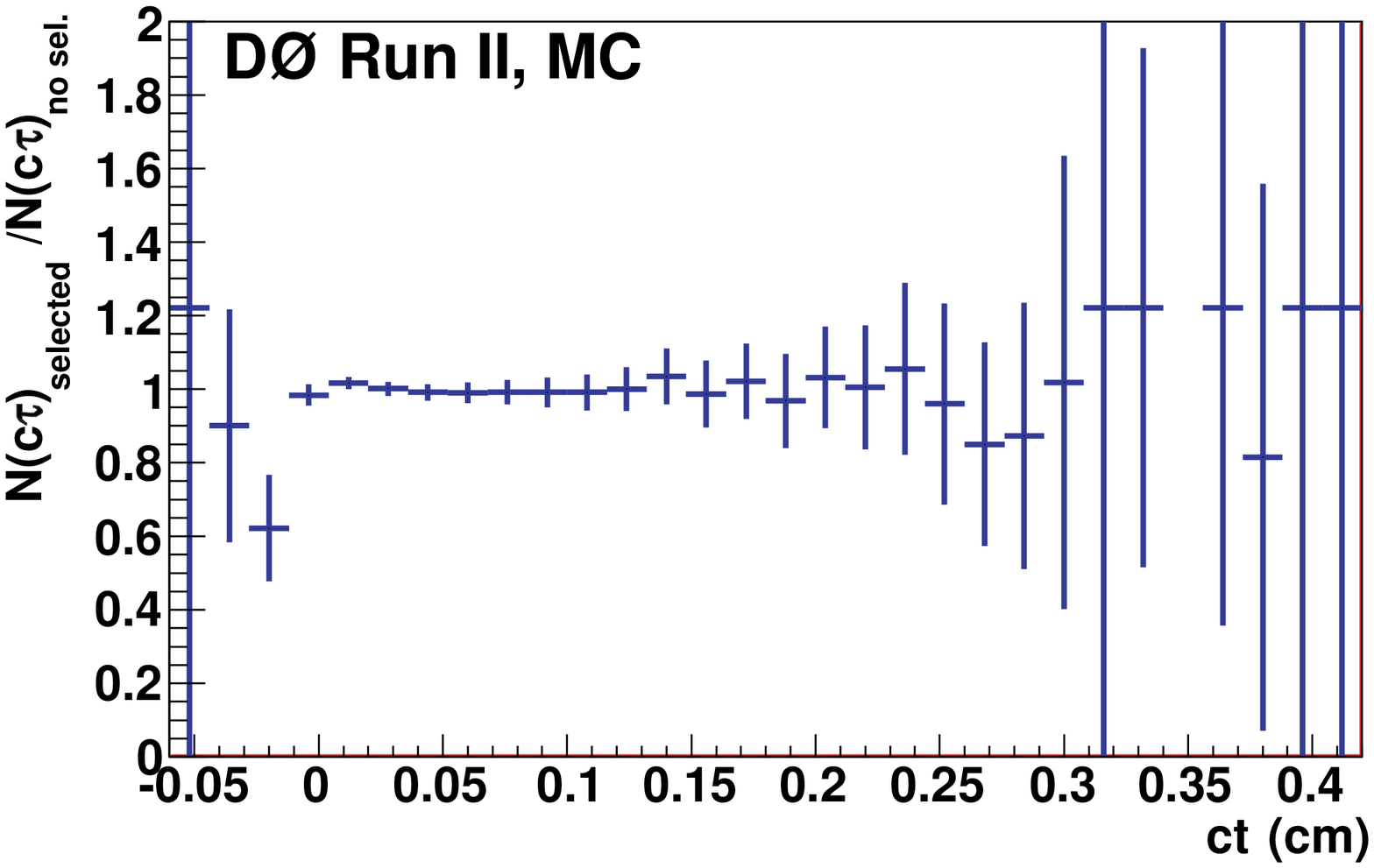}} 
 \caption{Test of uniformity of the efficiencies of the BDT selection using 
a MC sample with $\phi_s=-0.5$. The  figure shows the ratios of the
normalized distributions of
(a -- c) the three angles 
and (d)  the proper decay length, before and after the BDT selection.
} 
\label{eff}
\end{center}
\end{figure}

\end{widetext}

\clearpage

\section{Detector acceptance} 
\label{appmcdata}

\begin{figure}[htbp]
\begin{center}
\subfigure[]
{\includegraphics*[width=0.24\textwidth]{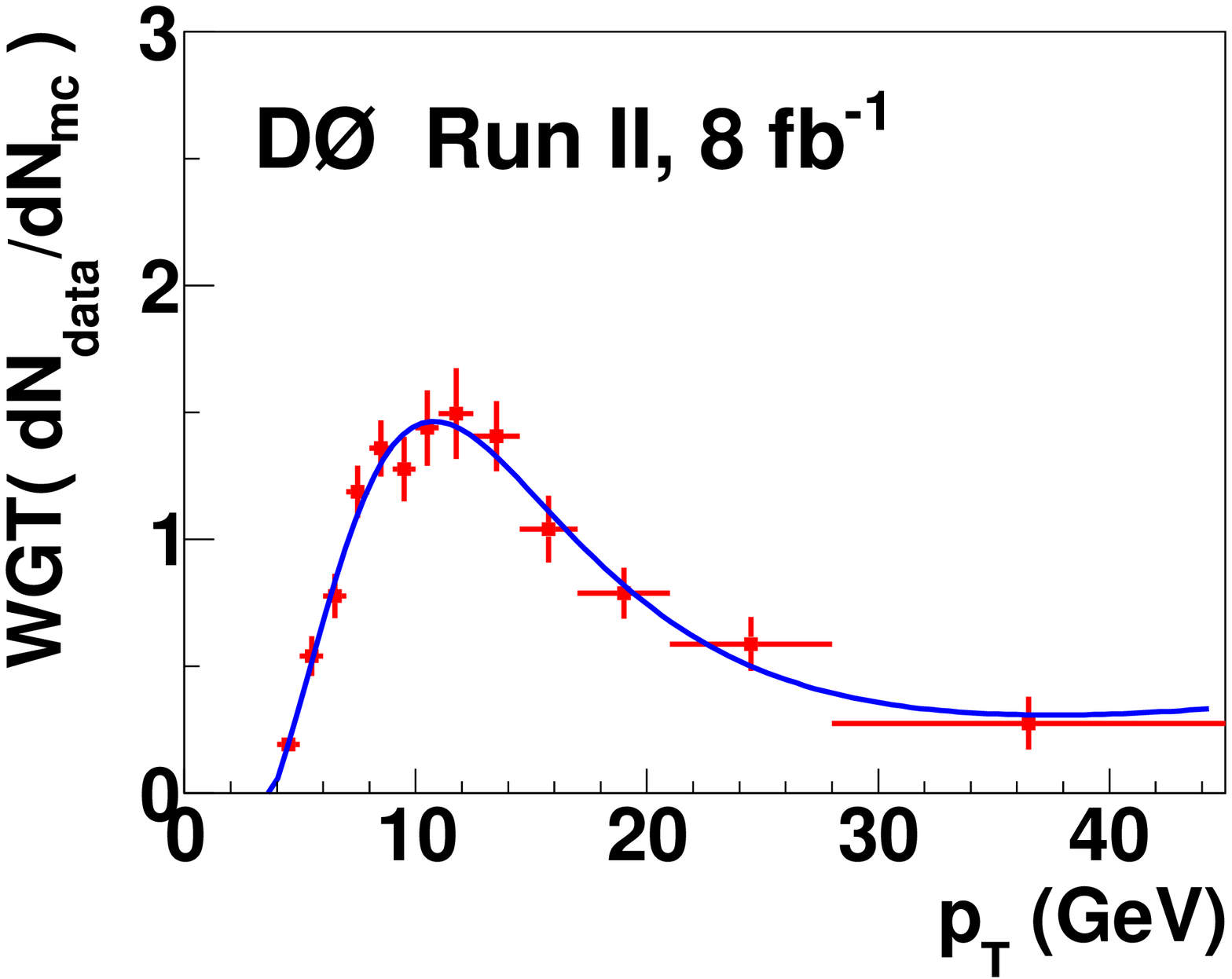}}\hfill
\subfigure[]
{\includegraphics*[width=0.24\textwidth]{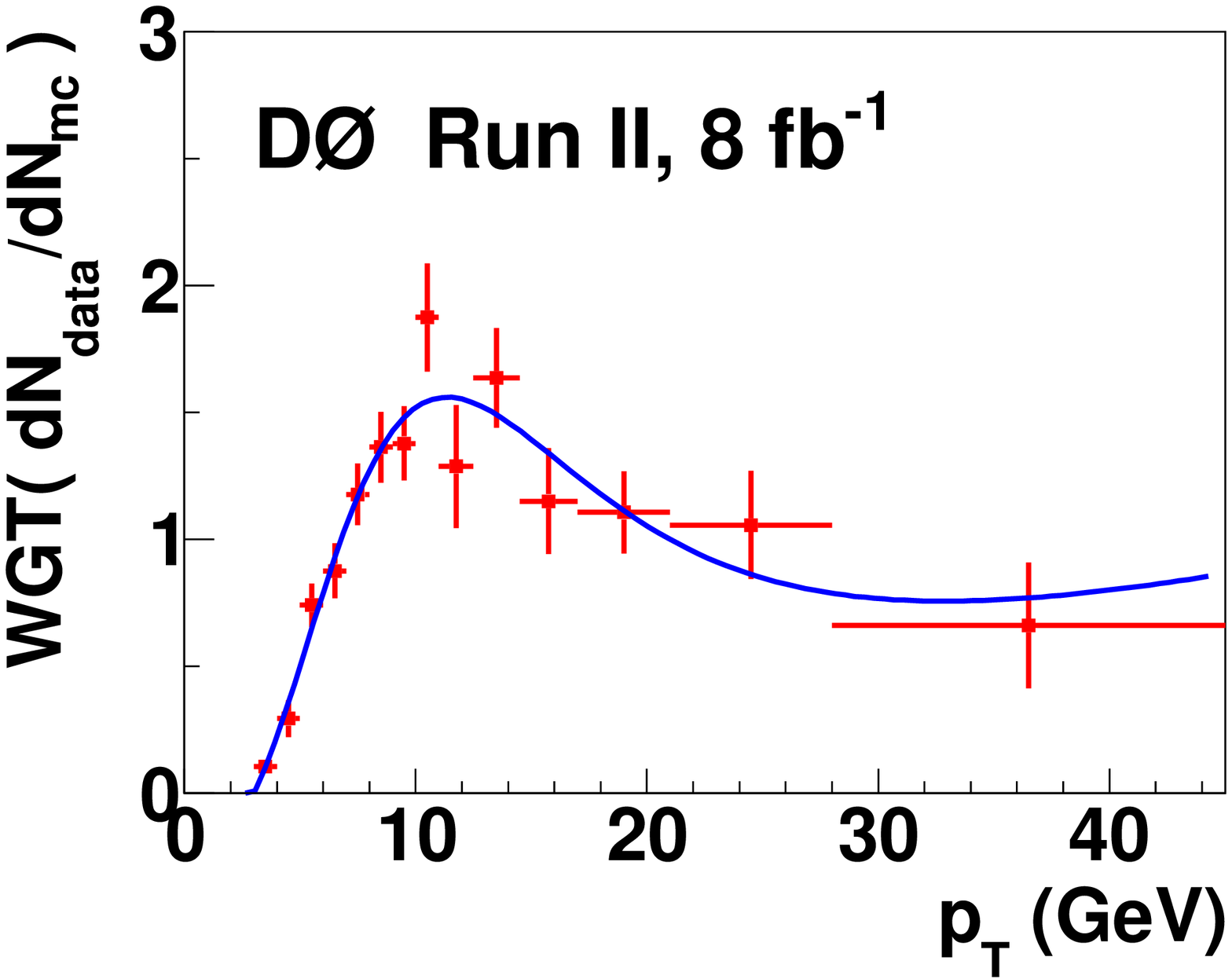}}
\caption{
Weight factor as a function of $p_T(J/\psi)$  used to correct MC $p_T$
distribution of \bs\ and \bd\ decay objects for  
(a) central region,  and (b) forward region. The curves are empirical fits
to a sum of a Landau function and a polynomial.}
\label{fig:psiwt}
\end{center}
\end{figure}

We take into account the shaping of 
the signal distribution by the detector acceptance and kinematic selection
by introducing  acceptance functions in the three angles
of the transversity basis .
The acceptance functions are derived from Monte Carlo simulation.
Due to the event triggering  effects,
the momentum spectra of final-state objects in data are harder than in MC.
We take into account the difference in the $p_T$ distribution of the
final-state objects in data and MC by introducing a weight factor
as a function of  $p_T(J/\psi)$, separately for the central 
($|\eta(\mu_{\rm leading})| <1$) 
and forward
regions.  
The weight factor is derived by forcing an agreement 
between the $J/\psi$  transverse momentum  spectra  in  data and MC.
The behavior of the weight factor as a function of  $p_T(J/\psi)$
for the BDT-based selection, 
for the central and forward regions, is shown in Fig.~\ref{fig:psiwt}.

\begin{widetext}

 \begin{figure}[htbp]
 \begin{center}
\subfigure[Leading muon]
{\includegraphics*[width=0.34\textwidth]{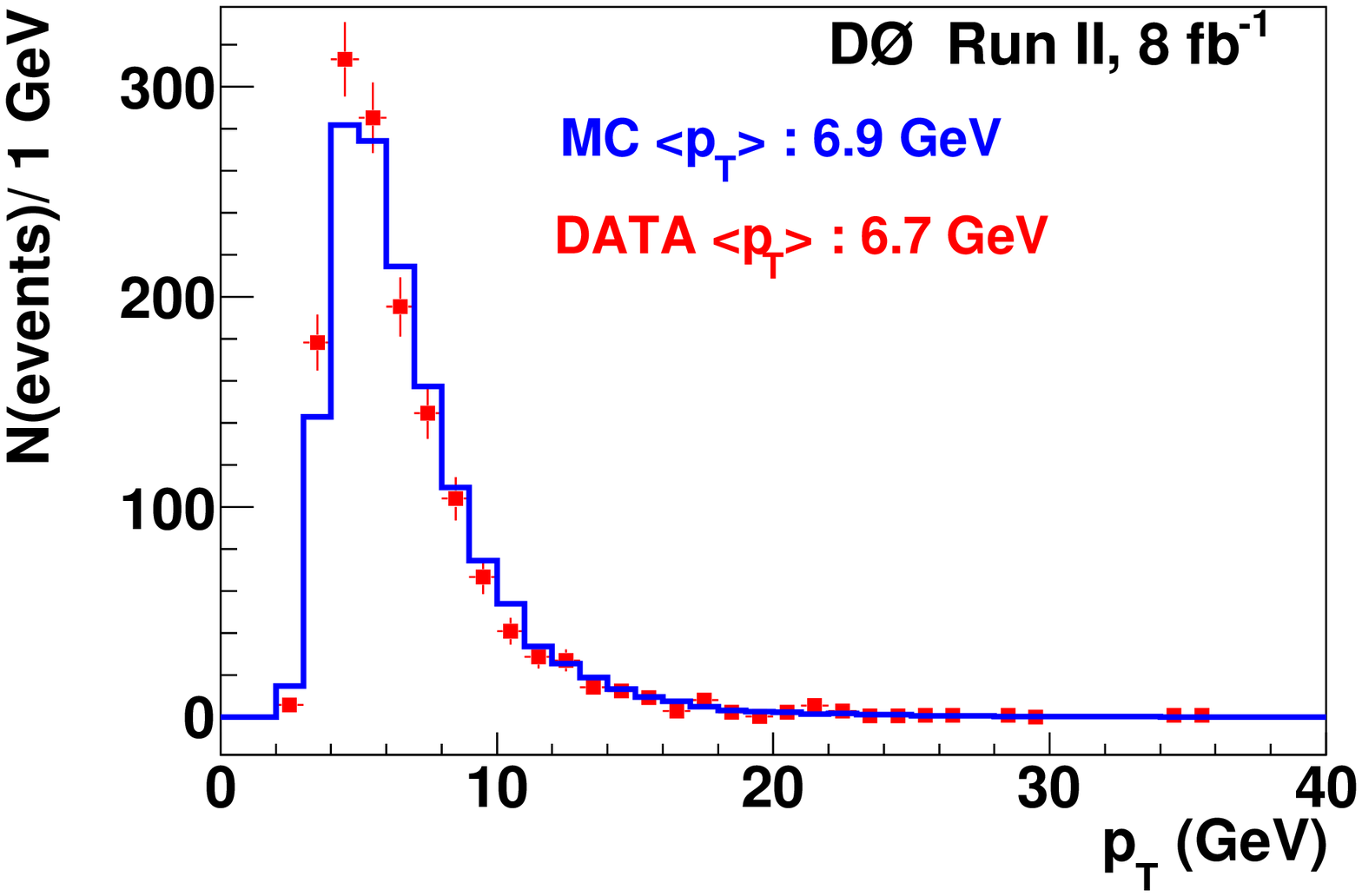}}
\subfigure[Trailing muon]
{\includegraphics*[width=0.34\textwidth]{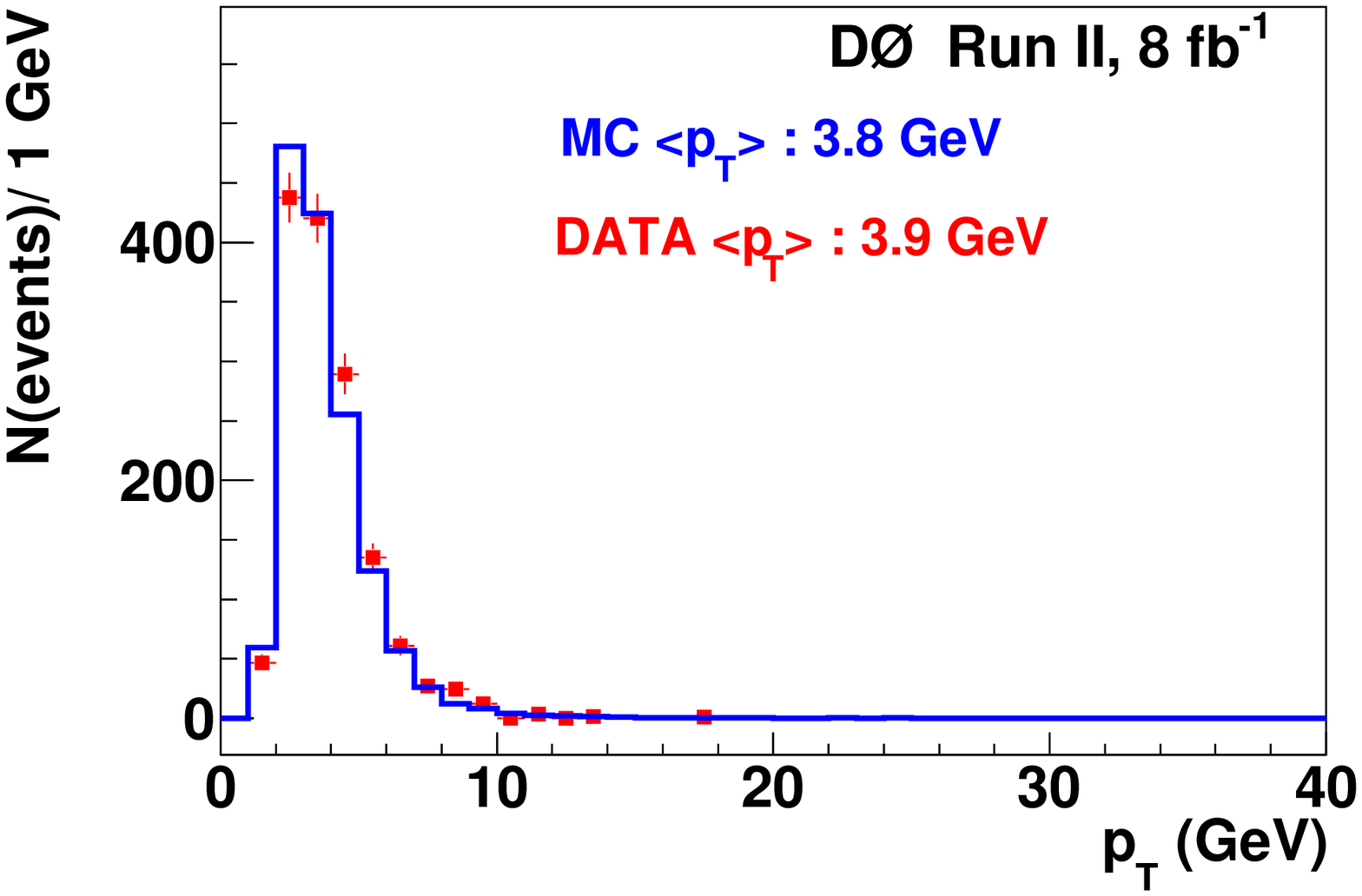}}
\subfigure[Leading kaon]
{\includegraphics*[width=0.34\textwidth]{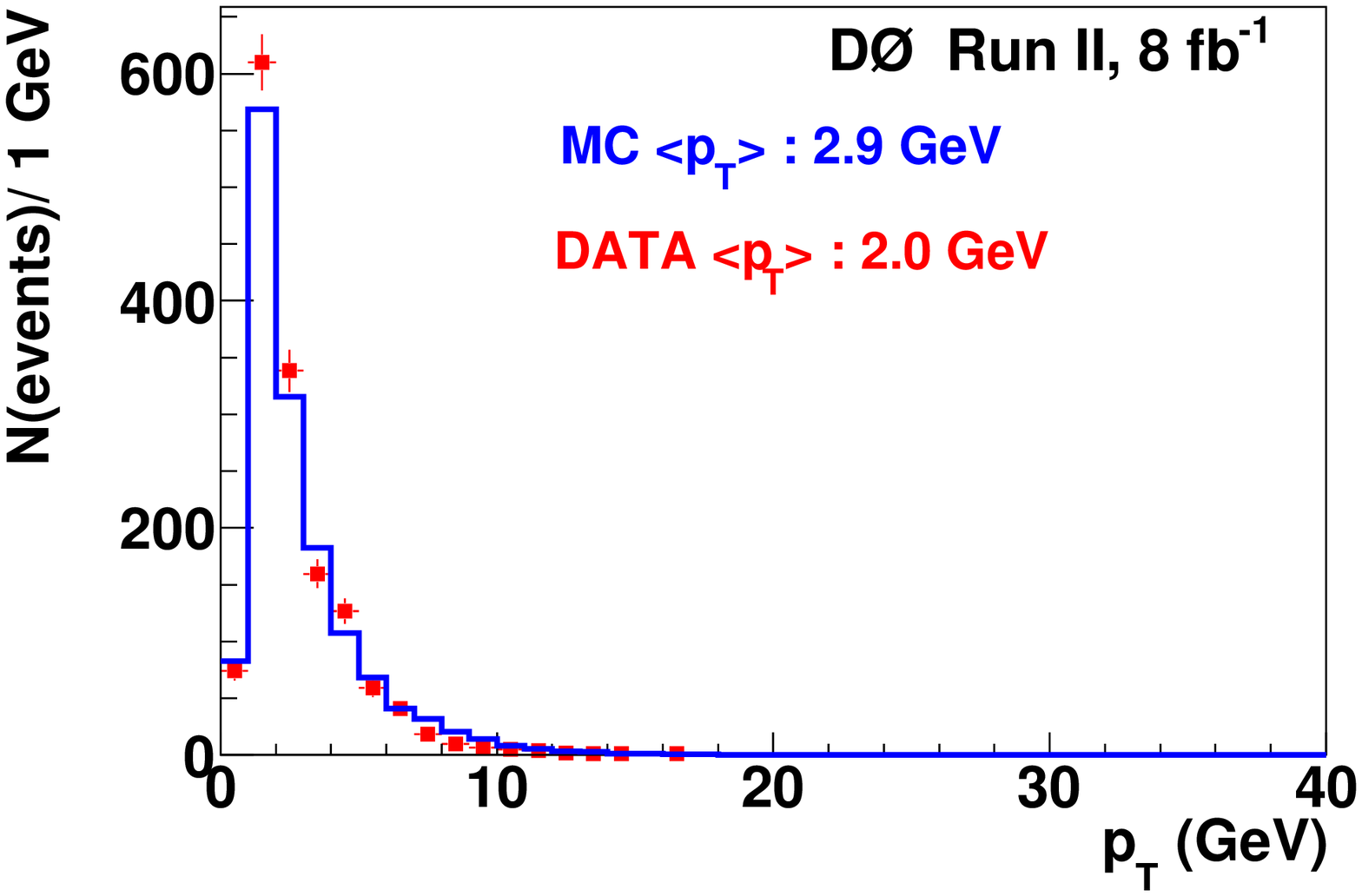}}
\subfigure[Trailing kaon]
{\includegraphics*[width=0.34\textwidth]{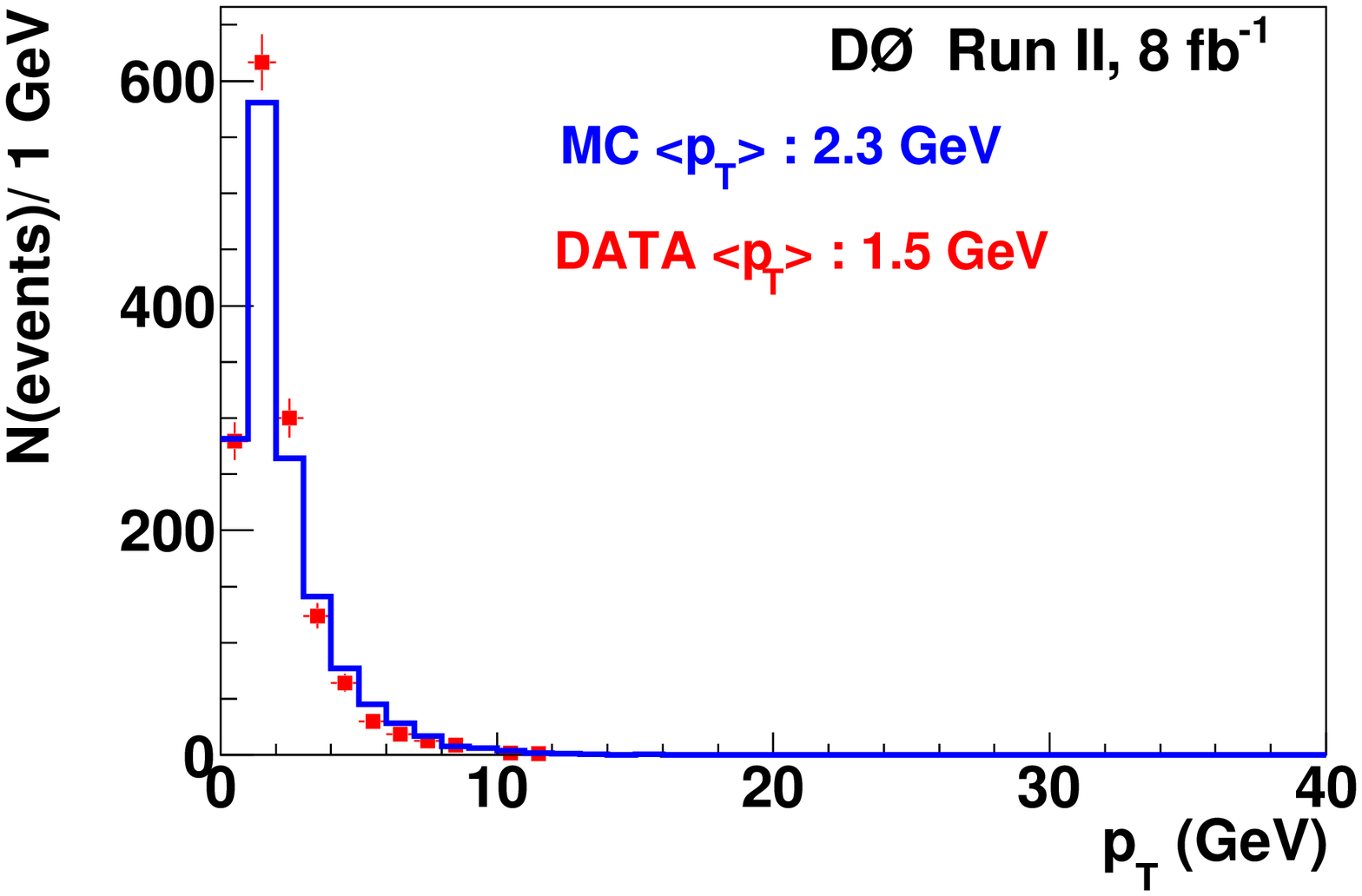}}
\caption{Transverse momentum 
 distributions  of the four final-state particles
in data (points) 
and weighted MC (solid histogram), for the BDT-based event selection.
 }
 \label{fig:mu1}
 \end{center}
 \end{figure}

\end{widetext}

 Figure~\ref{fig:mu1} shows the background-subtracted 
$p_T$ distributions 
of the leading and trailing
muon
 and leading and trailing kaon, 
in the central region.
There is a good agreement between data and MC for all
final-state particles after applying 
the  weight factor.
The acceptance in 
$\varphi$ and $\theta$ is shown in Fig.~\ref{fig:acceptance_BDT}.
The acceptance in  $\psi$ is shown in Fig.~\ref{fig:cos_psi_eff}.


\begin{widetext}

\begin{figure}[htbp]
\begin{center}
\includegraphics*[width=0.56\textwidth]{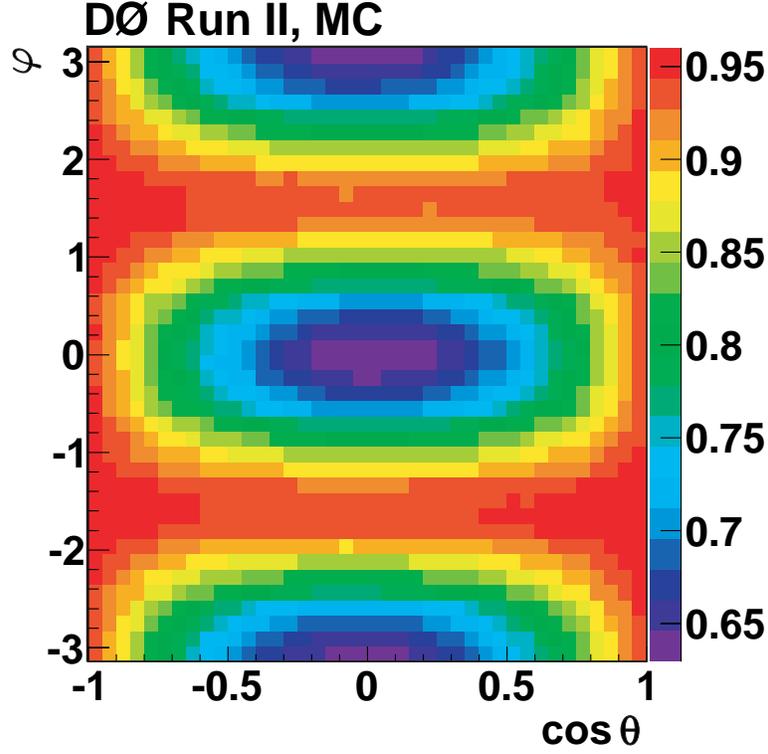}
\caption{(color online). Map of the detector acceptance on the plane 
$\varphi$ -- $\cos \theta$.
}
\label{fig:acceptance_BDT}
\end{center}
\end{figure}

\begin{figure}[htbp]
\begin{center}
\includegraphics*[width=0.4\textwidth]{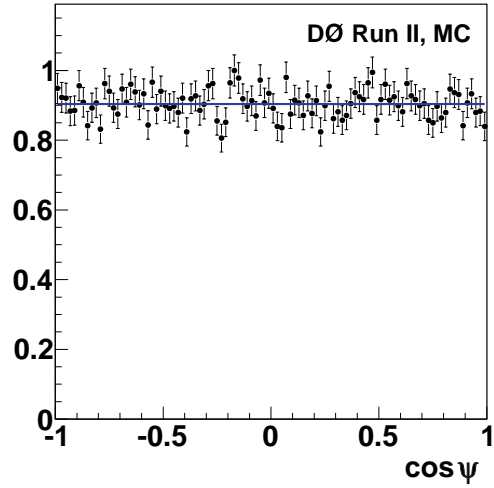}
\caption{Detector acceptance as a function of $\cos \psi$.
The acceptance is uniform in  $\cos \psi$.
}
\label{fig:cos_psi_eff}
\end{center}
\end{figure}

\end{widetext}

\clearpage

\section{Independent estimate of {\boldmath  $F_S$ } } 
\label{appfs}

In the Maximum Likelihood fit, the invariant mass
of the $K^+K^-$ pair is not used. To do so would require a good model of the $M(K^+K^-)$ dependence
of background, including a small $\phi(1020)$ component,  as a function of the 
$B_s^0$ candidate 
mass and proper time.
However, we can use the  $M(K^+K^-)$  mass information to make an independent
estimate of the non-resonant $K^+K^-$ contribution in the final state.

For this study, we use the ``Square-cuts'' sample, for which the event selection is
not biased in $M(K^+K^-)$.
Using events with decay length $ct>0.02$ cm to suppress background, we 
extract the $B_s^0$ signal in two  ranges of $M(K^+K^-)$:
$1.01 <M(KK)<1.03$ GeV and $1.03 <M(KK)<1.05$ GeV. 
The first range is that used by both selections, and contains the bulk of
the $\phi \rightarrow K^+K^-$ signal.  The second range will still contain a small
Breit-Wigner tail of $\phi \rightarrow K^+K^-$. From the simulated $M(K^+K^-)$ distribution of the
$B^0_s \rightarrow J/\psi \phi$ decay, shown in Fig.~\ref{fig:kkmc}, we obtain the fraction of the $K^+K^-$
decay products in the upper mass range to be $0.061 \pm 0.001$ of the total
range $1.01 < M(KK) < 1.05$ GeV. The $\cal{S}$-wave component is assumed to be a flat
distribution in $M(KK)$ across this range. Given that the widths of the ranges
are the same, the number of candidates due to the $\cal{S}$-wave contribution should
be the same for both.

The $B^0_s$ signal in each mass range is extracted by fitting the $B^0_s$ candidate
mass distribution to a Gaussian function representing the signal, a linear
function for the background, and MC simulation-based templates for the $B^0 \rightarrow J/\psi K^*$
reflection where the pion from the $K^*$ decay is assumed to be a kaon.
The two shape templates used, one for each mass range, are shown in Fig.~\ref{fig:psikst}.
The mass distributions, with fits using
 the above templates, are shown in Fig.~\ref{fig:bsvsmkk}.
The fits result in the $B^0_s$ yield of $3027 \pm 93$ events for
$1.01 < M(KK) < 1.03$ GeV and $547 \pm 94$ events for
$1.03 < M(KK) < 1.05$ GeV. In the mass range $1.01 < M(KK) < 1.03$ GeV, we
extract the fraction of $B^0_s$ candidates decaying into non-resonant $KK$ to be
$0.12 \pm 0.03$. The error includes the uncertainties in the signal and
background modelling. 
This excess  may be due to an $\cal S$-wave, or a non-resonant $\cal P$-wave,
or a combination of both. If we assign it entirely to the $\cal S$-wave, 
and assume it to be independent of $M(KK)$, 
we obtain the measured $\cal S$-wave fraction 
in the range  $1.01<M(K^+K^-)<1.03$ GeV  to be
$F_S = 0.12 \pm 0.03$.

\begin{widetext}

 \begin{figure}[h]
 \begin{center}
 \includegraphics*[width=0.45\textwidth]{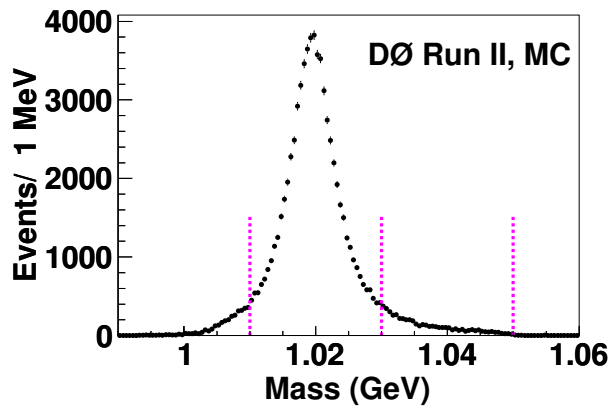}
\caption{
Invariant mass distribution of kaon pairs from the full simulation of the
decay $\phi \rightarrow K^+K^-$. Vertical dashed lines delineate the two
$M(KK)$ invariant mass bins considered.
}
 \label{fig:kkmc}
 \end{center}
 \end{figure}

 \begin{figure}[h]
 \begin{center}
 \includegraphics*[width=0.45\textwidth]{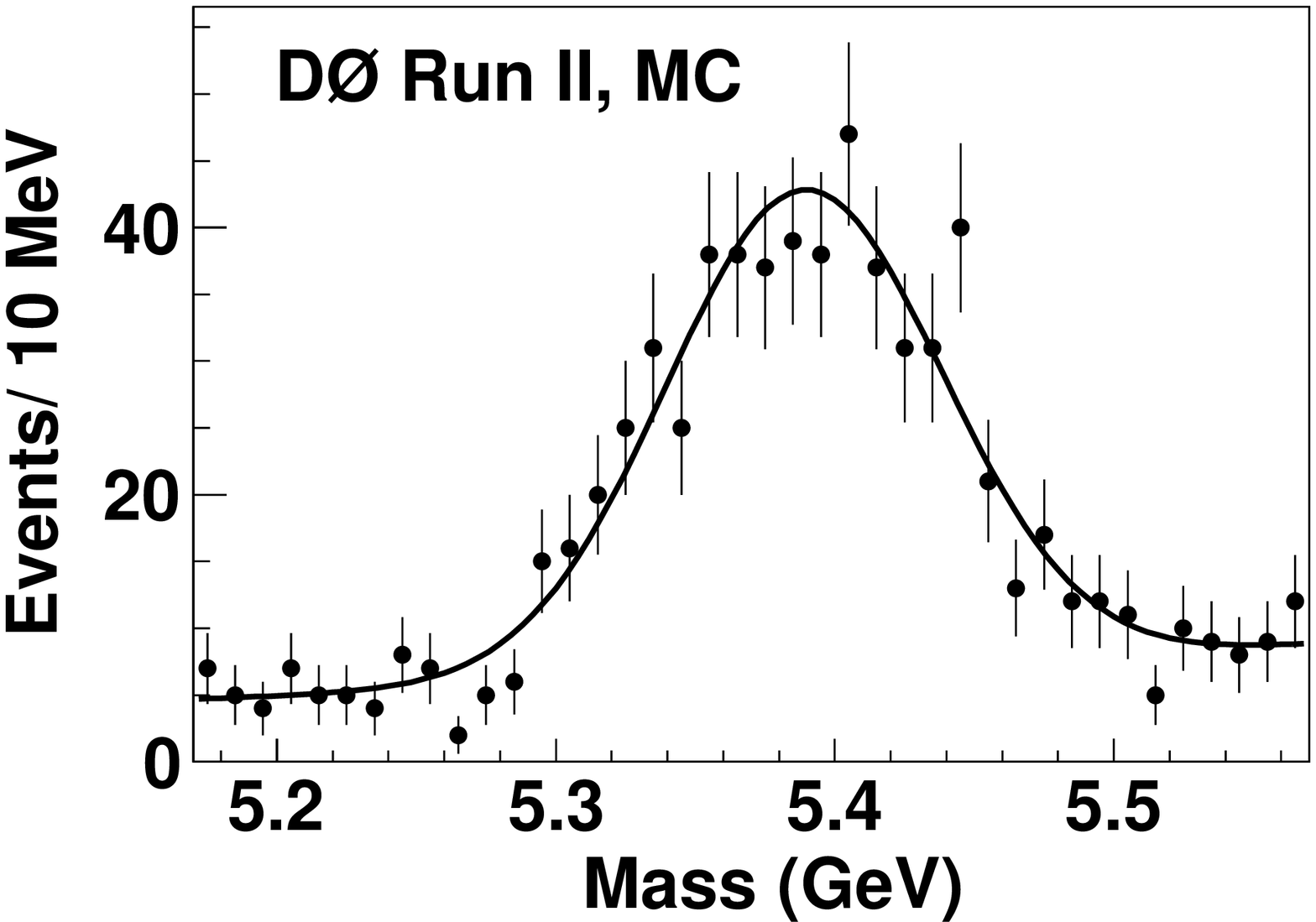}
\includegraphics*[width=0.45\textwidth]{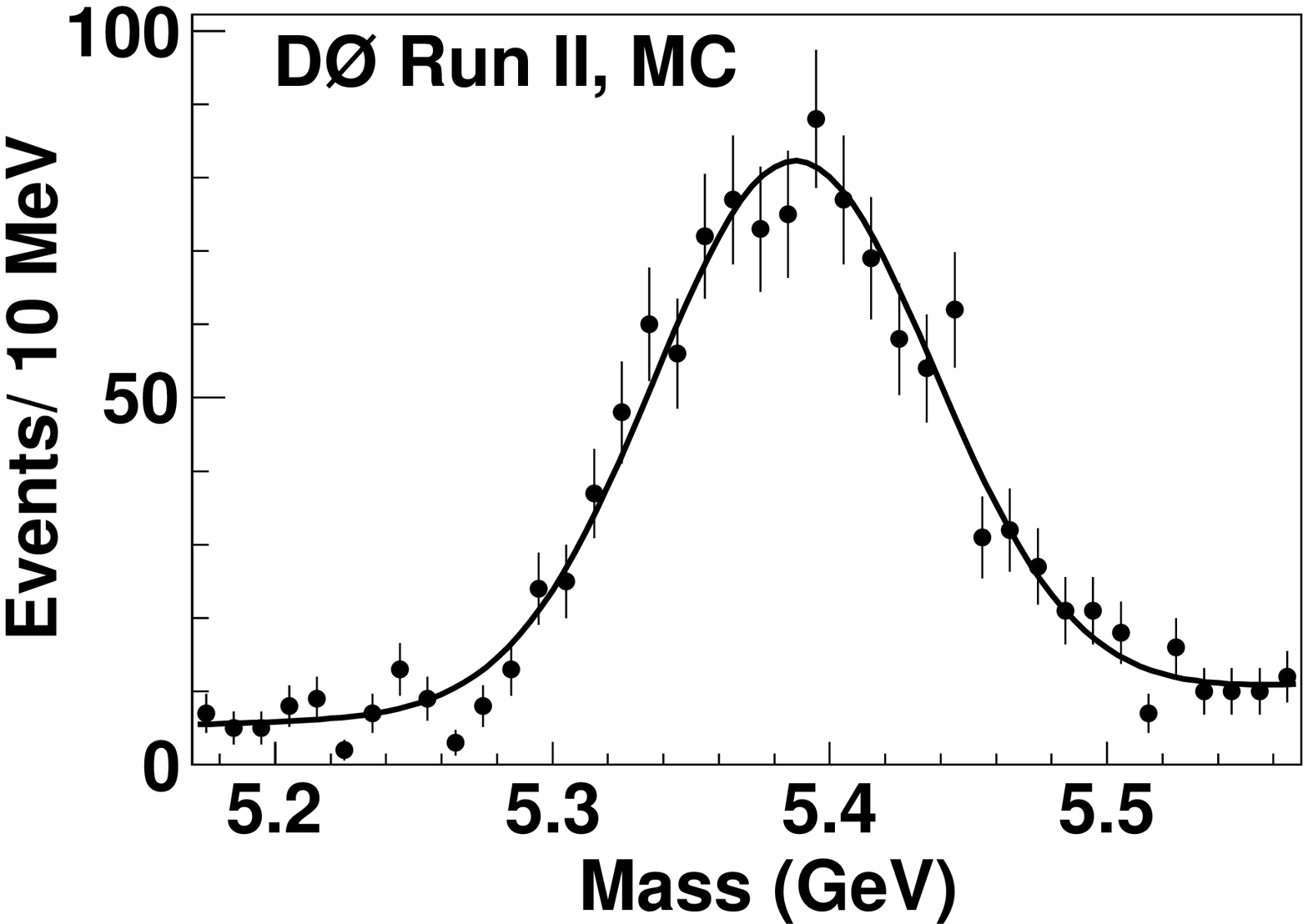}
\caption{The simulated distributions of the invariant mass 
of the \bddec\ decay products reconstructed under the \bsdec\ hypothesis
for $1.01 <M(KK)<1.03$ GeV (left) and  $1.03 <M(KK)<1.05$ GeV (right).
The curves are results of fits assuming a sum of two Gaussian functions.
}
 \label{fig:psikst}
 \end{center}
 \end{figure}

 \begin{figure}[h]
 \begin{center}
 \includegraphics*[width=0.45\textwidth]{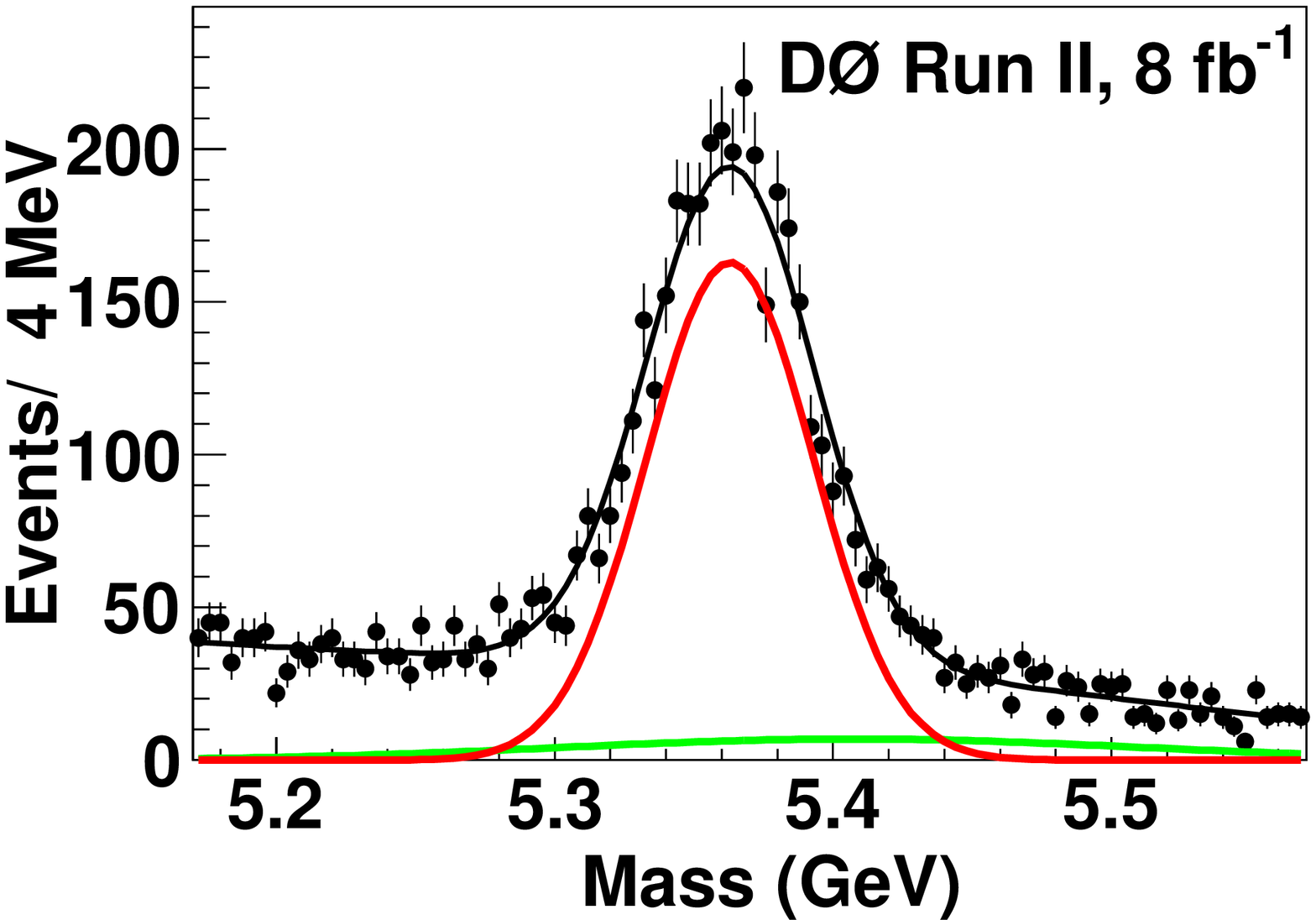}
 \includegraphics*[width=0.45\textwidth]{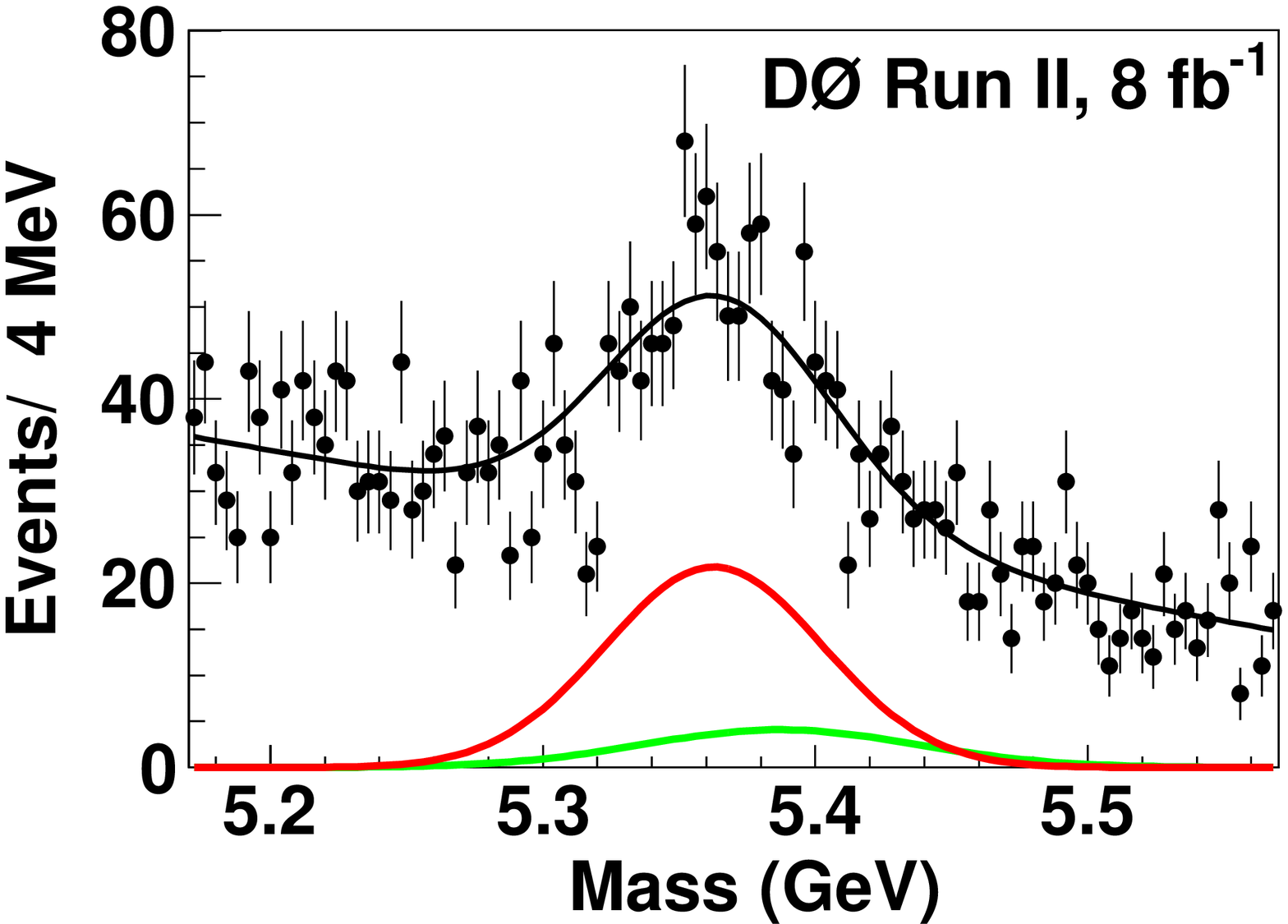}
\caption{
(color online). Invariant mass distributions of $B^0_s$ candidates with decay length $ct > 0.02$ cm
for $1.01 < M(KK) < 1.03$ GeV (left) and $1.03 < M(KK) < 1.05$ GeV (right).
Fits to a sum (black line) of a Gaussian function representing the signal
(red), an MC simulation-based template for the $B^0 \rightarrow J/\psi K^*$ reflection (green line), and
a linear function representing the background are used to extract the $B^0_s$
yield.
}
 \label{fig:bsvsmkk}
 \end{center}
 \end{figure}

\end{widetext}

\section{ $B_s^0 -\overline B_s^0$ oscillation  } 
\label{apposcill}

Under the hypothesis of  {\sl CP} conservation in the
$B_s^0$ decay, and a possible mixing-induced {\sl CP} violation,
the non-vanishing {\sl CP}-violating mixing angle should manifest itself
as a  $B_s^0 -\overline B_s^0$ oscillation with the
amplitude proportional to  $ \sin(\phi_s^{J/\psi \phi})$.
The observed time-dependent asymmetry
$\Delta N \equiv N( B_s^0) - N(\overline B_s^0)  = N_S \cdot C \cdot \sin(\phi_s^{J/\psi \phi})$,
  is  diluted by a product $C$ of several factors:
 (i) a factor of $(1 -2|A_{\perp}|^2) \cdot (1-2F_s) \approx 0.6\cdot0.7$
due to the  presence of the $CP$-odd decay,
 (ii) a factor of $\epsilon\cdot {\cal{D}}^2 \approx 0.03$
due to  the flavor tagging efficiency and accuracy, 
and (iii) a factor of  $\exp(-(\Delta M_s \sigma)^2/2) \approx 0.2$
 due to the limited  time resolution.
Thus, with  $N_S \approx 6000$ events, and $C \approx 0.0025$, we expect
$ N_S \cdot C   \approx 15$.

\begin{figure}[h]
 \begin{center}
\includegraphics*[width=0.45\textwidth]{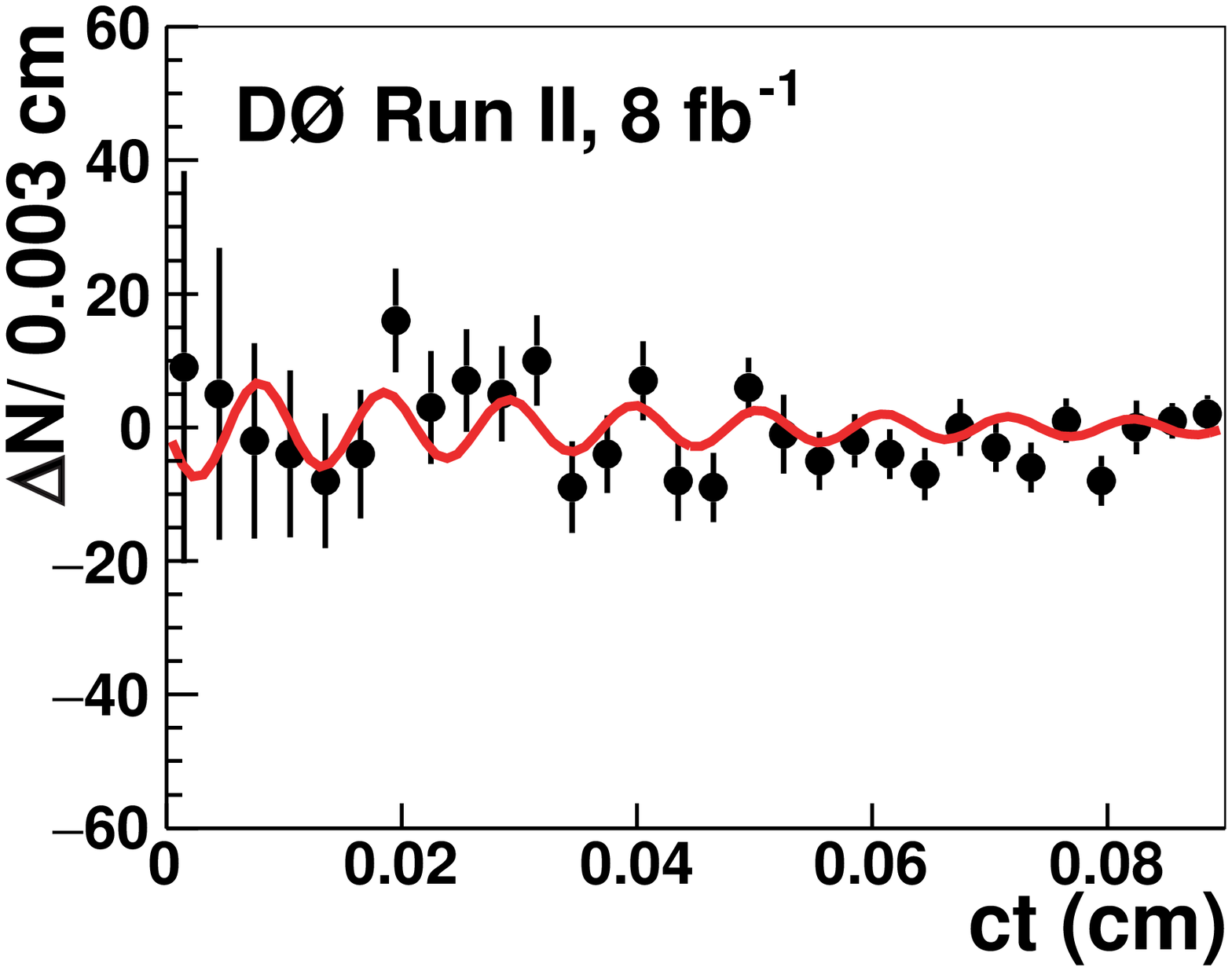}
\caption{
Proper decay length  evolution of the difference  $\Delta N = N( B_s^0) -  N({\overline B_s^0})$ 
 in the first 0.09 cm (3 ps) for the Square-cuts sample.
The curve represents the best fit to the oscillation with the frequency
of  $\Delta M_s = 17.77$ ps$^{-1}$.
}
 \label{fig:bs_osc}
 \end{center}
 \end{figure}

In Fig.~\ref{fig:bs_osc} we show the proper decay length  evolution of $\Delta N$ in the
first 90 $\mu$m, corresponding to approximately  twice the mean $B_s^0$ lifetime.
The curve represents a fit to the function
 $N_0\cdot \sin(\Delta M_s t)\cdot \exp(-t/\tau_s)$, with $N_0$ unconstrained and
with  $\Delta M_s \equiv17.77$ ps$^{-1}$.
The fit gives $N_0 = -6$ for the BDT-based sample and $-8$ for the Square-cuts
sample, with a statistical uncertainty of $\pm4$, 
corresponding to $\sin(\phi_s^{J/\psi \phi}) = N_0/N_S\cdot C \approx -0.4 \pm 0.3$.
This  one-dimensional analysis gives a result for  $\phi_s^{J/\psi \phi}$
that is consistent with the result of the full analysis.

\begin{widetext}

 \begin{figure}[h]
 \begin{center}
\subfigure[]
{\includegraphics*[width=0.40\textwidth]{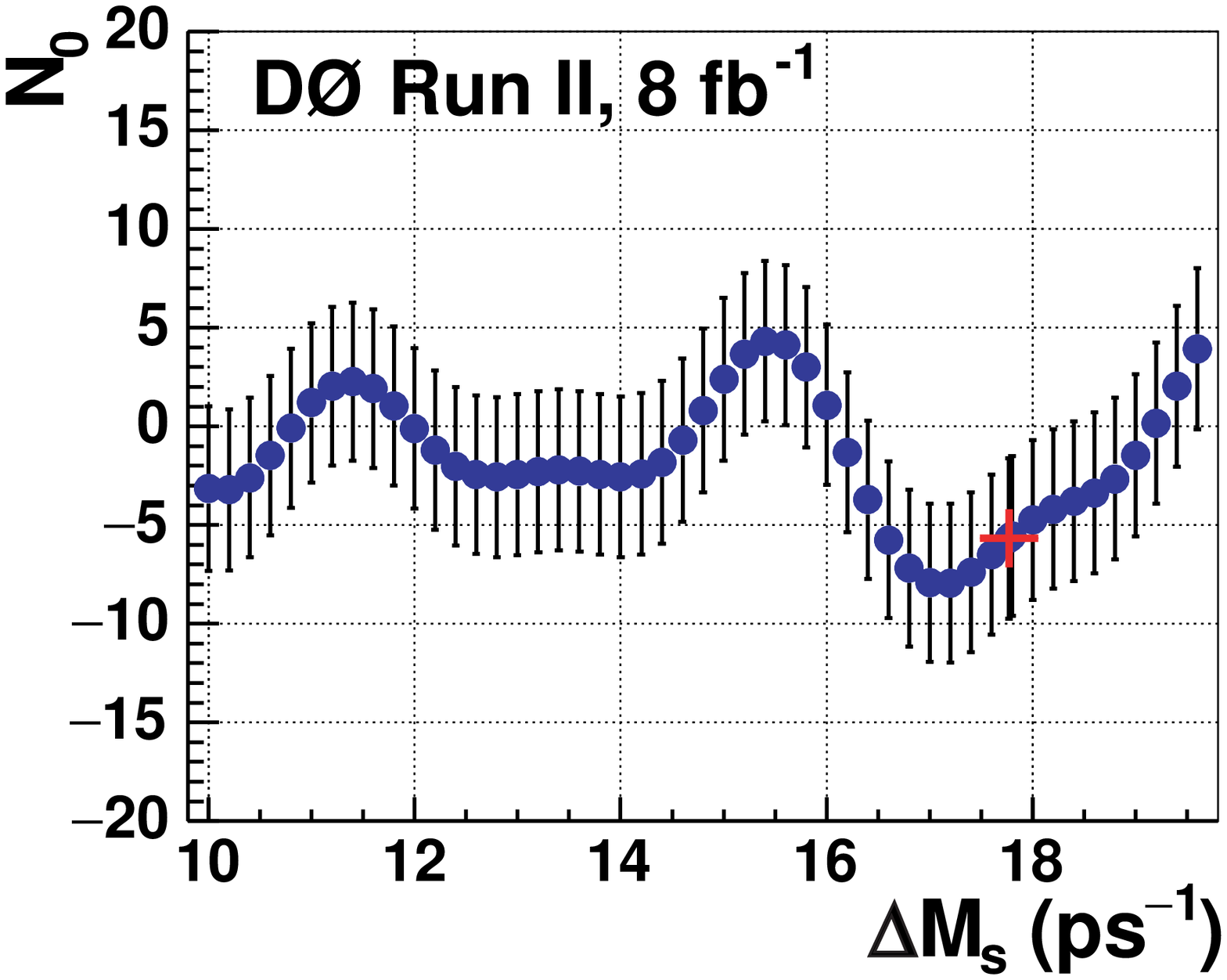}}
\subfigure[]
{\includegraphics*[width=0.40\textwidth]{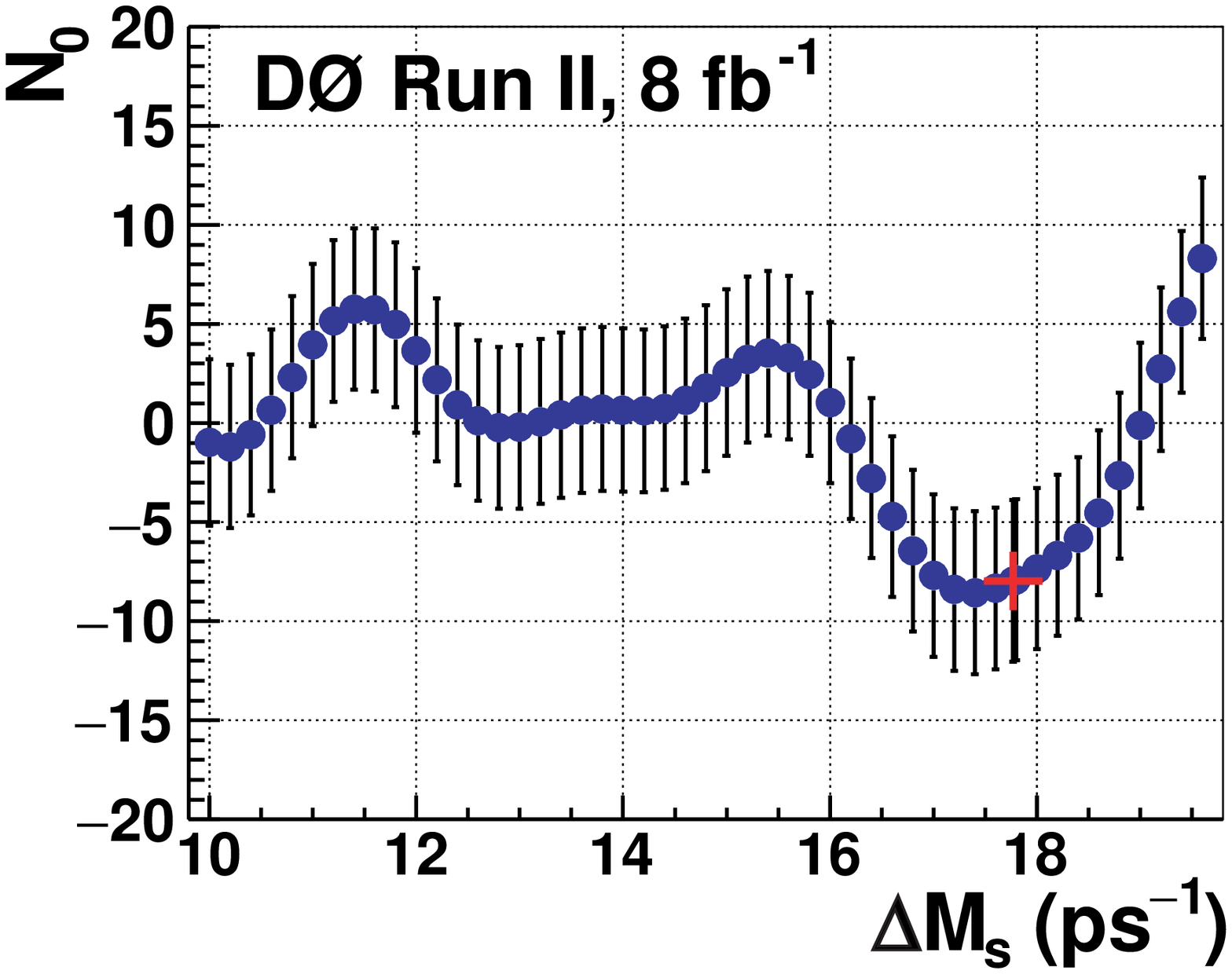}}
\caption{(color online).
The fitted magnitude of the $B_s^0 - \overline B_s^0$ oscillation
as a function of $\Delta M_s$ 
for (a) BDT selection and (b) Square cuts.
The red crosses 
correspond to  $\Delta M_s = 17.77$ ps$^{-1}$.
}
 \label{fig:bs_dms_fourier}
 \end{center}
 \end{figure}

\end{widetext}

Following the  Amplitude Method described in Ref.~\cite{moser}, we 
fit the above distributions at discrete values of
 $\Delta M_s$, and plot the fitted value of $N_0$
as a function of the probe frequency. The results are shown in  Fig.~\ref{fig:bs_dms_fourier}.
There is an undulating structure, with no significantly large  deviations
from zero. At $\Delta M_s$ near $17.77$ ps$^{-1}$
the data  prefer a negative  oscillation amplitude 
(and hence  a negative value of $\sin \phi_s^{J/\psi \phi}$).
The statistical uncertainty of the result of this simple approach
does not take into account uncertainties of the dilution factors,
related to the time resolution, CP-odd fraction, and the $\cal{S}$-wave fraction.

\end{document}